\documentclass[aps,prb,twocolumn,amsmath,amssymb,floatfix,footinbib,bibnotes,groupedaddress,longbibliography]{revtex4-1}

\usepackage{soul}
\usepackage{bm}
\usepackage{epsf}
\usepackage{amssymb}
\usepackage{esint}
\usepackage{amsmath}
\usepackage{comment}
\usepackage{graphicx}
\usepackage{rotating}
\usepackage{epsfig}
\usepackage{psfrag}
\usepackage{amsmath}
\usepackage{subfigure}
\usepackage{nicefrac}
\usepackage{bm}
\usepackage[usenames,dvipsnames]{color}
\usepackage{soul}
\usepackage{multirow}
\usepackage{titlesec}
\usepackage{tabularx}

\usepackage[colorlinks,citecolor=blue,urlcolor=blue,linkcolor=blue]{hyperref}
\newcommand{\vect}[1]{\boldsymbol{#1}}
\renewcommand{\a}{\alpha}

\newcommand*\xbar[1]{%
 \hbox{%
 \vbox{%
 \hrule height 0.3pt 
 \kern0.3ex
 \hbox{%
 \kern-0.1em
 \ensuremath{#1}%
 \kern-0.1em
 }%
 }%
 }%
}

\def \sgn#1{\text{sgn}\left( #1 \right)}
\def \up{\uparrow}
\def \down{\downarrow}

\newcommand {\apgt} {\ {\raise-.5ex\hbox{$\buildrel>\over\sim$}}\ }
\newcommand {\aplt} {\ {\raise-.5ex\hbox{$\buildrel<\over\sim$}}\ }

\begin{document}

\title{Unification of parton and coupled-wire approaches to\\ quantum magnetism in two dimensions
}

\author{\vspace{-1.5mm} Eyal Leviatan}
\affiliation{Department of Condensed Matter Physics, Weizmann Institute of Science, Rehovot 7610001, Israel}

\author{David F. Mross}
\affiliation{Department of Condensed Matter Physics, Weizmann Institute of Science, Rehovot 7610001, Israel}

\begin{abstract}
\vspace{-1.5mm}
The fractionalization of microscopic degrees of freedom is a remarkable manifestation of strong interactions in quantum many-body systems. Analytical studies of this phenomenon are primarily based on two distinct frameworks: field theories of partons and emergent gauge fields, or coupled arrays of one-dimensional quantum wires. We unify these approaches for two-dimensional spin systems. Via exact manipulations, we demonstrate how parton gauge theories arise in microscopic wire arrays and explicitly relate spin operators to emergent quasiparticles and gauge-field monopoles. This correspondence allows us to compute physical correlation functions within both formulations and leads to a straightforward algorithm for constructing parent Hamiltonians for a wide range of exotic phases. We exemplify this technique for several chiral and non-chiral quantum spin liquids.
\end{abstract}

\maketitle

\tableofcontents
\section{Introduction}
\label{sec.intro}
Determining the ground state of interacting spin systems is a quintessential problem in quantum condensed matter physics. Generic lattice Hamiltonians that are solely restricted by symmetries and locality typically yield ground states that spontaneously break one or more microscopic symmetries. Moreover, these phases exhibit only short-range entanglement and are therefore considered conventional. The tendency towards triviality can be avoided when additional ingredients, such as geometric frustration, prevent the formation of classical order and instead promote so-called quantum spin liquid (QSL) ground states~\cite{Lee2008,Balents2010,Savary+Balents2016,Zhou2017,Broholm2020}. These phases of matter are not characterized by any local order parameter. Instead, their principal feature is the existence of low-energy excitations that carry fractional quantum numbers and/or exhibit fractional statistics. In the cases of gapped QSLs, this definition can be sharpened into the notion of topological order~\cite{Wen2002}, which manifests itself in a universal nonlocal contribution to the ground-state entanglement~\cite{Hamma+Ionicioiu+Zanardi2005,Levin+Wen2006,Preskill+Kitaev2006} and a ground-state degeneracy on non-trivial manifolds.

Various solvable models have unambiguously demonstrated the possibility of QSL ground states~\cite{Kitaev2003,Kitaev2006,Yao+Kivelson2007,Chua+Yao+Fiete2011,Moessner+Raman2011}. However, the exact solution relies on rather fine-tuned interactions. In more generic models, strong evidence of QSL ground states has been found numerically using exact diagonalization~\cite{Sheng+Balents2005,Hickey2019}, quantum Monte Carlo~\cite{Isakov+Kim2006,Dang+Inglis2011,Kamiya2015}, variational Monte Carlo~\cite{Motrunich2005,Sheng2009,Block2011,Iqbal2011,Mishmash2013,Hu2016}, and density matrix renormalization group~\cite{Yan2011,Depenbrock2012,He2014,Gong2014,Zhu2015,Hu2015,Gong2015,He2017}. Finally, several recent experiments provide tantalizing evidence for QSLs. Notable examples include $\kappa-\text{ET}_2$~\cite{Shimizu2003,Kurosaki2005,Yamashita2008,Yamashita2009,Powell2011}, $\text{Pd}\left(\text{dmit}\right)_2$~\cite{Yamashita2010,Yamashita2011,Powell2011,Kato2014}, herbertsmithite~\cite{Helton2007,Han2012,Norman2016}, and $\alpha-\text{RuCl}_3$~\cite{Kasahara2018,Takagi2019}.

A major theoretical challenge towards studying QSL candidate Hamiltonians is the intrinsic nonlocality of their fractional excitations. Well-known examples of such quasiparticles are \textit{spinons}, neutral spin-$1/2$ excitation that may be bosons or fermions. In conventional magnets, two such spinons can only appear together and form spin-$1$ magnons. By contrast, individual spinons become liberated in QSLs. Both the nonlocality of spinons and their ability to appear in isolation can be encoded via an emergent gauge field under which they are charged. Individual spinons are not gauge invariant and thus not directly accessible to any (local) probe. Still, they may constitute bona fide quasiparticles when the gauge field is in a deconfined phase.

The primary workhorse for analytically describing such phases is known as parton construction (see, e.g., Refs.~\onlinecite{WenBook2004,ColemanBook2015}). Its starting point is a representation of lattice spins in terms of parton creation operators
$\psi^\dagger_{\vec r,\sigma}$, e.g., $S^+_{\vec r} = \psi^\dagger_{\vec r,\uparrow} \psi_{\vec r,\downarrow}$. In a constrained Hilbert space with exactly one parton per site, these operators can be used to faithfully represent any microscopic spin Hamiltonian.
However, the main utility of this representation, is the ability to determine possible long-wavelength theories that may---as a matter of principle---emerge from a lattice model with certain microscopic input, such as symmetries. The parton construction amounts to temporarily ignoring the constraint, which permits new mean-field \textit{Ans\"atze} that are highly non-trivial in terms of microscopic spins. Refining the mean-field theory to include fluctuations reveals the expected gauge structure. A key feature of this approach is its versatility in capturing various gapped and gapless QSLs as well as conventional phases. Its main drawback is an intrinsic difficulty to relate a given QSL phase to a specific spin model.
The most tangible connection between parton constructions and microscopic Hamiltonians is through projected wave functions~\cite{Gros1989}. However, such analyses are biased by the choice of mean-field and are computationally expensive.

Over the past years, an alternative technique that shines precisely at this Achilles' heel of parton approaches has gained in popularity. In `coupled-wire constructions', microscopic parent Hamiltonians for strongly correlated phases are constructed explicitly. Additionally, creation operators of fractional quasiparticles are expressible in terms of microscopic degrees of freedom. This approach was pioneered by Kane, Mukhopadhyay, and Lubensky for fractional quantum Hall states~\cite{Kane2002}. Recently, it has been extended to capture many other topological or strongly correlated phases, including: (i) a wider range of fractional quantum Hall~\cite{TeoKane2014,Klinovaja2014a,Meng2014a,Sagi2015a,Fuji2016,Kane2017,Fuji2019,Fontana2019,Imamura2019} and quantum spin Hall~\cite{Klinovaja2014b} states, (ii) Chern insulators~\cite{Sagi2014,Meng2014b,Santos2015} and superconductors~\cite{Mong2014,Seroussi2014,Vaezi2014,Sagi2017,Park2018,Laubscher2019}, (iii) exotic surface states of symmetry protected topological phases~\cite{Mross2015,Sahoo2016}, (iv) correlated states in twisted bilayer graphene~\cite{Wu2019},
(v) three-dimensional topological orders~\cite{Meng2015b,Sagi2015b,Iadecola2016,Sagi2018,Iadecola2019,Raza2019}, (vi) Weyl semimetals~\cite{Vazifeh2013,Meng2016}, and (vii) several QSLs~\cite{Nersesyan2003,Meng2015a,Gorohovsky2015,Patel+Chowdhury2016,Huang2016,Huang2017,Lecheminant2017,Chen2017,Rodrigo2018,Chen2019}. Furthermore, coupled-wire methods have been used to construct anyon models~\cite{Oreg2014,Stoudenmire2015}, classify symmetry protected topological phases~\cite{Lu2012,Neupert2014}, and derive field-theoretic dualities in $2+1$ dimensions~\cite{Mross2016,Mross2017}. For an introduction and overview of this approach, see also Ref.~\onlinecite{Meng2019}. 

A natural starting point for applying this technique to QSLs is given by a spin system where all couplings in the $\hat{y}$ direction, say, have been switched off. The resulting model may be profitably viewed as an array of one-dimensional spin chains (aka wires) along the $\hat{x}$ direction. Each spin chain is then taken to form a gapless one-dimensional QSL. The corresponding long-wavelength degrees of freedom form a coarse-grained basis for reintroducing interwire couplings. When these interactions are strongly relevant in the renormalization group sense, they may drive the system into a bona fide two-dimensional phase. Unfortunately, no general principle for the construction of parent Hamiltonians within this framework is known. Instead, it has to be done on a laborious case-by-case basis. Consequently, only a limited number of QSLs have been accessed in this manner~\cite{Nersesyan2003,Meng2015a,Gorohovsky2015,Patel+Chowdhury2016,Huang2016,Huang2017,Lecheminant2017,Chen2017,Rodrigo2018,Chen2019}.

\begin{figure} [tb]
\centering
 \includegraphics[width=\columnwidth]{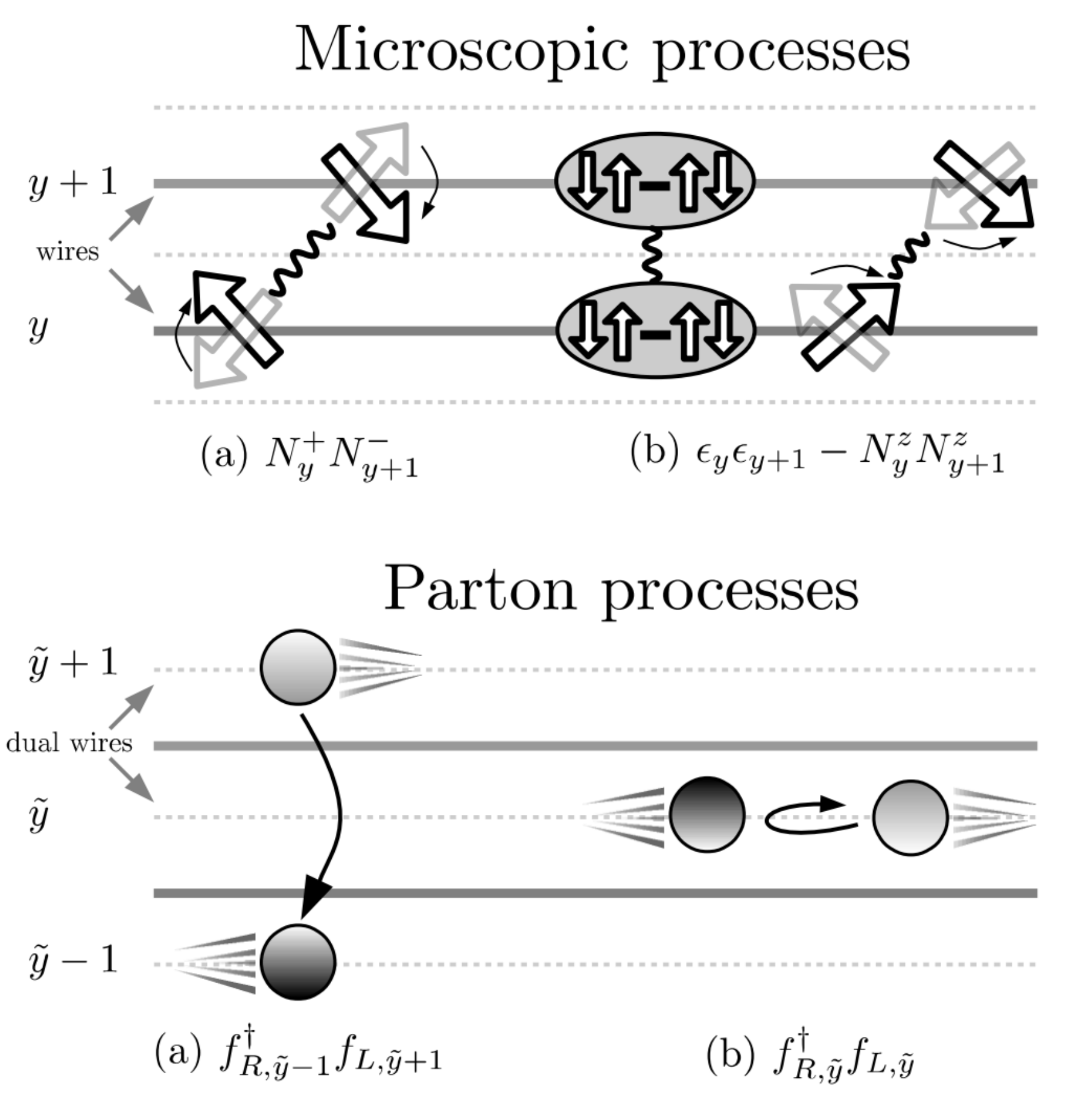}
 \caption{Microscopic interwire processes and the fermionic-parton counterparts onto which they map: (a) spin exchange between neighboring wires translates to fermion hopping by two dual wires, (b) the indicated combination of dimer-dimer and Ising interactions translates to fermion umklapp scattering on the corresponding dual wire.
 \label{fig:processes}}
\end{figure}

We present a framework that unifies the two approaches based on the well-known particle-vortex duality of bosons in $2+1$ dimensions~\cite{Peskin1978,Dasgupta+Halperin1981,Fisher+Lee1989}. Its recent implementation in the coupled-wire formalism~\cite{Mross2017} allows us to transcribe field-theoretic insights into explicit models and thereby achieve the desired connection between parton and coupled-wire methods. We find remarkably simple relationships between microscopic degrees of freedom, such as the N\'eel vector, $\vec N_y(x)$, or the valence-bond operator, $\epsilon_y(x)$, and parton operators in a suitable gauge. For example, we show that
\begin{align}
 & \sum_{\tilde y }f_{\tilde y ,\sigma,\chi_\sigma}^\dagger f_{\tilde y -2,\sigma,\overline{\chi}_\sigma} + \text{H.c.} = \sum_y N_{y+1}^+ N_{y}^- +\text{H.c.} ~,\\ 
 & \sum_{\tilde y } f_{\tilde y ,\sigma,R}^\dagger f_{\tilde y ,\sigma,L} + \text{H.c.} = \sum_y \left[\epsilon_{y+1} \epsilon_{y} - N_{y+1}^z N_{y}^z\right]~,
\end{align}
where $f^\dagger_{\tilde y ,\sigma,\chi}$ creates a fermionic parton with chirality $\chi=R/L$ and spin $\sigma=\up/\down$
on the `dual' wire $\tilde y =y+1/2$. In the first equation, $\chi_\sigma=R(L)$ for $\sigma=\up(\down)$ and $\overline{\chi}_\sigma = L(R)$ is the opposite chirality. These relations are represented graphically in Fig. \ref{fig:processes}.

\begin{figure} [h]
\centering
 \includegraphics[width=\columnwidth]{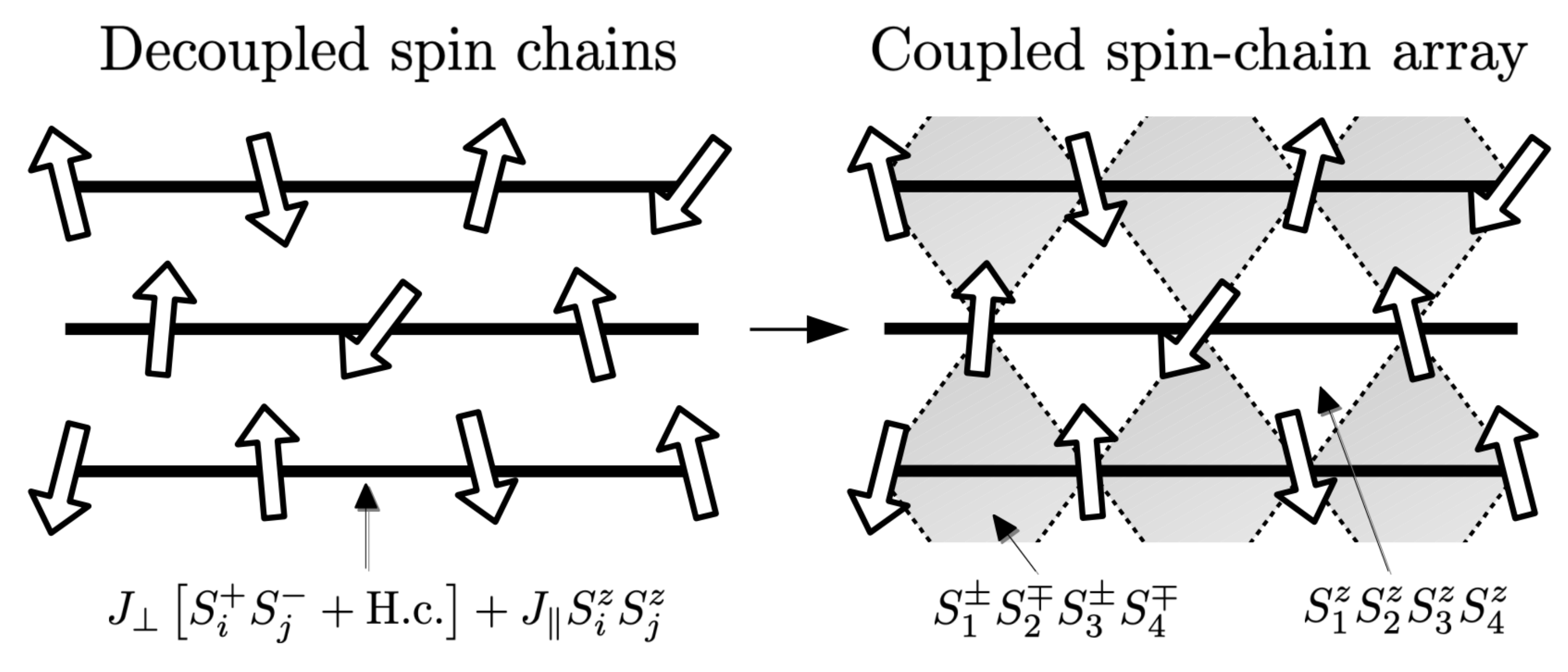}
 \caption{The parent Hamiltonian of a $\mathbb{Z}_2$ spin liquid can be obtained by translating the coupled-wire model that realizes a BCS superconductor of fermionic partons. The gray and white diamonds represent the indicated interaction between spins at their corners. These terms are reminiscent of those realizing Kitaev's toric code~\cite{Kitaev2003}. However, the present model conserves $S^z$ and realizes a phase with dynamical matter fields. See Sec.~\ref{subsubsec.FP_phases_BCS-Z2} for the derivation and a detailed discussion.
 \label{fig:z2}}
\end{figure}

Crucially, any coupled-wire model that separately conserves the two parton species maps onto a \textit{local} spin model. Parent spin Hamiltonians for a wide range of non-trivial phases can thus be generated by constructing weakly correlated two-dimensional band insulators or superconductors of partons. An example of such a model is shown in Fig.~\ref{fig:z2}, which illustrates the spin Hamiltonian for a $\mathbb{Z}_2$ QSL obtained from a superconductor of fermionic partons. Moreover, coupled-wire models for various strongly correlated states of bosons and fermions are also known in the literature~\cite{TeoKane2014,Klinovaja2014a,Sagi2015a,Fuji2016,Kane2017,Fuji2019,Imamura2019,Klinovaja2014b,Sagi2014,Meng2014b,Santos2015,Mong2014,Vaezi2014,Sagi2017,Park2018,Laubscher2019,Mross2015,Sahoo2016}. Each of these corresponds to a local spin model as well. These models realize QSLs that are not describable by a parton mean-field ansatz but require a further fractionalization of the partons. 

In addition to constructing parent Hamiltonians for specific gapped phases, we use the exact transformation between spins and partons to derive the gauge theory for the latter. We explicitly relate monopoles in the emergent gauge field to spin operators and determine the action of microscopic symmetries. These properties demonstrate the desired unification of partons and coupled wires. They, moreover, indicate that the same approach may be extended to gapless states in the future.

The rest of this paper is organized as follows: In Sec.~\ref{sec.PC}, we briefly summarize some key elements of parton constructions. In Sec.~\ref{sec.CW}, we review some well-known properties of spin-$1/2$ chains and their description using bosonization. We then describe how a two-dimensional easy-plane antiferromagnet (AFM), valence bond solid (VBS), and Ising-AFM are realized upon introducing interwire couplings and discuss their topological defects. In Sec.~\ref{sec.BP} and Sec.~\ref{sec.FP}, we introduce bosonic and fermionic partons, respectively, as combinations of the aforementioned defects. We derive the gauge theory that results when the spin-model is rewritten in terms of these nonlocal degrees of freedom and tabulate their symmetries. We then analyze several phases in both their spin and parton representations, with particular attention to the fate of the emergent gauge field in the latter. We conclude with a summary of our results and an outlook on possible extensions in Sec.~\ref{sec.discussion}. Finally, the Appendices discuss nonuniversal terms that are required for microscopically exact mappings but do not affect long-distance properties. They also reproduce several known properties of the pertinent gauge theories within concrete wire-based calculations. 

\section{Parton construction}
\label{sec.PC}
In this section, we briefly review partons in the context of interacting spin systems. Two widely-used parton constructions are based on the Schwinger-boson or Abrikosov-fermion representations of spin-1/2, i.e.,
\begin{align} 
\label{eq.PC_definitions}
 S^+ = \psi^\dagger_\up \psi_\down~, \qquad S^z = \frac{1}{2} \left(\psi^\dagger_\up \psi_\up - \psi^\dagger_\down \psi_\down \right)~.
\end{align}
Here $\psi_\sigma$ are either bosonic or fermionic annihilation operators subject to the constraint of a single parton per site, $ \hat n \equiv \sum_\sigma \psi^{\dagger}_\sigma \psi_\sigma = 1$. The expression of spins through partons has built-in redundancy, i.e., physical spin operators are invariant under phase rotations $\psi_\sigma \rightarrow \psi_\sigma e^{i \phi}$.

\subsection{Mean-field theory}
\label{subsec.PC_MF}
The representation Eq.~\eqref{eq.PC_definitions} allows expressing generic Hamiltonians with $S^z$-conserving two-spin interactions as
\begin{align}
\label{eq.PC_MF_Hamiltonian}
 \nonumber H = & \sum\nolimits_{\boldsymbol{r},\boldsymbol{r}'} J^\perp_{\boldsymbol{r},\boldsymbol{r}'} \psi^\dagger_{\up,\boldsymbol{r}} \psi_{\up,\boldsymbol{r}'} \psi^{\dagger}_{\down,\boldsymbol{r}'} \psi_{\down,\boldsymbol{r}} + \text{H.c.} \\
 & + \sum\nolimits_{\boldsymbol{r},\boldsymbol{r}'} J^z_{\boldsymbol{r},\boldsymbol{r}'}\sum\nolimits_\sigma \psi^\dagger_{\sigma,\boldsymbol{r}} \psi_{\sigma,\boldsymbol{r}'} \psi^\dagger_{\sigma,\boldsymbol{r}'} \psi_{\sigma,\boldsymbol{r}}~.
\end{align}
The absence of quadratic terms implies that $\hat n_{\vect r}$ on each site is conserved, i.e., individual partons are immobile. Still, parton-hopping may emerge at low energies, which can be captured by mean-field \textit{Ans\"atze} such as $\left< \psi^\dagger_{\sigma,\boldsymbol{r}} \psi_{\sigma',\boldsymbol{r}'} \right> = \chi_{\boldsymbol{r},\boldsymbol{r}'} \delta_{\sigma,\sigma'}$. The corresponding mean-field Hamiltonian is
\small
\begin{align}
\label{eq.PC_MF_MF0-Hamiltonian}
 H_{\text{MF}} = \sum\nolimits_{\boldsymbol{r},\boldsymbol{r}',\sigma} \left(\chi_{\boldsymbol{r},\boldsymbol{r}'} \psi^{\dagger}_{\sigma,\boldsymbol{r}'} \psi_{\sigma,\boldsymbol{r}} + \text{H.c.}\right) + \mu \sum\nolimits_{\boldsymbol{r}} \hat{n}_{\boldsymbol{r}}~,
\end{align}
\normalsize
where $J^\perp_{\boldsymbol{r},\boldsymbol{r}'}$ and $J^z_{\boldsymbol{r},\boldsymbol{r}'}$ were absorbed into the definition of $\chi_{\boldsymbol{r},\boldsymbol{r}'}$. We also included a chemical potential that enforces, on average, the now-violated single-occupancy constraint. The mean-field Hamiltonian also breaks the local redundancy: under $\psi_{\sigma,\boldsymbol{r}} \rightarrow \psi_{\sigma,\boldsymbol{r}}e^{i \phi_{\boldsymbol{r}}}$ the mean-field parameter acquires a phase 
\begin{align}
\label{eq.PC_gauge_chi-transformation}
 \chi_{\boldsymbol{r},\boldsymbol{r}'} \rightarrow \chi_{\boldsymbol{r},\boldsymbol{r}'} \exp\left[-i\left(\phi_{\boldsymbol{r}} - \phi_{\boldsymbol{r}'}\right)\right]~.
\end{align}
Both flaws can be remedied by allowing for local fluctuations in $\chi_{\boldsymbol{r},\boldsymbol{r}'}$ and $\mu$ on top of the mean-field values.

\subsection{Compact gauge fluctuations}
\label{subsec.PC_gauge}

The replacement $\chi_{\boldsymbol{r},\boldsymbol{r}'}\rightarrow\chi_{\boldsymbol{r},\boldsymbol{r}'}\exp\left[-i a_{\boldsymbol{r},\boldsymbol{r}'}\right]$ can restore the redundancy if $a_{\boldsymbol{r},\boldsymbol{r}'}$ acquires a shift under phase rotations. Specifically, it must transform like the spatial components of a gauge field, i.e., $a_{\boldsymbol{r},\boldsymbol{r}'} \rightarrow a_{\boldsymbol{r},\boldsymbol{r}'} + \phi_{\boldsymbol{r}} - \phi_{\boldsymbol{r}'}$. Allowing fluctuations of $a_{\boldsymbol{r},\boldsymbol{r}'}$ and encoding fluctuations of the chemical potential via a temporal component, $\mu \rightarrow \mu + i a_{0,\boldsymbol{r}}$, results in the (Euclidean) action
\begin{align}
\label{eq.PC_gauge_MF1-Lagrangian}
{\cal S}_\text{parton} &= \int_\tau\sum\nolimits_{\boldsymbol{r},\sigma} \psi^\dagger_{\sigma, \boldsymbol{r}} \left(\partial_\tau - i a_{0,\boldsymbol{r}} - \mu \right) \psi_{\sigma, \boldsymbol{r}} \\
 +&\int_\tau\sum\nolimits_{\boldsymbol{r},\boldsymbol{r}',\sigma} \left( \chi_{\boldsymbol{r},\boldsymbol{r}'} \psi^\dagger_{\sigma,\boldsymbol{r}'} e^{-i a_{\boldsymbol{r},\boldsymbol{r}'}} \psi_{\sigma,\boldsymbol{r}} + \text{H.c.}\right)~. \nonumber 
\end{align}

A crucial aspect of this theory is its periodicity in $a_{\vect r,\vect r'}$, which permits `monopole' events where the flux $\vect \Delta \times \vect a$ changes by $2\pi$. The gauge field is thus \textit{compact}~\cite{Polyakov1987}. Equivalently, any \textit{induced} Maxwell term generated by integrating out $\psi$ inherits the periodicity of the minimal coupling and is of the form $\cos\left(\vect \Delta \times \vect a \right)$. By contrast, a bare Maxwell term $\propto(\vect \Delta \times \vect a)^2$, which arises, e.g., in the particle-vortex duality~\cite{Peskin1978,Dasgupta+Halperin1981,Fisher+Lee1989}, would exclude isolated monopoles; for a recent review including modern developments, see Ref.~\onlinecite{Senthil+Son+Wang+Xu2019}. The same conclusion may be reached after taking the continuum limit by analyzing the microscopic operator that corresponds to the emergent gauge flux. In the present case, it is given by the scalar spin chirality~\cite{Wen+Zee1989} 
\begin{align}
\label{eq.PC_gauge_scalar-spin-chirality}
 \vec{S}_{\boldsymbol{r}_1} \cdot \left( \vec{S}_{\boldsymbol{r}_2} \times \vec{S}_{\boldsymbol{r}_3}\right) \underset{\text{cont. limit}}{\longrightarrow} \vect{\nabla} \times \vect{a}~,
\end{align}
which is not conserved in most microscopic spin models. Consequently, monopole events exist in the corresponding gauge theory. By contrast, in particle-vortex duality, the gauge flux is identified with the microscopically conserved boson density. Monopoles are thus absent in that case, i.e., the gauge field is non-compact.

To study compact gauge theories, it is often convenient to separate the gauge field into a monopole-free part, $\vec{a}^0$, and a singular part, $\vec{a}^{\mathcal{M}}$, that contains monopoles of strength $2\pi q_i$ at space-time points $\vec r_i = (\tau_i,\vect r_i)$, i.e.,
\begin{align}
\label{eq.PC_gauge_singular-gauge}
 \vec \nabla \cdot (\vec \nabla \times \vec a^{\mathcal{M}}) =2\pi \sum\nolimits_i q_i \delta(\vec r - \vec r_i)~.
\end{align}
This separation is useful, e.g., for assessing the relevance of monopoles in the presence of matter fields $\psi$. The monopole-monopole correlation function is
\begin{align}
\label{eq.PC_gauge_monopole-corr}
 \nonumber C^{\mathcal{M}}_{\vec r_1-\vec r_2} & = \frac{\int \mathcal{D} \left[\psi, a_\mu^{0}\right] e^{-S\left[\psi,a_\mu^0+a_\mu^{\mathcal{M}}\right]}}{\int \mathcal{D}\left[\psi, a_\mu^{0}\right] e^{-S\left[\psi,a_\mu^0\right]}} \\
 & = \left< e^{S\left[\psi,a_\mu^0\right]-S\left[\psi,a_\mu^0+a_\mu^{\mathcal{M}}\right]}\right>_0~.
\end{align}
Here, the singular gauge field $\vec{a}^{\mathcal{M}}$ must satisfy Eq.~\eqref{eq.PC_gauge_singular-gauge} with two opposite monopoles, $q_1 = -q_2 = 1$, but is otherwise arbitrary. 

While conceptually straightforward, the evaluation of $C^{\mathcal{M}}$ based on this formula is a formidable task, only achievable in specific limits. Extensive studies of compact gauge theories in $2+1$ dimensions have found that their infrared behavior falls into one of three categories:
\begin{enumerate}
\item \textit{Confining}: When all matter fields trivially gapped, the low-energy theory is a pure gauge theory. There, monopoles always proliferate and result in confinement~\cite{Polyakov1987}.
\item \textit{Gapped}: Confinement is avoided when the gauge field becomes massive due to the formation of a condensate (Higgs mass) or a topologically non-trivial insulator (Chern-Simons mass). 
\item \textit{Gapless}: The presence of gapless matter, e.g., in the form of a large number of Dirac-fermion species or a Fermi surface, can render monopoles irrelevant. Such systems behave like non-compact gauge theories~\cite{Ioffe+Larkin1989,Hermele2004,SSLee2008}.
\end{enumerate}

\section{Coupled-wire approach}
\label{sec.CW}
Consider a two-dimensional array of antiferromagnetic spin-1/2 chains (aka wires) with conserved $S^z$. The long-wavelength properties of each chain, labeled by an integer $y$ and extending along $\hat x$, can be efficiently described using Abelian bosonization~\cite{GiamarchiBook2004,GogolinBook2004}. This framework is, moreover, convenient for including interwire couplings and studying their effect. We thus introduce a pair of conjugate variables $\Theta_y(x) , \Phi_y(x)$ and describe the spin-chain array by a Euclidean path integral ${\cal Z} = \int {\cal D}\Phi {\cal D} \Theta e^{-\cal S}$. 
The action ${\cal S}$ contains both intrawire terms and couplings between different chains; both will be specified below.
In our convention, smooth and staggered components of the microscopic spins, $\vec S_{\vect r} = \vec J_{\vect r} +(-1)^{x+y} \vec N_{\vect r}$, are encoded as
\begin{equation}
\begin{alignedat}{2}
\label{eq.CW_Bosonization-identities}
 J^z_{\vect r} & = \frac{1}{\pi}\partial_x \Theta_y~, && N^z_{\vect r} \sim\sin \left(2 \Theta_y\right), \\
 J^+_{\vect r}&\sim e^{i\Phi_y} \sin\left(2\Theta_y\right)~,\qquad && N^+_{\vect r}\sim e^{i\Phi_y}~.
\end{alignedat}
\end{equation}
The transformation of $\Theta$ and $\Phi$ under microscopic symmetries can be readily deduced from these expressions; we summarize the ones pertinent to this work in Table \ref{tab.CW_coupled_symmetries}. 

\begin{center}
\begin{table} [tb]
 \caption{Symmetry action on bosonized degrees of freedom, with asterisks denoting anti-unitary symmetries ($i\rightarrow -i$).}
 \begin{tabularx}{\linewidth}{l l r@{}l r@{}l}
 \hline\hline
 \textbf{Symmetry}& & &$\Phi$ & &$\Theta$ \tabularnewline
 \hline
 $U(\alpha)$ & U(1) spin rot. & &$\Phi + \alpha$ & &$\Theta$ \tabularnewline
 $\Pi_y$ & $\pi$ rot. around $S_y$&\hspace{2pt} $-$&$\Phi+ \pi$ & $-$&$\Theta$ \tabularnewline
 $T_x:x\rightarrow x+1$ & $x$-translation& &$\Phi + \pi$ & &$\Theta + \frac{\pi}{2}$ \tabularnewline
 $T_y:y\rightarrow y+1$ & $y$-translation & &$\Phi+ \pi$ & &$\Theta+ \frac{\pi}{2} $ \tabularnewline
 $\mathcal{I}_x:x\rightarrow-x$ & $x$-inversion (site) & &$\Phi$ & \quad$-$ &$\Theta + \frac{\pi}{2}$ \tabularnewline
 $\mathcal{I}_y:y\rightarrow -y+1$ \hspace{2pt} & $\tilde{y}$-inversion & & $\Phi+ \pi$ & &$\Theta+ \frac{\pi}{2}$ \tabularnewline
 $\mathcal{T}^*: \tau \rightarrow -\tau$ & Time-reversal (TR) & &$\Phi - \pi$ & $-$&$\Theta$ \tabularnewline
 $\mathcal{T}'^* \equiv T_y \cdot \mathcal{T}$ & AFM-TR & &$\Phi$ & $-$&$\Theta-\frac{\pi}{2}$ \tabularnewline
 \hline\hline
 \end{tabularx}
 \label{tab.CW_coupled_symmetries}
\end{table}
\end{center}
\subsection{Decoupled spin-1/2 chains} 
\label{subsec.CW_decoupled}
We describe the long-wavelength properties of each spin chain by ${\cal L}_\text{chain}={\cal L}_\text{LL}+{\cal L}_{4\pi}$, where \small
\begin{align}
\label{eq.CW_decoupled_LL}
 {\cal L}_\text{LL} &=\frac{i}{\pi}\partial_x \Theta_y \partial_\tau \Phi_y+ \frac{v K}{2\pi}\left(\partial_x \Phi_y \right)^2 + \frac{v}{2\pi K}\left(\partial_x \Theta_y \right)^2~, \\
\label{eq.CW_4pi-phase-slip}
 {\cal L}_{4\pi} &= g_{4\pi} \cos\left(4\Theta_y\right)~.
\end{align}
\normalsize
The Luttinger-liquid Lagrangian ${\cal L}_\text{LL}$ is perturbed by the non-linear $ {\cal L}_{4\pi}$, which introduces $4\pi$ phase slips into $S^+$. Its scaling dimension at the Gaussian fixed point, $\Delta_{4\pi} = 4K$, determines the nature of the phase. For $K > 1/2$ phase slips are irrelevant, and the ground state is gapless, with power-law correlations in $S^{x,y,z}$ and the VBS order parameter
\begin{align}
\label{eq.CW_decoupled_dimerization}
 \epsilon_{\vect r} = \left(-1\right)^x S^+_{\vect r} S^-_{\vect r + \hat x} + \text{H.c.} \sim \left(-1\right)^y \cos\left(2\Theta_y\right)~.
\end{align}
An array of spin-chains in this phase is easily destabilized by various types of interwire couplings and will thus be our starting point for accessing two-dimensional phases.

\begin{figure}[tb]
\centering
 \includegraphics[width=\columnwidth]{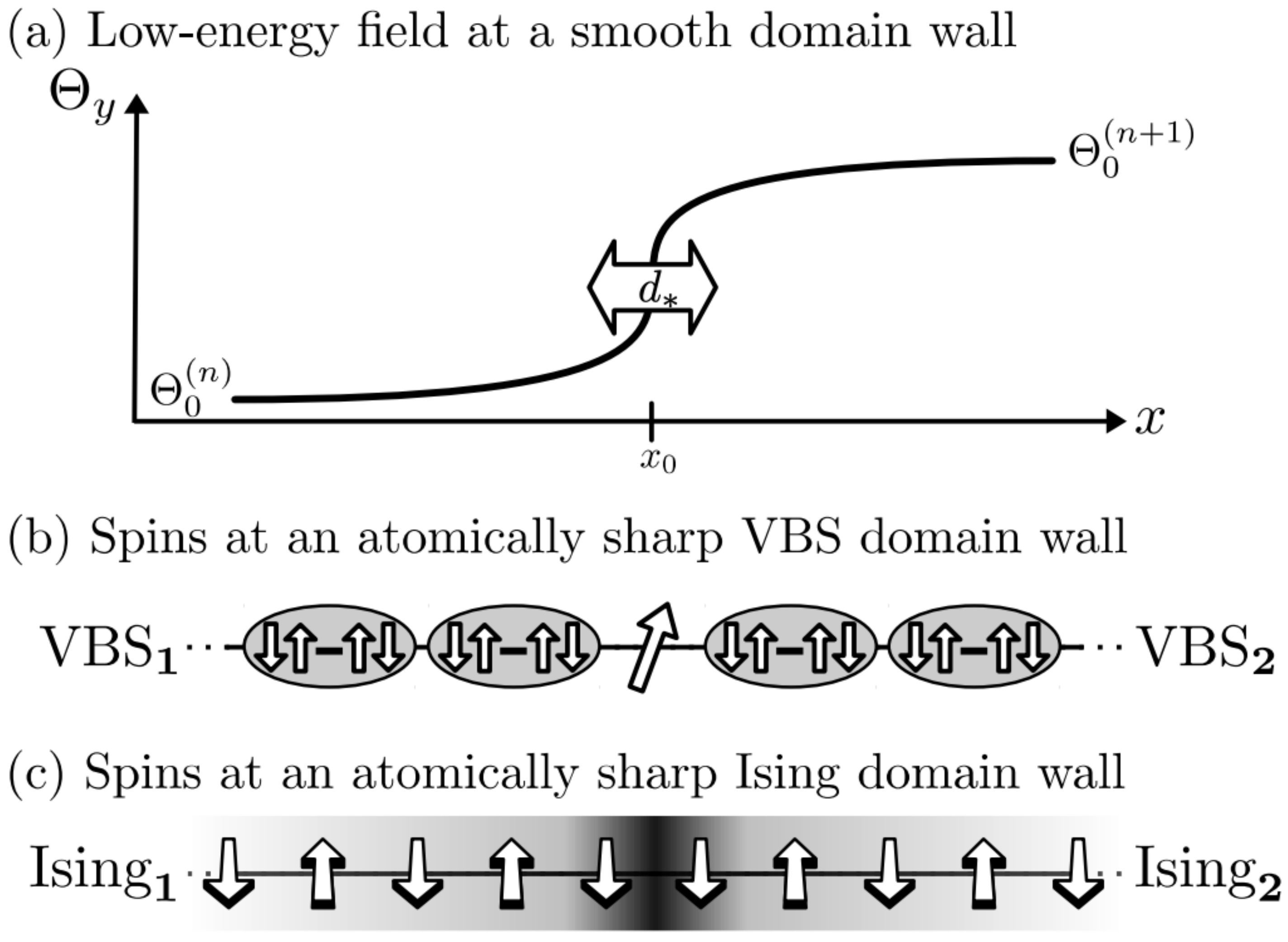}
 \caption{(a) The fundamental quasiparticle excitations of a one-dimensional VBS and an Ising-AFM are domain walls between the two degenerate ground states. In both cases, the bosonic variable $\Theta(x)$ changes by $\pi/2$, from one minimum to the next, over a distance set by the correlation length $d_*$. Consequently, such defects carry spin-$1/2$ and cost a finite energy $\propto 1/d_*$. When the correlation length is comparable to the lattice spacing, the domain wall can be depicted in terms of microscopic spins as shown in (b) for a VBS, and (c) for an Ising-AFM. The spin-$1/2$ associated with this excitation becomes readily apparent in this limit.
 \label{fig:dw}}
\end{figure}

It is, however, useful to briefly review the opposite case of relevant ${\cal L}_{4\pi}$. We begin by introducing a dimensionless coupling constant, whose bare value at the microscopic length $d_0$ is $\tilde g_{4\pi}\equiv 16 \pi K d_0^2 g_{4\pi}/v$. For $K < 1/2$ it grows under renormalization and reaches order unity at a length $d_*$. For small $|\tilde g_{4\pi}|$ and $K$, the scaling dimension of the cosine implies $d_* \simeq d_0 |\tilde g_{4\pi}|^{1/(4K-2)}$. Beyond this scale, each field $\Theta_y$ become trapped around a minimum of the cosine. To describe the low-energy fluctuations we, therefore, expand the cosine to quadratic order and write
\begin{align}
 {\cal L}_{4\pi} &= \frac{\tilde g_{4\pi} v}{16 \pi K d_0^2} \cos\left(4\Theta_y\right)\rightarrow \frac{ v}{2 \pi K d_*^2} \left(\Theta_y-\Theta_{y,0}^{(n_y)}\right)^2.
\end{align}
Here $\Theta^{(n_y)}_{y,0}$ denote the minima of the cosines, labeled by the integers $n_y$. To identify the ground state, it is sufficient to replace $\Theta_y\rightarrow \Theta^{(n_y)}_{y,0}$ in all observables. For negative $\tilde g_{4\pi}$, the minima are at $\Theta^{(n_y)}_{y,0}= \pi n_y/2$, and there is VBS order $\epsilon_{\vect r} \propto (-1)^{n_y+y}$. For positive $\tilde g_{4\pi}$ we instead have $\Theta_{y,0}^{(n_y)}= \pi n_y/2 + \pi/4$ reflecting Ising-N\'eel order, i.e., $N^z \propto (-1)^{n_y}$. We denote the two possible ground states, $n_y$ even and $n_y$ odd, by VBS$_{1}$(Ising$_{1}$) and VBS$_{2}$(Ising$_{2}$), respectively.
In both cases, they are related by $x$-translations (cf.~Table \ref{tab.CW_coupled_symmetries}) and one is selected spontaneously when that symmetry is broken. The universal properties of these gapped phases are insensitive to the value of $d_*$, which may be viewed as a new parameter that replaces $\tilde g_{4\pi}$.

Consider now a domain wall where the state of the $y_0$th wire is characterized by $n_{y_0}$ for $x<x_0$ and by $n_{y_0}+1$ for $x>x_0$. Near the domain wall, $\Theta_y$ changes smoothly by $\pi/2$ over a distance $\sim d_*$ to avoid incurring a large elastic energy cost (see Fig.~\ref{fig:dw}). The precise form of this interpolation is not essential for our purposes; a sample function is $\delta\Theta_\text{DW}\left(x_0\right)= \tan^{-1}\left[ e^{\left(x-x_0\right)/d_*}\right]$. The total spin associated with introducing an $N$-fold domain wall, $\Theta_{y_0}(x) \rightarrow \Theta_{y_0}(x)+N \delta \Theta_\text{DW}(x_0)$, is
\begin{align}
\label{eq.CW_decoupled_DW-spin}
 S^z_{\text{tot}} \equiv \frac{1}{\pi}\int_x \sum\nolimits_y\partial_x \Theta_y = \frac{N}{\pi}\int_x \partial_x \delta\Theta_\text{DW} = \frac{N}{2}~.
\end{align}
Crucially, this value is universal and only depends on the asymptotic behavior of $\Theta$. Moreover, the associated energy cost takes a finite nonuniversal value proportional to $d_*^{-1}$ (see Appendix \ref{app.defect-energies_1d-VBS-defect} for details). By contrast, a `half domain wall,' i.e., $N=1/2$, which formally carries spin $1/4$, costs a finite energy density for all $x>x_0$. Consequently, the total energy diverges linearly with the system size, i.e., such configurations are confined.

\subsection{Coupled spin-chain arrays} 
\label{subsec.CW_coupled}
\begin{figure}[tb]
\centering
 \includegraphics[width=\columnwidth]{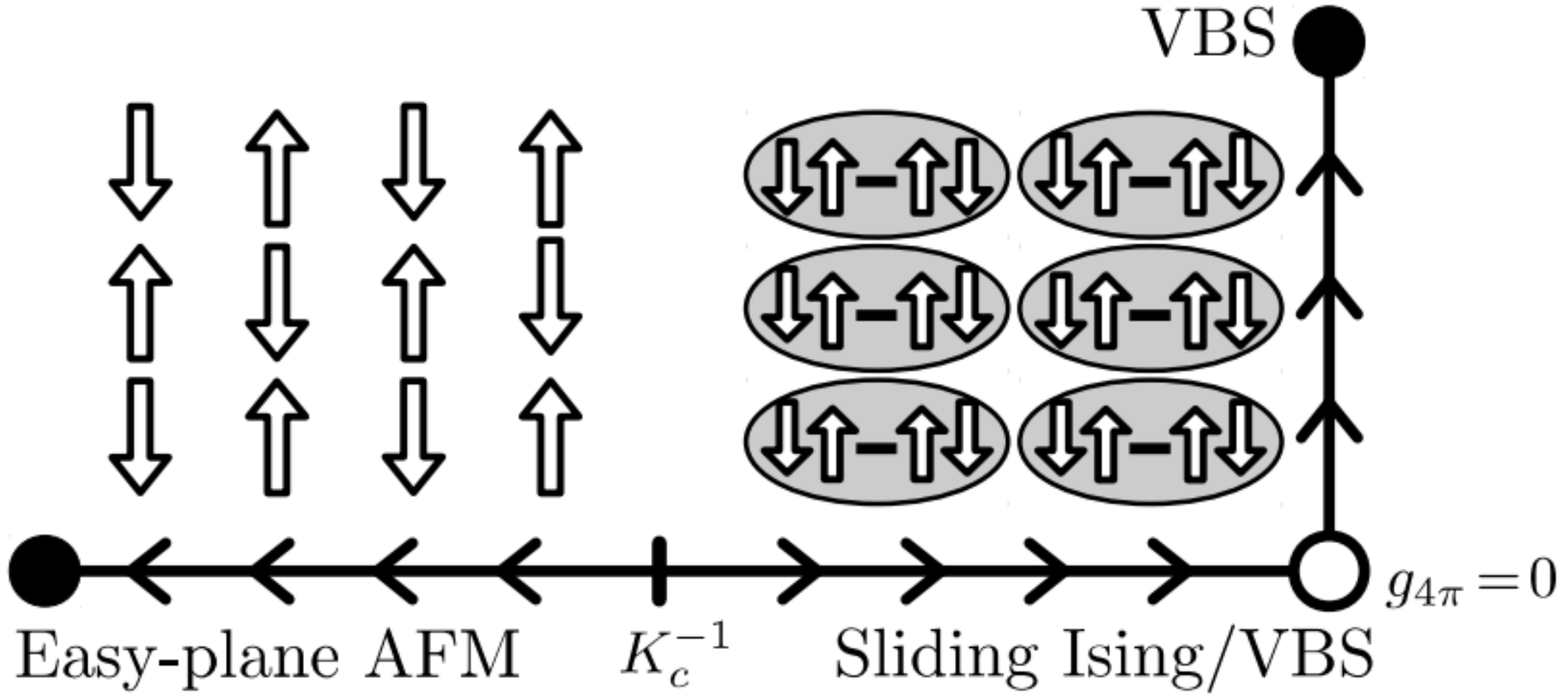}
 \caption{The lowest-order coupling terms between neighboring spin chains drive the two-dimensional array either into an easy-plane AFM or into a gapless `sliding Ising/VBS' state. Phase slips are irrelevant in the former, but strongly relevant in the latter, where they lead to a fully gapped VBS phase. 
 \label{fig:phases}}
\end{figure}

To describe two-dimensional phases, we initially neglect $4\pi$ phase slips in the action of the decoupled array
\begin{align}
\label{eq.CW_coupled_S-decoupled}
 {\cal S}_\text{decoupled} = \int_{x,\tau}\sum\nolimits_y \mathcal{ L}_\text{chain}~.
\end{align}
Instead, we perturb ${\cal S}_\text{decoupled}$ by interwire couplings that drive the system to a new fixed-point and analyze the effect of $4\pi$ phase slips there. The leading coupling terms between neighboring wires that are compatible with the symmetries in Table \ref{tab.CW_coupled_symmetries} are 
\begin{subequations} 
\label{eq.CW_coupled_couplings}
\begin{align}
\label{eq.CW_coupled_couplings-tunnel}
 {\cal L}_{t} = & g_t \cos\left(\Phi_{y+1} - \Phi_{y}\right)~,\\
\label{eq.CW_coupled_couplings-backscatter}
 {\cal L}_{u} = & g_u \cos\left(2 \Theta_{y+1} + 2 \Theta_{y}\right)~.
\end{align}
\end{subequations}
A third cosine, $\cos\left(2 \Theta_{y+1} - 2 \Theta_{y}\right)$, has the same scaling dimension at the decoupled fixed point as the one in $\mathcal{L}_{u}$. However, it can be obtained by combining the latter with $4\pi$ phase slips and thus need not be treated independently. 
The cosines in Eq.~\eqref{eq.CW_coupled_couplings} compete to drive the wire array into different symmetry-broken states (see Fig.~\ref{fig:phases}. Their topological defects are crucial for relating spins to the bosonic or fermionic partons. We, therefore, briefly discuss how their key properties arise within the coupled-wire framework. 

\subsubsection{Easy-plane antiferromagnet (AFM)} 
\label{subsubsec.CW_coupled_AFM}
Consider $K \gg 1$ such that $\mathcal{L}_t$ of Eq.~\eqref{eq.CW_coupled_couplings-tunnel} is strongly relevant while ${\cal L}_u$ of Eq.~\eqref{eq.CW_coupled_couplings-backscatter} flows to zero. As in Sec.~\ref{subsec.CW_decoupled}, we introduce a dimensionless coupling constant with bare value $\tilde{g}_t \equiv \pi d_0^2 g_t / vK <0$. The flow of $\tilde g_t$ to strong coupling permits us to replace
\begin{align}
\label{eq.CW_coupled_AFM_Lt-quad}
 {\cal L}_t=\frac{v K \tilde g_t}{\pi d_0^2} \cos\left(\Phi_{y+1} - \Phi_{y}\right)\rightarrow \frac{ v K}{2\pi d_*^2} \left[\Delta \Phi\right]_{\tilde{y}}^2~,
\end{align} 
where $\Delta_{\tilde{y},y'} \equiv \left(\delta_{y+1,y'}-\delta_{y,y'}\right)$ is the discrete $y$-derivative, naturally centered on a dual wire $\tilde{y}=y+1/2$. The scaling dimension of the cosine at the decoupled fixed point implies $d_* \simeq d_0 \left|\tilde{g}_t\right|^{1/(K^{-1}-4)}$ for small $\tilde g_t$, but that is not essential for our purposes.

On a finite array of $N_\text{w}$ wires, only $N_{\text{w}}-1$ of the differences $\Delta \Phi$ are linearly independent, and the system remains gapless. The missing linear combination $N_{\text{w}}^{-1}\sum\nolimits_y \Phi_y$ cannot be pinned due to the global $U(1)$ spin-rotation symmetry (cf.~Table \ref{tab.CW_coupled_symmetries}). 
This property reflects the presence of a Goldstone mode due to spontaneous $U(1)$ symmetry breaking. The effective action
\begin{align}
\label{eq.CW_coupled_AFM_Leff}
 \mathcal{L}_{\text{SF}} = \mathcal{L}_{\text{LL}} + \frac{vK}{2\pi d_*^2}\left[\Delta\Phi\right]_{\tilde{y}}^2~,
\end{align}
can be brought to a more familiar form by performing the Gaussian integral over $\Theta_y$. Additionally taking the continuum limit, $d_*^{-1}\Delta \rightarrow \partial_y$, results in
\begin{align}
\label{eq.CW_coupled_AFM_Goldstone}
 {\cal S}_{\text{SF}} = \frac{K}{2\pi v}\int_{\tau,x,y}\left[(\partial_\tau \Phi)^2 +v^2\left( \vect \nabla\Phi\right)^2 \right]~.
\end{align} 
It is straightforward to verify that $4\pi$ phase slips are irrelevant at this new fixed point and that it exhibits N\'eel order, $\left< N^+_{\vect r} \right> \neq 0$.

\begin{figure}[tb]
\centering
 \includegraphics[width=\columnwidth]{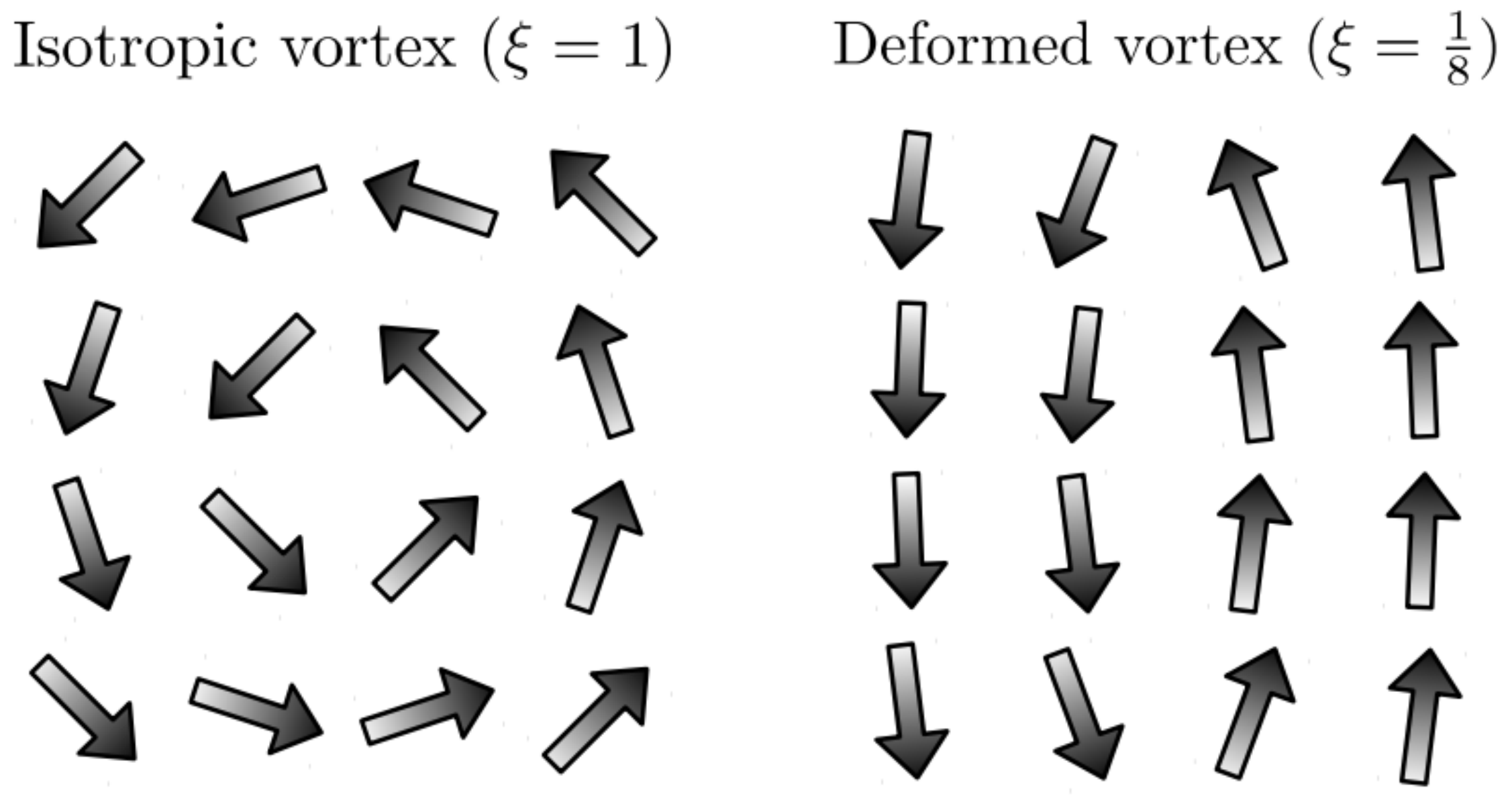}
 \caption{Two vortex configurations, shown in terms of the N\'eel vector $\vec N$, may have significant differences in trial energies but are topologically equivalent. For formal manipulations the precise choice is unimportant, and the more anisotropic limit $\xi \rightarrow 0$ turns out to be the most convenient in the present case.
 \label{fig:vortices}}
\end{figure}

For the topological defects, the periodicity of $\mathcal{L}_{t}$ is paramount. The number of domain walls in a given cosine is $N_{\tilde{y}} \equiv \frac{1}{2\pi} \int_x \left[\Delta \partial_x\Phi\right]_{\tilde{y}} \sim\frac{1}{2\pi}
\varointctrclockwise_\Gamma d\vect l\cdot \vect \nabla \Phi$, where $\Gamma$ encloses the plaquette containing $\tilde{y}$. Consequently, $N_{\tilde{y}}$ is precisely the number of magnetic vortices contained within this plaquette or, equivalently, on the dual wire. 
A sample configuration that contains an isolated vortex is $\delta\Phi^{\text{vortex},\xi}_y(x) =\arg\left[\left(x-x_0\right)/\xi + i \left(y-\tilde y_0\right)\right]$. According to Eq.~\eqref{eq.CW_coupled_AFM_Leff}, its energy exhibits the familiar logarithmic divergence with system size (see Appendix \ref{app.defect-energies_AFM-vortices}).
For the formal manipulations below, it is convenient to use the $\xi \rightarrow 0$ limit of the above expression, i.e.,
\begin{align}
\label{eq.CW_coupled_AFM_vortex-wire}
 \delta\Phi^{\text{vortex}}_y(x)= \pi \sgn{\tilde{y}_0-y}H\left(x-x_0\right)~,
\end{align}
with $H(x)$ the Heaviside step function. In this configuration, all phase winding is concentrated along one-dimensional lines rather than uniformly spread as in the isotropic one. A vortex in the form of Eq.~\eqref{eq.CW_coupled_AFM_vortex-wire} is created by the operator
\begin{align}
\label{eq.CW_coupled_AFM_vortex-operator}
 V^\dagger_{\tilde{y}_0}(x_0) \equiv \exp\left[i\sum\nolimits_y \sgn{y-\tilde{y}_0} \Theta_y\left(x_0\right)\right]~,
\end{align} 
which satisfies $V^\dagger \Phi V = \Phi + \delta\Phi^{\text{vortex}}$. The same form of this vortex operator was used in Ref.~\onlinecite{Mross2017} to implement dualities between bosons and Dirac fermions on coupled-wire arrays. Vortex configurations in both limits are illustrated in Fig.~\ref{fig:vortices}.

\subsubsection{intrawire valence bond solid (VBS)}
\label{subsubsec.CW_coupled_VBS}
For $K \ll 1/4$, it is $\mathcal{L}_{u}$ of Eq.~\eqref{eq.CW_coupled_couplings-backscatter} that is strongly relevant while $\mathcal{L}_t$ of Eq.~\eqref{eq.CW_coupled_couplings-tunnel} flows to zero. We focus on positive $g_u$ and express it in terms of a dimensionless coupling constant. To describe low energies, we replace
\begin{align}
\label{eq.CW_coupled_AFM_Lu-quad}
 \nonumber {\cal L}_u=&\frac{v \tilde g_u}{4 \pi K d_0^2} \cos\left(2\Theta_{\tilde y + 1}+ 2\Theta_{\tilde y}\right) \\
 &\rightarrow \frac{ v }{2\pi K d_*^2} \left(\left[S \Theta\right]_{\tilde{y}}-\textstyle\frac{\pi}{2}\right)^2~,
\end{align}
where $S_{\tilde{y},y'} \equiv \left(\delta_{y+1,y'}+\delta_{y,y'}\right)$ is naturally centered on a dual wire. This Lagrangian is identical to Eq.~\eqref{eq.CW_coupled_AFM_Lt-quad} upon replacing $\Theta_y \leftrightarrow \left(-1\right)^y \left(\Phi_y + \pi y/2\right)$ and $K\leftrightarrow K^{-1}$, which does not affect the kinetic interwire terms. Consequently, the analysis of the easy-plane AFM carries over.

\begin{figure*}[tb]
\centering
 \includegraphics[width=\textwidth]{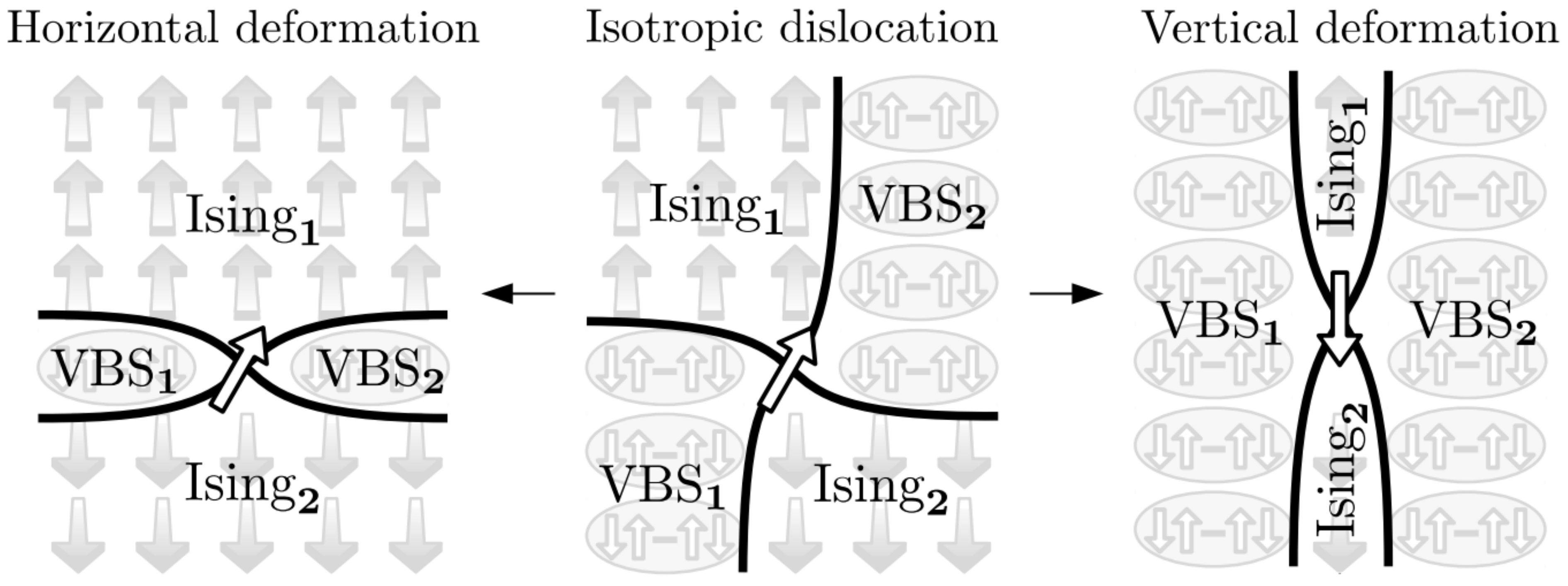}
 \caption{Four fundamental domain boundaries between VBS$_{1}$, Ising$_{1}$, VBS$_{2}$, and Ising$_{2}$ regions terminate in a dislocation that carries spin $1/2$. This topological property holds irrespective of the detailed configuration. It becomes apparent in the strongly anisotropic limits, where either the Ising or the VBS phases extend only along a one-dimensional line. There, the properties of the dislocation can be inferred from the one-dimensional domain walls illustrated in Fig.~\ref{fig:dw}. In our formulation, bosonic partons in a specific gauge are introduced as the extreme vertically deformed dislocations; they carry spin $1/2$ and live on dual wires.
 \label{fig:dislocations}}
\end{figure*}

The ground state exhibits Ising-N\'eel and/or VBS orders according to 
\begin{align}
\label{eq.CW_coupled_VBS_order-parameters}
 \langle N^z\rangle \propto \sin \left(2 \Theta_0\right)~, \qquad \langle\epsilon\rangle \propto \cos \left(2 \Theta_0\right)~,
\end{align}
with a spontaneously chosen $\Theta_0$. Around a topological defect, $\Theta_0$ winds smoothly by $\pi$. Such defects are illustrated in Fig.~\eqref{fig:dislocations} and will be referred to as dislocations. Around them, the state of the system changes from Ising$_1$ first to VBS$_1$, then to Ising$_2$, next to VBS$_2$, and finally returns to the Ising$_1$ configuration \footnote{These defects share several features with $\mathbb{Z}_4$ vortices in columnar VBS states \cite{Levin2004}. There, the two Ising phases are replaced by interwire valence-bond states VBS$_{3,4}$ that arise as $\pi/2$ rotations of the VBS$_{1,2}$. The VBS$_{3,4}$ configurations interchange under $T_y$ and are invariant under ${\cal T}'$, just like Ising$_{1,2}$. However, their transformations under $T_x$ and ${\cal T}$ are different. interwire VBS states in wire models are described in Sec.~\ref{subsubsec.BP_phases_SF+Mott_VBSy}. Unfortunately, operators that would create $\mathbb{Z}_4$ VBS vortices are not readily accessible in this framework.}.
The creation operator in the extreme vertically-deformed limit is
\small
\begin{align}
\label{eq.CW_coupled_VBS_dislocation-operator}
 b_{\tilde{y}_0}^\dagger(x_0) \equiv \exp\left[\frac{i}{2}\sum\nolimits_y \sgn{y-\tilde{y}_0}\left(-1\right)^{y}\Phi_{y}\left(x_0\right)\right]~,
\end{align} 
\normalsize
which satisfies $b^\dagger \Theta b = \Theta - \left(-1\right)^y \delta \Theta^{\text{vortex}}$ with $\delta \Theta^{\text{vortex}}$ as in Eq.~\eqref{eq.CW_coupled_AFM_vortex-wire}. 

At this fixed point [referred to as sliding Ising/VBS in Fig.~\eqref{fig:phases}], the previously neglected $4\pi$ phase-slip term, Eq.~\eqref{eq.CW_4pi-phase-slip}, is strongly relevant. Its flow to strong coupling eliminates the zero-energy mode $\Theta_y(x) \rightarrow\Theta_y(x)+(-1)^y\alpha $ and results in a gapped phase. As in the case of decoupled chains, a VBS is realized for negative $g_{4\pi}$, while positive $g_{4\pi}$ leads to an Ising-AFM. In both phases, the energy cost of creating an isolated dislocation diverges linearly with the system size (see Appendix \ref{app.defect-energies_VBS-dislocations/partons}). Tightly bound dislocation---anti-dislocation pairs are thus the fundamental low-energy excitations. When the two are located on neighboring plaquettes, they form precisely the staggered component of the microscopic spin operators 
\begin{align}
\label{eq.CW_coupled_VBS_dislocations-as-partons}
 b^\dagger_{y-1/2}b_{y+1/2} = \begin{cases}
 N^{+}_{y}\quad \text{$y$ even}~,\\
 N^{-}_{y}\quad \text{$y$ odd}~.
 \end{cases}
\end{align}
Notice the similarity to the parton decomposition $S^+ = \psi^\dagger_\uparrow \psi_\downarrow$ in Eq.~\eqref{eq.PC_definitions}
\footnote{There are also notable differences: (i) The dislocation operators $b_{\uparrow(\downarrow)}$ reside on \textit{half} the wires while Eq.~\eqref{eq.PC_definitions} yields operators, $\psi_\sigma$, on \textit{all} wires. Nevertheless, since $\psi_\sigma$ are \textit{constrained} while $b_{\sigma}$ are \textit{unconstrained}, the number of degrees of freedom agrees. (ii) interwire hopping of $b_{\sigma}$ is represented by a local operator [cf.~Eq.~\eqref{eq.BP_couplings-tunnel}] while for $\psi_\sigma$ such hopping is absent on the lattice scale. There, parton kinetic terms may emerge at long wavelengths, described by fluctuations around a non-trivial saddle point. Dislocations should thus be viewed as representing the low-energy progeny of the lattice operators $\psi_\sigma$. (iii) The operators $b_\sigma$ are formally assigned to reside between wires while $\psi_\sigma$ live on the original lattice sites (different decompositions are of course possible, see, e.g., Ref.~\onlinecite{Savary+Balents2012}). However, since hopping of individual $\psi_\sigma$ at the lattice scale is a mean-field artifact [see also (i) and (ii)], this is not a significant difference.}. To make this connection manifest, we introduce the redundant label $\sigma = \uparrow,\downarrow$ for odd and even $\tilde y$, respectively. We then define $b^\dagger_{\tilde y,\sigma} = e^{i \varphi_{\tilde y,\sigma}}$ and the associated dislocation density $\rho_{\tilde y,\sigma} = \frac{1}{\pi}\partial_x \theta_{\tilde y,\sigma}$ with
\begin{subequations}
\label{eq.CW_coupled_VBS_partons}
 \begin{align}
 \label{eq.CW_coupled_VBS_partons-phi}
 \varphi_{\tilde y,\sigma} = & \textstyle\frac{1}{2} \sum\nolimits_{y} \sgn{y-\tilde y}\left(-1\right)^{y}\Phi_{y}~,\\
 \label{eq.CW_coupled_VBS-partons-theta}
 \theta_{\tilde y,\sigma} = & \sigma \left( \Theta_{y +1} + \Theta_{ y} \right)~.
 \end{align}
\end{subequations}
The total number of $b_\sigma$--bosons (dislocations centered on even or odd dual wires) is
\begin{align}
\label{eq.CW_coupled_VBS_partons-number}
 Q_\sigma \equiv \int_x \sum\nolimits_{\tilde y} \rho_{\tilde y,\sigma} = \frac{1}{\pi} \int_x \sum\nolimits_{\tilde y}\partial_x \theta_{\tilde y,\sigma} = \sigma S^z_\text{total}~,
\end{align} 
with $S^z_\text{total}$ as defined in Eq.~\eqref{eq.CW_decoupled_DW-spin}. Crucially, when $S^z_\text{total}$ is microscopically conserved, then the number of each of these boson species is separately conserved. 

To conclude the discussion of this phase, we want to point out a close connection between dislocations and magnetic vortices. Consider a vortex in $b_\sigma$ only, but not in $b_{\bar \sigma}$. To construct its creation operator, one need only replace $\Theta \rightarrow \theta_\sigma$ in Eq.~\eqref{eq.CW_coupled_AFM_vortex-operator}. Explicitly, we introduce
\begin{align}
\label{eq.BP_bosonic-parton-vortices}
 \tilde{\varphi}_{\sigma,\tilde{y}} \equiv \begin{cases} 
 - \sum_{\tilde{y}'\text{ even}} \sgn{\tilde{y}-\tilde{y}'}\theta_{\sigma,\tilde{y}'} &\sigma = \downarrow(\tilde y \text{ odd})~, \\
 - \sum_{\tilde{y}'\text{ odd}} \sgn{\tilde{y}-\tilde{y}'}\theta_{\sigma,\tilde{y}'} &\sigma = \uparrow(\tilde y \text{ even})~.
 \end{cases}
\end{align}
Using Eqs.~\eqref{eq.CW_coupled_AFM_vortex-operator} and \eqref{eq.CW_coupled_VBS-partons-theta} we find that $V_{2\tilde y}=e^{i\tilde{\varphi}_{\uparrow,2\tilde{y}}}$ and $V_{2\tilde y+1}=e^{i\tilde{\varphi}_{\downarrow,2\tilde{y}+1}}$. Dislocations are thus dual to magnetic vortices in this precise sense.
 
\section{Bosonic partons from coupled wires}
\label{sec.BP}
To access a wider range of phases, including QSLs, we now develop a dual description of the coupled-wire model in terms of the topological defects described above. Such a reformulation of spin-$1/2$ models in terms of magnetic vortices has already been used in Ref.~\onlinecite{Senthil+Vishwanath+Balents+Sachdev+Fisher2004} to access, e.g., a putative `deconfined' quantum critical point between N\'eel and VBS orders. To connect coupled-wire and parton techniques, we instead focus on dislocations and argue that they form a bosonic-parton representation of the microscopic spins. 

As discussed at the end of the previous section, $b_\sigma$ possesses several of the relevant bosonic-parton properties. Moreover, while the combination $b_\uparrow^\dagger b_\downarrow$ is local in terms of microscopic operators, and carries spin $1$, individual $b_\sigma$ are nonlocal and do not have well-defined spin. This property closely relates to the local gauge redundancy of conventional parton decompositions discussed in Sec.~\ref{subsec.PC_gauge}. 
Even though $b_\sigma$ are nonlocal, the interwire couplings introduced in Eq.~\eqref{eq.CW_coupled_couplings} remain local under the mapping. Translating them via Eq.~\eqref{eq.CW_coupled_VBS_partons}, we find an intraspecies nearest-neighbor tunneling term and an umklapp term 
\begin{subequations}
\label{eq.BP_couplings}
 \begin{align}
 \label{eq.BP_couplings-tunnel}
 \mathcal{L}_{t} &= g_t \cos\left(\varphi_{\tilde{y}+2,\sigma} - \varphi_{\tilde{y},\sigma}\right) \sim b^\dagger_{\tilde{y}+2,\sigma} b_{\tilde{y},\sigma}+\text{H.c.}~,\\
 \label{eq.BP_couplings-backscatter}
 \mathcal{L}_{u} &= g_u \cos\left(2\theta_{\tilde{y},\sigma}\right) \sim b^\dagger_{\tilde{y},\sigma} b_{\tilde{y},\sigma} U(x)~,
 \end{align}
\end{subequations}
where $U(x) = U(x+1)$ is a weak periodic potential (see, e.g., Ref.~\onlinecite{GiamarchiBook2004}). The periodicity of $\mathcal{L}_u$ implies that $b_\sigma$ are both at unit-filling. Consequently, the densities of $b_\uparrow$ and $b_\downarrow$ are the same as in conventional parton constructions (cf.~Sec.~\ref{sec.PC}). By contrast, the intrawire interactions of the microscopic spin chains are nonlocal in bosonic-parton variables. They have a natural interpretation in terms of an emergent gauge field, as we will now explain. 

\subsection{Gauge theory}
\label{subsec.BP_symm+theory}
To derive the action governing the bosonic partons, we invert the mapping in Eq.~\eqref{eq.CW_coupled_VBS_partons} and insert them into the microscopic wire-array model in Eq.~\eqref{eq.CW_coupled_S-decoupled}. It is convenient to include, on top of $\mathcal{L}_{\text{chain}}$, the symmetry-allowed quadratic interwire couplings
\begin{align}
\label{eq.BP_symm+theory_inter}
 \mathcal{L}_\text{inter} = \frac{u_B}{2\pi}\left[S \partial_x\Theta\right]_{\tilde{y}}^2+\frac{u_V}{8\pi}\left[\Delta \partial_x\Phi\right]_{\tilde{y}}^2~.
\end{align}
In the limit of weakly coupled wires, these arise as the leading renormalizations of the kinetic energy due to Eq.~\eqref{eq.CW_coupled_couplings}, but in generic cases, $u_B$ and $u_V$ should be viewed as independent parameters.

While $\mathcal{L}_{t}$, $\mathcal{L}_{u}$, and $\mathcal{L}_{\text{inter}}$ [Eqs.~\eqref{eq.CW_coupled_couplings} and \eqref{eq.BP_symm+theory_inter}] are local in terms of partons, the intrawire interactions $\mathcal{L}_{\text{chain}}$ are highly nonlocal. This nonlocality can be encoded \textit{exactly} through an emergent gauge field $\vec{a}$, just as in the case of the boson-vortex duality (see Appendix \ref{app.gauge_bosons} for details)~\cite{Mross2017}. We express the resulting gauge theory as
$\mathcal{S}_b = \int_{x,\tau} \sum\nolimits_{\tilde{y}} \left[ \mathcal{L}_{b}+ \mathcal{L}_{\text{Maxwell}}+\mathcal{L}_\text{int} \right]$. The first two contributions contain parton and gauge-field kinetic terms as well as the coupling between the two, i.e.,
\begin{align}
\label{eq.BP_symm+theory_L-0}
 \nonumber &\mathcal{L}_{b} = \frac{i}{\pi} \partial_x \theta_{\tilde{y},\sigma} (\partial_{\tau} \varphi_{\tilde{y},\sigma}-a_{0,\tilde{y}}) + \frac{v_B}{2\pi} \left(\partial_x \varphi_{\tilde{y},\sigma}-a_{1,\tilde{y}}\right)^2 \\&\qquad+ \frac{u_B}{2\pi}\left(\partial_x \theta_{\tilde{y},\sigma}\right)^2 ~,\\
\label{eq.BP_symm+theory_L-Maxwell}
 &\mathcal{L}_\text{Maxwell} = \frac{\kappa\tilde{v}}{4\pi} \left[\Delta a_1\right]_{y}^2 + \frac{\kappa}{4\pi \tilde{v}} \left[\Delta a_0\right]_{y}^2~.
\end{align}
The parameters $v_B,$ $u_B,$ $\tilde v,$ and $\kappa$ are nonuniversal. For their expression in terms of microscopic spin-chains parameters, see Appendix \ref{app.gauge_bosons}. There, we also specify the last term, ${\cal L}_\text{int}$, which contains exponentially decaying interwire density-density and current-current interactions. 

The gauge-field Lagrangian ${\cal L}_\text{Maxwell}$ describes an anisotropic Maxwell term in the $a_2=0$ gauge but missing the $\propto(\partial_\tau a_{1,\tilde{y}} - \partial_x a_{0,\tilde{y}})^2$ contribution. Such a term will be generated upon integrating out matter fields at high energies. We demonstrate this in Sec.~\ref{subsubsec.BP_phases_Mott+Mott-VBSx} for the case of trivially-gapped partons. Finally, it is instructive to express the bosonized model in terms of $b_\sigma$ as
\small
\begin{align}
\label{eq.BP_symm+theory_L-Kin(b)}
 \mathcal{L}_b' = & b^\dagger_{\tilde{y},\sigma}\left(\partial_\tau-ia_0- \mu\right)b_{\tilde{y},\sigma}+ \frac{v_B}{2\rho_0} \left|(\partial_x- i a_x) b_{\tilde{y},\sigma}\right|^2 ,
\end{align}
\normalsize
where $\rho_0=\pi$ is the boson density. This Lagrangian has the expected structure for bosonic partons [cf.~Eq.~\eqref{eq.PC_gauge_MF1-Lagrangian}]. 

\subsubsection*{Monopoles}
\label{subsubsec.BP_symm+theory_monopole}
The $4\pi$ phase slips in Eq.~\eqref{eq.CW_4pi-phase-slip} are also nonlocal in bosonic parton variables. 
To interpret them, we introduce the operator
\begin{align}
\label{eqn.monopoleoperator}
 \widehat{\mathcal{M}}_{\vect{r}} &\equiv \exp\left[2i\left(-1\right)^{y}\Theta_y\right] = \exp\left[i\sum\nolimits_{\tilde{y}}\text{sgn}(y-\tilde{y})\theta_{\tilde{y}}\right]\nonumber \\&= \exp\left[-i \tilde{\varphi}_{\sigma,\tilde{y}} - i\tilde{\varphi}_{\bar{\sigma},\tilde{y}-1}\right]~.
\end{align}
For both bosonic partons, $\widehat{\mathcal{M}}_{\vect{r}_0} b^\dagger_\sigma \widehat{\mathcal{M}}^\dagger_{\vect{r_0}}\rightarrow b^\dagger_\sigma e^{i\alpha\left(\vect{r}-\vect{r}_0\right)}$, where $\alpha(\vect{r})$ winds counter-clockwise by $2\pi$ around the origin. 
$\widehat{\mathcal{M}}_{\vect{r}}$ can thus be viewed as the insertion of a $2\pi$ monopole in the emergent gauge field at position ${\vect{r}}$.
Since it is odd under lattice translations (cf.~Table \ref{tab.CW_coupled_symmetries}), $4\pi$ monopoles created by $\widehat{\mathcal{M}}^2_{\vect{r}}$ are the minimal ones allowed by symmetries.

It is useful to disentangle the monopoles from the matter fields. We, therefore, write $\mathcal{M}^\dagger = e^{i \phi_{\mathcal{M}}}$ and replace Eq.~\eqref{eq.CW_4pi-phase-slip} by
\begin{align} 
\label{eq.BP_symm+theory_monopoles1}
 \nonumber \mathcal{L}_{4\pi} = & \frac{g_{4\pi}}{2} \left[\mathcal{M}_{\vect{r}}^2 + \mathcal{M}^{2\dagger}_{\vect{r}}\right] \\
 & - \frac{i}{\pi} \partial_x\lambda_y (\phi_{\mathcal{M},y} - \tilde \varphi_{\sigma,\tilde{y}} - \tilde \varphi_{\bar{\sigma},\tilde{y}-1})~,
\end{align} 
where $\lambda$ is a Lagrange multiplier, and $\phi_{\mathcal{M}}$ is now an independent variable in the functional integral. A simple shift $a_{0,\tilde{y}} \rightarrow a_{0,\tilde{y}} - \sum_{y'} \sgn{y'-\tilde{y}}\lambda_y$ decouples the Lagrange multiplier from the matter fields and moves it into the gauge-field action. Integrating it out then yields our final form of the monopole Lagrangian
\begin{align} 
\label{eq.BP_symm+theory_monopoles2}
 \mathcal{L}_\mathcal{M} = \frac{g_{\mathcal{M}}}{2} \left[\mathcal{M}_{\vec{r}}^2 + \mathcal{M}^{2\dagger}_{\vec{r}}\right] -\frac{\kappa}{4\pi \tilde{v}} \left[\Delta a_0 -  \frac{i\tilde{v}}{\kappa} \partial_x \phi_{\mathcal{M}}\right]_{y}^2~,
\end{align}
with parameters $\kappa,\tilde v$ as in Eq.~\eqref{eq.BP_symm+theory_L-Maxwell} and where we have relabeled $g_{4\pi}\rightarrow g_{\mathcal{M}}$. For $g_{\mathcal{M}}=0$, the Gaussian integral over $\phi_{\mathcal{M}}$ is a complete square and does not affect any gauge-field or matter correlation function. Alternatively, the second term in $\mathcal{L}_\mathcal{M}$ may be viewed as a `bare' kinetic energy for the monopoles. Their interaction with the gauge field, which is \textit{not} of minimal coupling form, introduces additional terms  $\sim\left(\partial_\tau\phi_{\mathcal{M}}\right)^2$ and $\sim \left(\Delta\phi_{\mathcal{M}}\right)^2$ upon integrating out short-distance gauge fluctuations. The low-energy behavior of the monopoles can, of course, be qualitatively different, depending on the phase of the gauge theory [see, e.g., Eq.~\eqref{eq.BP_phases_SF+SF-AFM_eff-monopole-L}].

\subsubsection*{Symmetries}
\label{subsubsec.BP_symm+theory_symm}
To complete the description of the parton-gauge theory, we now specify how the microscopic symmetries of Table \ref{tab.CW_coupled_symmetries} are implemented. The straightforward application of the mapping between spins and partons leads to the symmetry properties summarized in Table \ref{tab.BP_symm+theory_symmetries}. Additionally, we introduce an external probing field $\vec{A}$ that minimally couples to the conserved $S^z$ of the microscopic spins. In its presence, the theory for the decoupled spin-chain is augmented to $\mathcal{L}_{\text{chain}} \rightarrow \mathcal{L}_{\text{chain}} + \mathcal{L}_A$ with
\begin{align}
\label{eq.BP_symm+theory_A-microscopic}
 \mathcal{L}_A = -\frac{i}{\pi}\partial_x\Theta_y A_{0,y} - \frac{v K}{\pi} \partial_x \Phi_y A_{1,y} + \frac{v K}{2 \pi}A_{1,y}^2~.
\end{align}
Re-deriving the bosonic parton action with these terms (see Appendix \ref{app.gauge_bosons}) amounts, at lowest order in $\Delta$, to replacing $a_{\mu,\tilde y} \rightarrow a_{\mu,\tilde y} - \frac{1}{4}(-1)^{\tilde{y}} \left[S A_\mu\right]_{\tilde y}$ in Eqs.~\eqref{eq.BP_symm+theory_L-0} and \eqref{eq.BP_symm+theory_L-Kin(b)}.

\begin{center}
\begin{table} [ht]
 \caption{The action of microscopic symmetries on bosonic partons and their interpretation in the parton gauge theory. Anti-unitary symmetries ($i \rightarrow -i$) are indicated by asterisks.}
 \begin{tabularx}{\linewidth}{l r@{}l r@{}l r@{}l l}
 \hline\hline
 \textbf{Microscopic} & & \multirow{2}{*}{$b^\dagger_{\up}b_{\down}$\quad \hspace{-1.3pt}} & & \multirow{2}{*}{$\rho_{\sigma}$\quad \hspace{-1.3pt}} & & \multirow{2}{*}{$\mathcal{M}$\ } & \textbf{Parton} \\
 \textbf{symmetry} & & & & & & & \textbf{interpretation} \tabularnewline
 \hline
 $U\left(\alpha\right)$ & $e^{i\alpha}$ & $b^\dagger_{\up}b_{\down}$ & & $\rho_{\sigma}$ & & $\mathcal{M}$ & 
 Global $Q_\uparrow - Q_\downarrow$ \tabularnewline
 & & & & & & & gauge transform. \tabularnewline
 $\Pi_y$ & $-$ & $b^\dagger_{\down}b_{\up}$ & $-$ & $\rho_{\sigma}$ & & $\mathcal{M}^\dagger\quad$ & Particle-hole (PH) \tabularnewline
 $T_x:x \rightarrow x+1$\quad & $-$ & $b^\dagger_{\up}b_{\down}$ & & $\rho_{\sigma}$ & $-$ & $\mathcal{M}$ & $x$-translation \tabularnewline
 $T_y: \tilde{y} \rightarrow \tilde{y}+1$ & $-$ & $b^\dagger_{\up}b_{\down}$ & $-$ & $\rho_{\bar{\sigma}}$ & $-$ & $\mathcal{M}^\dagger$ & PH + spin-flip \tabularnewline
 $\mathcal{I}_x:x \rightarrow -x$ & & $b^\dagger_{\up}b_{\down}$ & $-$ & $\rho_{\sigma}$ & $-$ & $\mathcal{M}^\dagger$ & $x$-inversion \tabularnewline
 $\mathcal{I}_y: \tilde{y} \rightarrow -\tilde{y}$ & $-$ & $b^\dagger_{\up}b_{\down}$ & & $\rho_{\sigma}$ & $-$ & $\mathcal{M}^\dagger$ & $\tilde{y}$-inversion \tabularnewline
 $\mathcal{T}^*: \tau \rightarrow -\tau$ & $-$ & $b^\dagger_{\down}b_{\up}$ & $-$ & $\rho_{\sigma}$ & & $\mathcal{M}$ & PH$^*$ \tabularnewline
 $\mathcal{T}^{\prime *}: \tau \rightarrow -\tau$ & & $b^\dagger_{\down}b_{\up}$ & & $\rho_{\bar{\sigma}}$ & $-$ & $\mathcal{M}^\dagger$ & TR \tabularnewline
 \hline\hline
 \end{tabularx}
\label{tab.BP_symm+theory_symmetries}
\end{table}
\end{center}

\subsubsection*{Alternative perspective}
\label{subsubsec.BP_symm+theory_alt}
An alternative route to the parton gauge theory begins with rewriting the wire array in terms of magnetic vortices [see Eq.~\eqref{eq.CW_coupled_AFM_vortex-operator}]. On a lattice, these vortices experience an average flux of $\pi$ per plaquette~\cite{Lannert2001,Sachdev+Park2002}. Their band structure thus exhibits two valleys, which amounts to two vortex flavors at low energies. To see how this is reflected in the wire framework, consider the interwire couplings of Eq.~\eqref{eq.CW_coupled_couplings}. The XY spin exchange $\mathcal{L}_t$ translates into $2\pi$ phase slips for the vortex $V_{\tilde y}$ while
\begin{align}
\label{eq.BP_AFMvortex-hopping}
 \mathcal{L}_{u} &= g_u V^\dagger_{\tilde{y}+2} V_{\tilde{y}}+\text{H.c.}~
\end{align}
The wire-array model ${\cal L}_\text{LL} +{\cal L}_u+{\cal L}_t$ can thus be equivalently expressed in terms of two separately conserved vortex flavors that reside on even or odd dual wires. Moreover, all vortices are coupled to the same non-compact gauge field $\vec{a}_{\text{vortex}}$ whose flux represents the conserved $S^z$. (The derivation of this gauge theory within the wire framework is identical to the one performed above for bosonic partons.) Performing separate dualities for the two vortex flavors results in two species of bosonic partons (cf. the final paragraph of Sec.~\ref{subsubsec.CW_coupled_VBS}) coupled to a single gauge field $\vec{a}$ that is, likewise, non-compact. Its flux corresponds to the difference between the numbers of $V_\text{even}$ and $V_\text{odd}$ vortices. This difference ceases to be conserved in the presence of the $4\pi$ phase slips. Indeed,
\begin{align}
\label{eq.BP_AFMvortex-spicies-tunneling}
 \mathcal{L}_{4\pi} &= g_{4\pi}V_{2\tilde{y}}V_{2\tilde{y}}V^\dagger_{2\tilde{y}+1}V^\dagger_{2\tilde{y}+1} +\text{H.c.}~
\end{align}
allows vortex \textit{pairs} to switch their flavor. Such processes change the flux of the gauge field $\vec{a}$ by $4\pi$ and, thereby, render it compact.

\subsection{Phases of bosonic partons / spins}
\label{subsec.BP_phases}
To demonstrate the generality of the formalism developed above, we now apply it to several concrete examples. In the spirit of most parton constructions, we primarily consider phases that are trivial at the mean-field level, i.e., in the absence of gauge fluctuations. These include Mott insulators, superfluids, and integer quantum Hall states. 
Coupled-wire models that realize such phases are either known in the literature or can be constructed relatively easily~\cite{TeoKane2014,Lu2012,Mross2015,Fuji2016}. The mapping in Eq.~\eqref{eq.CW_coupled_VBS_partons} then immediately provides a corresponding coupled-wire model in terms of the microscopic spin variables. We carefully examine both representations of each phase---in terms of partons and of spins---and demonstrate their equivalence.

We analyze the parton models as follows: First, we determine the ground state and excitations of ${\cal L}_{b}+ \delta {\cal L}$ at the mean-field level, i.e., by treating $\vec a$ as static. Second, we reintroduce the gauge-field dynamics and determine how they are affected by the matter fields. If the gauge field remains massless, we analyze the effect of monopoles. Third, we examine the quasiparticle content of the gauge theory. Finally, we perform a conventional analysis of the equivalent coupled-wire model in terms of the microscopic spin variables, $\mathcal{L} = \mathcal{L}_{\text{LL}} +\delta \mathcal{L}\left[\Phi,\Theta\right]$, and verify that its properties match the ones obtained from the parton gauge theory. 

\subsubsection{Mott insulator / intrawire VBS}
\label{subsubsec.BP_phases_Mott+Mott-VBSx}
\textit{Mean field}---A trivial Mott insulator of partons forms when the umklapp term of Eq.~\eqref{eq.BP_couplings-backscatter} gaps the bosons on each wire separately. This interaction does not contain $\varphi_{\tilde y,\sigma}$, which can, therefore, be integrated out trivially. The parton Mott insulator is thus described by
\begin{align}
\label{eq.BP_phases_Mott+Mott-VBSx_L}
 \nonumber \mathcal{L}_\text{Mott}=&\frac{1}{2\pi v_B}\left(\partial_\tau \theta_{\tilde{y},\sigma}\right)^2 + \frac{u_B}{2\pi}\left(\partial_x \theta_{\tilde{y},\sigma}\right)^2 + g_u \cos \left(2\theta_{\tilde{y},\sigma}\right) \\
 & + \frac{i}{\pi} \theta_{\tilde{y},\sigma} \left(\partial_x a_{0,\tilde{y}}-\partial_{\tau} a_{1,\tilde{y}}\right) ~.
\end{align}
At sufficiently long length scales, each field $\theta_{\tilde{y},\sigma}$ becomes trapped around the minima of the cosine, as in Sec.~\ref{subsec.CW_decoupled}. Integrating out the massive fluctuations around these minima yields ${\cal L}_\text{ind}\propto (\partial_{\tau} a_{1,\tilde{y}}-\partial_x a_{0,\tilde{y}})^2$, which modifies the dielectric constant of the (static) gauge field $\vec a$.
The periodicity of the cosine implies that
$\theta_{\tilde{y},\sigma}$ winds by $\pi$ at a fundamental domain wall. Such a configuration describes an isolated parton, created by the operator $b^\dagger_\sigma$. At the mean-field level, these spin-$1/2$ excitations are the elementary quasiparticles.

\textit{Gauge fluctuations}---We now reinstate the dynamics of the gauge field and supplement its bare action, Eq.~\eqref{eq.BP_symm+theory_L-Maxwell}, by ${\cal L}_\text{ind}$. The effective Lagrangian is thus
\small
\begin{align}
\label{eq.BP_phases_Mott+Mott-VBSx_L-gauge}
 {\cal L }_\text{MW} = \frac{\kappa}{4\pi \tilde{v}} \left\{\left[\Delta a_{0}\right]_{y}^2+ \tilde{v}^2 \left[\Delta a_{1}\right]_{y}^2 + d_*^2 \left(\partial_x a_{0,\tilde{y}} - \partial_\tau a_{1,\tilde{y}}\right)^2\right\}~,
\end{align}
\normalsize
where the nonuniversal length scale $d_*$ encodes the flow of $g_u$ to strong coupling. Recall that in our formulation, $\vec{a}$ is a non-compact gauge field, and monopoles are included through $\mathcal{L}_\mathcal{M}$ [cf.~Eq.~\eqref{eq.BP_symm+theory_monopoles2}]. In their absence, bosonic partons would interact logarithmically via the gauge field. (The interaction potential is readily obtained from $\langle a_0 a_0\rangle$ at $\omega=0$.)

However, it is well-known that \textit{compact} $U(1)$ gauge theories may be unstable to monopole proliferation, i.e., confinement. The relevance of monopoles can be assessed from their correlation function. For an isolated monopole---anti-monopole pair at locations $\vect R_i = (x_i,d_* y_i)$ and equal time, it is given by
\small
\begin{align}
\label{eq.BP_phases_Mott+Mott-VBSx_Monopole-corr-def}
 &\nonumber C^\mathcal{M}_{\vect{R}_1- \vect{R}_2} \equiv \left<\mathcal{M}^\dagger_{y_1}(x_1)\mathcal{M}_{y_2}(x_2)\right> \\
 & = \frac{\int \mathcal{D}\left[a,\phi_{\mathcal{M}}\right] e^{i \phi_{\mathcal{M},y_1}(x_1) - i\phi_{\mathcal{M},y_2}(x_2) } e^{-\mathcal{S}_{\text{gauge}}-\mathcal{S}_{\mathcal{M}}}}{\int \mathcal{D}\left[a,\phi_{\mathcal{M}}\right] e^{-\mathcal{S}_{\text{gauge}}-\mathcal{S}_{\mathcal{M}}}}~.
\end{align}
\normalsize
For $g_{\mathcal{M}}=0$, the theory is Gaussian and performing the integral over $\phi_{\mathcal{M}}$ yields
\small
\begin{align}
\label{eq.BP_phases_Mott+Mott-VBSx_Monopole-corr}
 C^\mathcal{M}_{\vect R} = \exp\left\{\frac{\kappa d_*}{4\pi^2\tilde{v}}\int_{\vect{k},\omega} \frac{\cos\left(\vect{k}\cdot\vect{R}\right)-1}{k_x^{2}}\left[1-\frac{\kappa d_*^2}{2\pi \tilde{v}}\left\langle \left|\varepsilon_2\right|^{2}\right\rangle \right]\right\}~.
\end{align}
\normalsize
Here $\varepsilon_2=\Delta a_{0} /d_*$ is the $y$-component of the emergent electric field, 
$\left<\ldots\right>$ denotes the Gaussian average over $\vec a$, according to $\mathcal{L}_{\text{MW}}$, and $\vect k =(k_x,k_y)$. The momentum $k_y$ is expressed in units of $d_*^{-1}$, i.e., the Fourier transform is defined as $f_y\left(x,\tau\right) = \frac{\sqrt{d_*}}{\left(2\pi\right)^{3/2}}\int_{\vect{k},\omega} e^{i \vect k \cdot \vect R+i\omega \tau}\tilde{F}(\vect k,\omega)$. 
Inserting the $\varepsilon_2$ correlation function and evaluating the integral for large $|\vect R |$, we find $C^{\mathcal{M}}_{ \vect R}\sim \exp\left[\frac{\kappa d_*}{2\left|\vect R\right|}\right]$.

Since $C^{\mathcal{M}}_{\vect{ R}}$ approaches a non-zero constant at long distances, monopoles are strongly relevant. Beyond a length scale $l_*$, we thus expand the cosine in ${\mathcal L}_{\mathcal{M}}$ to quadratic order. Integrating out $\phi_{\mathcal{M}}$ then results in a modified theory for $\vec{a}$ given by
\begin{align}
\label{eq.BP_phases_Mott+Mott-VBSx_L-gauge-w/monopole} 
 {\cal L }_{\text{confining}} = \mathcal{L}_{\text{MW}} + \frac{\kappa}{4\pi \tilde{v}}\left[\Delta a_{0}\right]_{y}\frac{1}{l_*^2\partial_x^2 - 1} \left[\Delta a_{0}\right]_{y}~,
\end{align}
with $\mathcal{L}_{\text{MW}}$ as in Eq.~\eqref{eq.BP_phases_Mott+Mott-VBSx_L-gauge}. The corresponding analytically continued gauge-field propagator has poles at real frequencies $\omega =\pm \tilde{v}\sqrt{k_x^{2}+ k_y^{2}+l_*^{-2}}$, i.e., there are no gapless gauge-field modes for finite $l_*$.

\begin{figure}[tb]
\centering
 \includegraphics[width=\columnwidth]{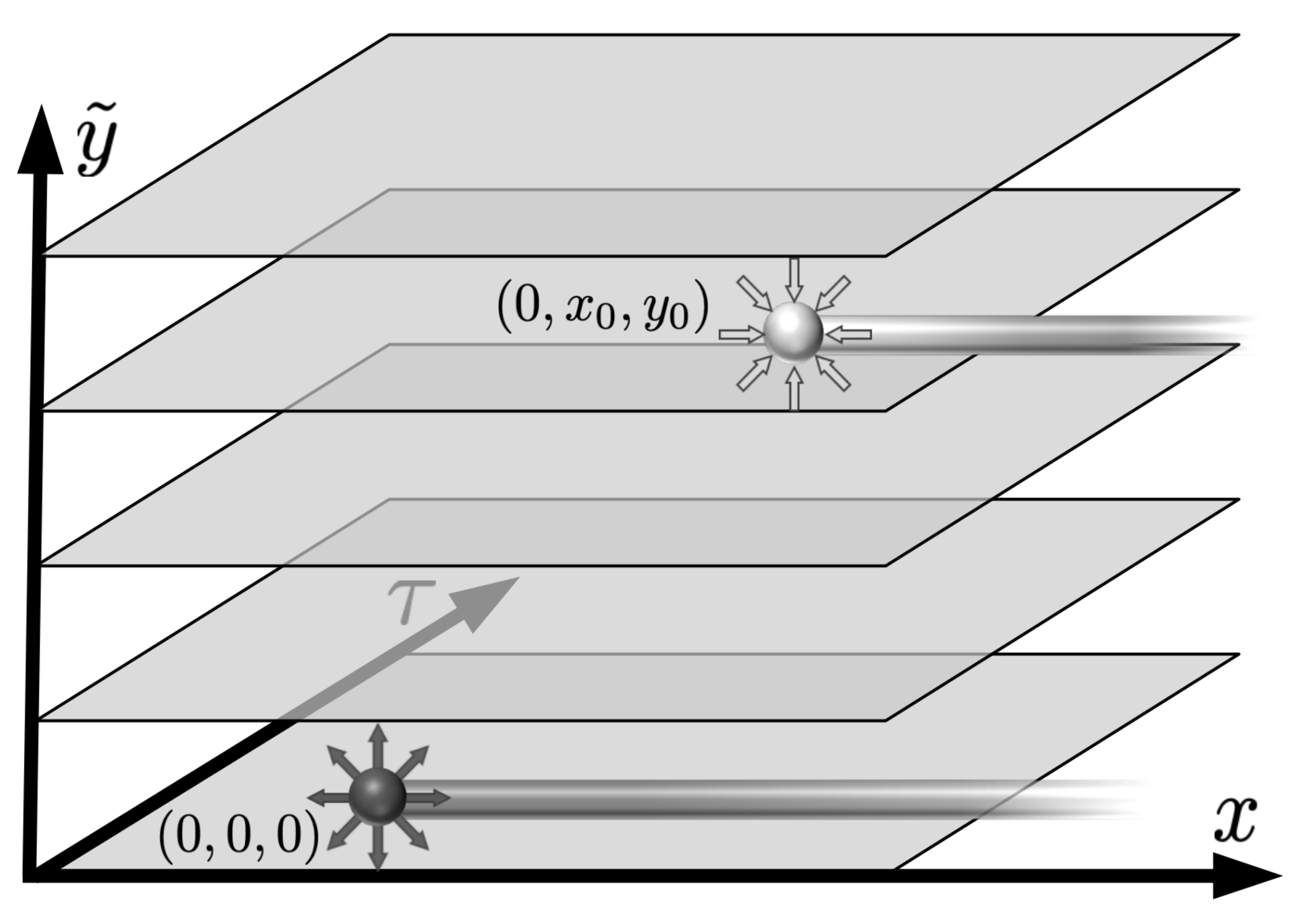}
 \caption{A singular gauge-field configuration with monopoles of opposite signs can be realized via two flux tubes that extend from infinity at fixed $\tau$ and $y$. Each $x$--$\tau$ plane represents a dual wire at coordinate $\tilde y=y+1/2$, and monopoles are located at $y$, i.e., between wires.
 \label{fig:monopoles}}
\end{figure}

Alternatively, $C^{\mathcal{M}}_{\vect{ R}}$ can be obtained from ${\cal L}_\text{MW}$ alone via Eq.~\eqref{eq.PC_gauge_monopole-corr}, i.e., by externally imposing the desired singularities on $\vec{a}$. For specificity, consider $\vec a^{\mathcal{M}}$ such that $(\partial_{\tau} a^{\mathcal{M}}_1 - \partial_x a^{\mathcal{M}}_0) = \Delta a^{\mathcal{M}}_1 =0$ and
\begin{align}
 \Delta a^{\mathcal{M}}_0 = 2\pi \left[\delta_{y,0}H\left(x\right) - \delta_{y,y_0}H\left(x-x_0\right)\right]\delta\left(\tau\right)~.
\end{align}
This configuration is depicted in Fig.~\ref{fig:monopoles}. It describes two flux tubes emanating from $x = \infty$ and extending, at fixed $y=0$ and $y=y_0$, to $x=0$ and $x=x_0$, respectively. Inserting this expression into Eq.~\eqref{eq.PC_gauge_monopole-corr}, with the gauge-field action of Eq.~\eqref{eq.BP_phases_Mott+Mott-VBSx_L-gauge}, we reproduce $C^{\mathcal{M}}_{\vect{R}}$ in Eq.~\eqref{eq.BP_phases_Mott+Mott-VBSx_Monopole-corr}. 

\textit{Quasiparticles}---In the presence of a confining gauge field, individual partons seize to be finite-energy quasiparticles. It would be tempting, but incorrect, to compute their interaction potential directly from the effective gauge theory in Eq.~\eqref{eq.BP_phases_Mott+Mott-VBSx_L-gauge-w/monopole}. To obtain ${\cal L}_\text{confining}$, we expanded $g_{\mathcal{M}} \cos \left(2 \phi_{\mathcal{M}}\right)$ around a specific minimum. This restriction to a single topological sector is innocent in the special case of partons on the same dual wire. There, only the trivial sector contributes, and we can indeed use ${\cal L }_\text{confining}$ to find
\begin{align}
 V(x) -V(0) \sim & \frac{\tilde{v}}{\kappa d_*} \int_{\vect{k}} \frac{ 1- \cos\left(k_x x\right) }{k_x^2\left(1 +l_*^2k_y^2\right)} \sim \frac{\tilde{v}}{\beta d_{*}l_{*}} \left|x\right| ~.
\end{align} 
In the generic case, a careful sum over different sectors, as performed in Appendix \ref{app.defect-energies_VBS-dislocations/partons}, is required. The result is linear confinement in all directions: It is impossible to isolate any excitation charged under the emergent gauge field without incurring a diverging energy cost. Consequently, finite-energy excitations can only be created as combinations of the gauge-neutral quasiparticles $b_{\sigma}^\dagger b_{\sigma}$, carrying spin 0, and $b_{\sigma}^\dagger b_{\bar{\sigma}}$ carrying spin 1.

\textit{Spin model}---There are two complementary routes to identifying the microscopic phase: through symmetry considerations and by direct translation to a microscopic model. 
For the former, recall how monopoles transform under the microscopic symmetries (see Table \ref{tab.BP_symm+theory_symmetries}). The non-zero expectation value acquired by $\mathcal{M}$ implies that $x$-translation symmetry is reduced to translations by two sites. Furthermore, for $g_{\mathcal{M}}<0$, the site-centered $x$ inversion is also broken, while bond-centered inversion is preserved. Other symmetries, in particular time reversal and $U(1)$ spin rotations, remain intact. These properties identify the microscopic phase as an intrawire VBS. We arrive at the same conclusion by using the transformation from parton to spin variables. The gauge theory maps onto ${\cal L}_\text{LL}+{\cal L}_{u}$, which we already analyzed in Sec.~\ref{subsubsec.CW_coupled_VBS}. Its gapped ground state exhibits VBS order, $\left<\varepsilon_{\vect r}\right> \neq 0$, with integer-spin excitations, exactly as we found in the gauge theory above. 
\subsubsection{Superfluid / Easy-plane AFM}
\label{subsubsec.BP_phases_SF+SF-AFM}
\textit{Mean field}---Consider now a superfluid phase where both partons condense. Recall that the parton number is separately conserved for both species. The corresponding $U(1)$ symmetries are spontaneously broken when the tunneling term of Eq.~\eqref{eq.BP_couplings-tunnel} flows to strong coupling. We proceed as before and expand
\begin{align}
\label{eq.BP_phases_SF+SF-AFM_deltaL-quad}
 \mathcal{L}_t = \frac{v_B \tilde g_t}{4 \pi d_0^2}\cos\left(\varphi_{\tilde y+2,\sigma}-\varphi_{\tilde y,\sigma}\right) \rightarrow \frac{v_B}{8\pi d_*^2} \left[\Delta_2 \varphi_{\sigma}\right]^2_{\tilde{y}}~,
\end{align}
where $\left(\Delta_2\right)_{y,y'}=\delta_{y+2,y'}-\delta_{y,y'}$. The field $\theta_{\tilde{y},\sigma}$ does not enter ${\cal L}_t$ and can thus be integrated out trivially. Additionally taking the long-wavelength limit, $d_*^{-1} \Delta_2 \rightarrow 2 \partial_y$, we arrive at the action 
$\mathcal{S}_b^{\text{SF}} = \int_{\tau,x,y} \sum_\sigma \mathcal{L}_{\text{SF},\sigma}$, with 
\begin{align}
\label{eq.BP_phases_SF+SF-AFM_L-eff} 
 \mathcal{L}_{\text{SF},\sigma} = \frac{v_B}{2\pi }\left[\frac{1}{c^2}\left(\partial_\tau \varphi_{\sigma} - a_{0}\right)^2 + \left(\vect \nabla \varphi_{\sigma} - \vect a\right)^2 \right]~.
\end{align}
This low-energy theory describes two fields that disperse linearly with velocity $c = \sqrt{u_B v_B}$---Goldstone modes associated with the two condensed parton species. Integrating out $\varphi_\sigma$ results in the familiar Meissner response
\begin{align}
\label{eq.BP_phases_SF+SF-AFM_Meissner}
 \mathcal{L}_{\text{Meissner}} =\frac{v_B}{2\pi } \bar{a}_{\mu} \left[\delta^{\mu\nu} - \frac{p^\mu p^\nu}{\vec p \cdot \vec p}\right]\bar{a}_{\nu}~,
\end{align}
where $\vec p = (\frac{\omega}{c},k_x,k_y)$ and $\vec {\bar{a}} = (\frac{a_0}{c},a_1,a_2)$.
Finally, the periodicity of the original cosine in Eq.~\eqref{eq.BP_phases_SF+SF-AFM_deltaL-quad} permits $2\pi$ vortices in either condensate, which are logarithmically confined as in Sec.~\ref{subsubsec.CW_coupled_AFM}.

\textit{Gauge fluctuations}---We reinstate the gauge-field dynamics, governed by $\mathcal{L}_{\text{MW}} + \mathcal{L}_{\mathcal{M}}$ of Eqs.~\eqref{eq.BP_symm+theory_L-Maxwell} and \eqref{eq.BP_symm+theory_monopoles2}. The induced Meissner term renders the gauge field massive and, thereby, monopoles strongly irrelevant. We verify this explicitly by integrating out $\vec a$ to obtain the effective monopole Lagrangian. In the limit of small frequencies and momenta, we find
\begin{align}
\label{eq.BP_phases_SF+SF-AFM_eff-monopole-L}
 \mathcal{L}_{\mathcal{M},\text{eff}} = \frac{ \tilde{v}}{4\pi \kappa}\left[1+\frac{\kappa c^2 d_{*}^{2} }{2 v_B\tilde{v}} \left(\frac{\omega^2}{c^2}+k_y^{2}\right)\right]k_x^{2} \left|\phi_{\mathcal{M}}\right|^2~.
\end{align}
The corresponding monopole-monopole correlation function decays faster than exponentially, i.e.,
\begin{align}
\label{eq.BP_phases_SF+SF-AFM_monopole-corr-result}
 C^{\mathcal{M}}_{\vect{R}} \propto \begin{cases}
 e^{-\nicefrac{x^2}{\xi_1^2}}\qquad &y = 0~,\\
 e^{-\nicefrac{L}{\xi_2}}\qquad\qquad &y \neq 0~,
 \end{cases}
\end{align}
where $L$ is the wire length, and $\xi_{1,2}$ are nonuniversal length scales.

\textit{Quasiparticles}---To determine the fate of the partons, we focus on one species and integrate out the other. Recall that the gauge field $\vec{a}$ lives on all dual wires, while each parton species resides on wires with a specific parity. We, therefore, integrate out $b_\downarrow$ and the gauge field on even wires to obtain the effective gauge theory for $b_\uparrow$ on odd wires. The long-wavelength expansion of the gauge-field action reproduces the form of $\mathcal{L}_{\text{Meissner}}$, in Eq.~\eqref{eq.BP_phases_SF+SF-AFM_Meissner}, with rescaled fields and momenta (see Appendix \ref{app.gauge-calc_CondBosons} for the exact wire-based calculation). Integrating out the massive gauge fluctuations does not qualitatively change the low-energy theory for $b_\uparrow$. In the present case, it is of the form ${\cal L}_\text{SF}$ in Eq.~\eqref{eq.CW_coupled_AFM_Leff}, and vortices in the phase of $b_\uparrow$ are logarithmically confined.

\begin{figure}[tb]
\centering
 \includegraphics[width=\columnwidth]{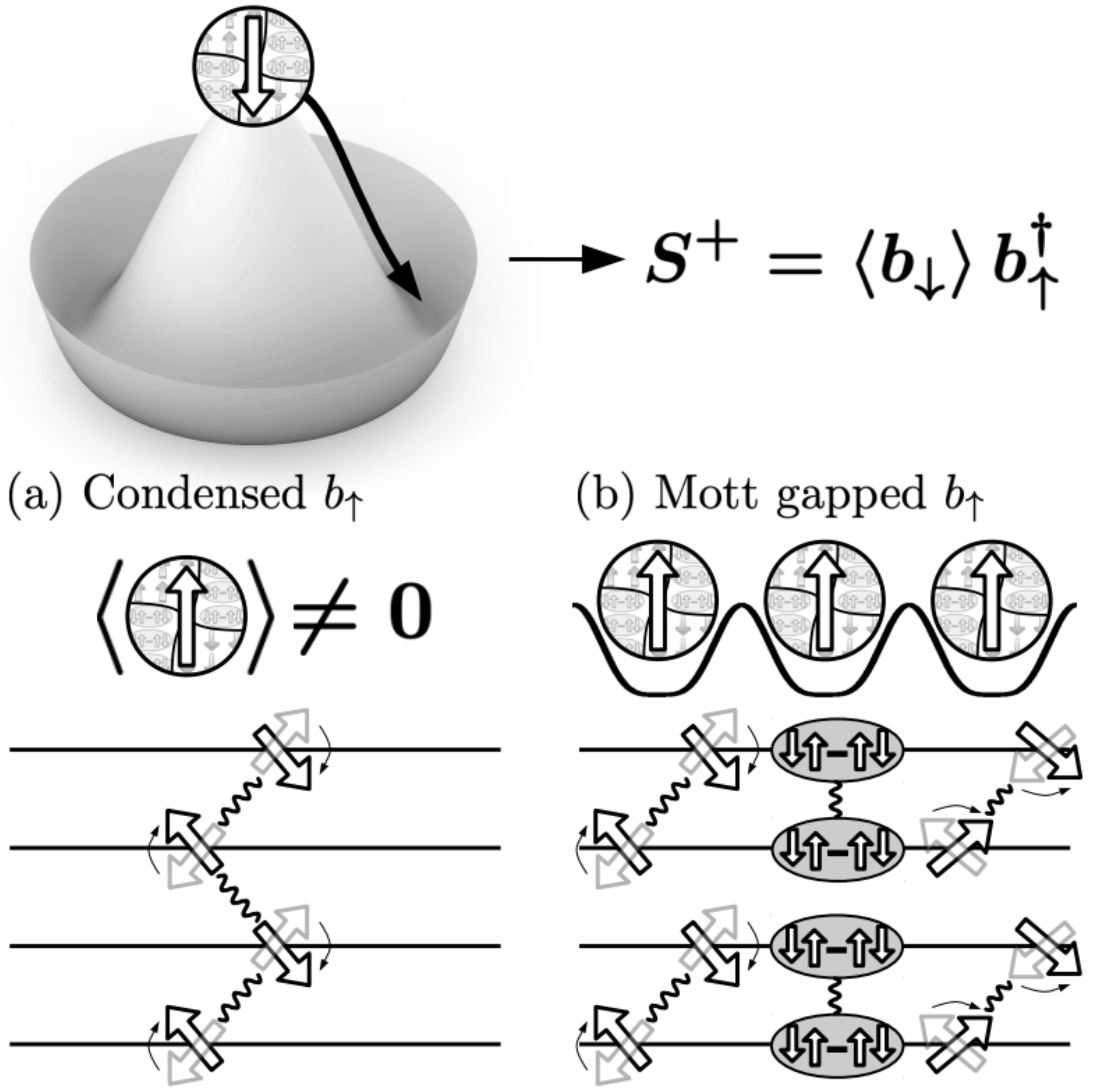}
 \caption{When one species of bosonic partons condenses, the gauge field acquires a Higgs mass. Moreover, the condensate renders the second parton species local and identifies it with the microscopic spin-raising operator. (a) Condensation of the second species results in a magnetically ordered phase with $\langle S^+\rangle\neq 0$. The corresponding interwire terms are precisely the ones discussed for the easy plane AFM in Sec.~\ref{subsubsec.CW_coupled_AFM}. (b) When the second parton species forms a Mott insulator, a non-magnetic state arises. The parton theory that realizes this phase maps onto the coupled-wire model of an interwire VBS.
 \label{fig:condensed}}
\end{figure}

It is instructive to analyze the role of the external probing field as introduced in Sec.~\ref{subsec.BP_symm+theory}. At the mean-field level, the partons couple to $\vec{A}$ with charges $(e_{\uparrow},e_{\downarrow})=(1/2,-1/2)$. Condensation of either species forces the flux of $\vec A$ to be quantized in units of $4\pi$. However, in the gauge theory, a simple shift $a_{\mu,\tilde{y}} \rightarrow a_{\mu,\tilde{y}} + \frac{1}{4} \left[S A_{\mu}\right]_{\tilde{y}}$ leads to $(e_{\uparrow},e_{\downarrow})=(1,0)$ and, consequently, $2\pi$ quantization. This apparent ambiguity disappears when gauge fluctuations are accounted for. Indeed, integrating out $\vec a$ and $b_\downarrow$ in the presence of $\vec{A}$, we find an effective field theory for $b_\up$ with $e_\uparrow=1$. The same conclusion can be reached by noting that, once $b_\downarrow$ has a non-zero expectation value, $b_\uparrow$ is identified with the spin raising operator $S^+$ [see Fig.~\ref{fig:condensed}(a)].
Further integrating out the remaining parton, $b_\up$, yields a Meissner response in the form of Eq.~\eqref{eq.BP_phases_SF+SF-AFM_Meissner} for $\vec{A}$ (see Appendix \ref{app.gauge-calc_CondBosons} for details). Consequently, $U(1)$ spin-rotation symmetry is spontaneously broken. 

\textit{Spin model}---The microscopic phase breaks $U(1)$ spin-rotation as well as time-reversal symmetries (see Table \ref{tab.BP_symm+theory_symmetries}). Moreover, the discrete translation symmetries in the $\hat x$ and $\hat y$ directions are both reduced to steps of two. These properties identify the microscopic phase as the easy-plane AFM described in Sec.~\ref{subsubsec.CW_coupled_AFM}. Indeed, the parton gauge theory maps onto ${\cal L}_\text{LL} + \mathcal{L}_t$, which was studied there in detail. We found the same gapless ground state, topological excitations, and, implicitly, the same $2\pi$ quantization of flux. 

\subsubsection{Correlated Mott insul. / interwire VBS}
\label{subsubsec.BP_phases_SF+Mott_VBSy}
\textit{Mean field}---Consider now a superfluid phase of one parton species and a Mott insulator of the other, i.e., $\delta \mathcal{L} = g_t\cos\left(\varphi_{2\tilde y+2,\downarrow}-\varphi_{2\tilde y,\downarrow}\right) + g_u \cos \left(2 \theta_{2\tilde{y}+1,\uparrow} \right)$. The mean-field analysis is the same as in the two previous cases; the low energy theory contains the Goldstone mode of the condensed $b_\downarrow$ and individual $b_\uparrow$ as finite energy excitation. Vortices in the $b_\downarrow$--condensate, created by $V_\text{odd} \sim e^{-i \tilde \varphi_\downarrow}$ [cf.~Eq.~\eqref{eq.BP_bosonic-parton-vortices}], are logarithmically confined. 

\textit{Gauge fluctuations}---As in the parton superfluid phase, Sec.~\ref{subsubsec.BP_phases_SF+SF-AFM}, the gauge field acquires a Higgs mass through the $b_\downarrow$ condensate. Consequently, monopoles can again be safely discarded. 

\textit{Quasiparticles}---The effect of gauge fluctuations on the mean-field excitation $b_\uparrow$ can be inferred, as in the last section, by successively integrating out $b_\downarrow$ and $\vec a$.
Exciting a single $b_\uparrow$ boson above the Mott gap thus corresponds microscopically to a spin-1 excitation [see Fig.~\ref{fig:condensed}(b)]. In addition, vortex excitations $V_\text{odd}$ turn into local spin-0 quasiparticles. Formally, this follows from the qualitatively different behavior of the correlation function $\langle[\Delta_2\varphi_\down]_{2\tilde{y}+1}[\Delta_2\varphi_\down]_{2\tilde{y}'+1}\rangle$ at zero frequency: at the mean-field level, it falls off quadratically with distance, but in the gauge theory, it decays exponentially. (For an explicit calculation of the vortex energy, see Appendix \ref{app.defect-energies_Higgs-vortices}.)

\textit{Spin model}---In this phase, $y$-translation symmetry is broken, but $\tilde{y}$ inversion is preserved (cf.~Table \ref{tab.BP_symm+theory_symmetries}). Moreover, $U(1)$ spin-rotation, time-reversal, and $x$-translation symmetries all remain intact. These symmetry properties, along with the quasiparticle content, imply an interwire VBS. To verify this explicitly, we transcribe the cosines in $\delta \mathcal{L}$ to spin variables
\small
\begin{align}
\label{eq.BP_phases_SF+Mott_VBSy_L-microscopic}
 \delta \mathcal{L}_{\text{inter}-\text{VBS}} = g_t \cos{\bigl(\left[\Delta \Phi\right]_{2\tilde{y}+1}\bigr)} + g_u \cos{\bigl(2\left[S\Theta\right]_{2\tilde{y}+1}\bigr)}~.
\end{align}
\normalsize
One readily verifies that the arguments of the two cosines commute. Consequently, the two terms can simultaneously reach strong coupling. In the resulting fixed-point Hamiltonian, only pairs of wires are coupled. To characterize the ground state, it is thus sufficient to analyze a two-leg ladder. 

We diagonalize the interaction by introducing new conjugate variables $\Phi_\pm =\frac{1}{2} \left(\Phi_{2}\pm\Phi_{1}\right)$ and $\Theta_{\pm}=\left(\Theta_{2}\pm\Theta_{1}\right)$. The fields $\Phi_-$ and $\Theta_+$ get trapped around the minima of their respective cosines, and small fluctuations are massive. Fundamental domain walls in the two are created by $D_+ = e^{i \Phi_+}$ and $D_- = e^{i \Theta_-}$. The former is identified with the spin-raising operator, i.e., $D_+ \propto S^+_{1/2} $, with a proportionality factor determined by the pinned $\Phi_-$. The latter similarly describes $2\pi$ phase slips in $S^+$, i.e., $D_- \propto e^{ \mp 2 i \Theta_{1,2}}$. Consequently, the two types of defects carry spin 1 and spin 0, respectively. Alternatively, the spin can be computed via the general expression
\begin{align}
\label{eq.BP_phases_SF+Mott_VBSy_CalcSz-excitation}
 \delta S^z_{\text{tot.}} \left[D\right] = \int_x \sum\nolimits_y \left< D \left[S^z_y, D^\dagger \right] \right>~.
\end{align}
In the present case, $\sum\nolimits_y S^z_y =\frac{1}{\pi}\partial_x \Theta_+$ and we again find $\delta S^z_{\text{tot.}}\left[D_-\right]=0$ and $\delta S^z_{\text{tot.}}\left[D_+\right]=1$.

\subsubsection{Quantum Hall states / Chiral spin liquids}
\label{subsubsec.BP_phases_QH-CSL}
\textit{Mean field}---As the first example of a fractionalized phase, consider a bilayer quantum Hall state of bosonic partons. At filling factor $\nu = 2/n$, it can be realized with the interwire coupling
\begin{align}
\label{eq.BP_phases_QH-CSL_L}
 \delta \mathcal{L}_{\text{QH}} = g_{\text{QH}}\cos\left(\varphi_{\tilde{y}+1,\sigma} - \varphi_{\tilde{y}-1,\sigma} - 2n\theta_{\tilde{y},\bar{\sigma}}\right)~.
\end{align}
This wire construction was proposed in Ref.~\onlinecite{TeoKane2014} for microscopic bosons $e^{i \varphi}$ and analyzed in detail.
Adapted to the present context, the resulting phase hosts two species of spin-$1/{2n}$ excitations that are self-bosons but exhibit mutual statistics $\pi/n$, i.e., the corresponding $K$-matrix is $n$ times the Pauli matrix $\sigma_x$. To find $K$, we calculate the response to $\vec{a}$ by replacing $\delta \mathcal{L}_\text{QH}$ with its quadratic expansion and integrating out the matter fields (see Appendix \ref{app.gauge-calc_BIQH}). The leading contribution at long wavelength is
\begin{align}
\label{eq.BP_phases_QH-CSL_L-gauge}
 \mathcal{L}_{\text{CS}} = \frac{i}{2\pi n} \left[S a_{1}\right]_{y}\left[\Delta a_{0}\right]_{y} \sim \frac{i}{2\pi n} \epsilon_{\mu\nu} a_\mu \partial_y a_\nu~,
\end{align}
where $\epsilon_{\mu\nu}$ is the antisymmetric tensor. Endowing $\vec{a}$ with a (redundant) spin label according to the dual-wire parity, the induced action for $\vec{a}_\sigma$ takes the expected form. In the continuum limit, we find 
\begin{align}
\label{eq.BP_phases_QH-CSL_L-CS}
 \mathcal{S}_{\text{CS}} = \frac{i}{4 \pi} \int_{x,y,\tau} \sum\limits_{\sigma,\sigma'} \epsilon_{\mu\nu\kappa}a_{\mu,\sigma} \left[K^{-1}\right]_{\sigma\sigma'} \partial_\nu a_{\kappa,\sigma'}~,
\end{align}
in the gauge $a_2=0$ and with $K=n\sigma_x$.

\textit{Gauge fluctuations}---We restore the status of $\vec a$ as dynamical, with fluctuations governed by the sum of the induced Chern-Simons and bare Maxwell terms. The latter contains the contribution $\propto \left[\Delta a_\mu\right]^2$, which translates into $\propto \left(a_{\mu,\uparrow}-a_{\mu,\downarrow}\right)^2$. 
This term renders the antisymmetric combination of $a_{\mu,\sigma}$ massive, while the Chern-Simons term results in a gap for the symmetric combination. Consequently, monopoles are strongly irrelevant and can be safely discarded. 

The Chern-Simons action, Eq.~\eqref{eq.BP_phases_QH-CSL_L-gauge}, implies that both microscopic and bosonic-parton time-reversal symmetries are broken. To see the latter, it is convenient to compute the response to the external probing field $\vec{A}$. Including it amounts to replacing $a_{\mu,\tilde y} \rightarrow a_{\mu,\tilde y} - \frac{1}{4}(-1)^{\tilde y} \left[S A_\mu\right]_{\tilde y}$ in the induced action (but not in the bare one). Integrating out the emergent gauge field we find the response 
\small
\begin{align}
\label{eq.BP_phases_QH-CSL_L-A-CS}
 \mathcal{L}_{A-\text{CS}} = -\frac{i}{8\pi n} \left[S A_1\right]_{\tilde{y}}\left[\Delta A_0\right]_{\tilde{y}} \sim -\frac{i}{8\pi n} \epsilon_{\mu\nu} A_\mu \partial_y A_\nu~.
\end{align}
\normalsize
Consequently, these phases are chiral and must exhibit topologically protected edge states.

\textit{Quasiparticles}---To identify the quasiparticles, we transform the partons into new composite bosons. Specifically, we attach to each parton $n$ fluxes of the opposite one. On the operator level, this procedure amounts to introducing bosons $\beta_{\tilde y,\sigma}=e^{-i\eta_{\tilde{y},\sigma}}$ with $\eta_{\tilde{y},\sigma} \equiv \varphi_{\tilde{y},\sigma} + n\tilde{\varphi}_{\bar{\sigma},\tilde{y}}$. Such manipulations often become more transparent in a schematic description that specifies only the couplings between particle currents, $\vec{j}_{b,\sigma}$, and gauge fields. The bosonic-parton theory in Eqs.~\eqref{eq.BP_symm+theory_L-0} and \eqref{eq.BP_symm+theory_L-Maxwell} is then expressed as
\begin{align}
 \mathcal {L}_{b} & = i \sum\nolimits_\sigma \vec{j}_{b,\sigma} \cdot \left(\vec{a} +\textstyle\frac{\sigma}{2}\vec{A}\right) + \cdots~,
\end{align}
where the ellipsis denotes kinetic terms for partons and dynamical gauge fields as well as short-range interactions ($\vec{A}$, as always, is an external probing field). Attaching $n$ mutual fluxes amounts to replacing $\mathcal{L}_{b} \rightarrow {\cal L}_\beta$ with
\begin{align}
\label{eq.BP_phases_QH-CSL_L-eta+a} \mathcal{L}_\beta= i \sum\nolimits_\sigma \vec{j}_{\beta,\sigma} \cdot \left(\vec{c}_\sigma + \vec{a}+\textstyle\frac{\sigma}{2}\vec{A}\right) - \frac{i}{2\pi n}c_\up d c_\down~.
\end{align}
Finally, we shift $\vec{c}_\sigma \rightarrow \vec{c}_\sigma -\vec{a}$ to decouple $\vec{a}$ from the matter fields and integrate it out to obtain
\begin{align}
\label{eq.BP_phases_QH-CSL_L-eta}
 \mathcal {L}^\prime_\beta & = i \sum\nolimits_\sigma \vec{j}_{\beta,\sigma} \cdot \left(\vec{c}_\sigma + \textstyle\frac{\sigma}{2}\vec{A}\right) + i \frac{(c_\uparrow - c_\downarrow)d(c_\uparrow - c_\downarrow)}{8\pi n}~.
\end{align}
In terms of composite bosons, the interwire coupling reads as $\delta {\cal L}_\text{QH} = g_\text{QH} \cos\left(\eta_{\tilde{y}+1,\sigma}-\eta_{\tilde{y}-1,\sigma}\right)$, i.e., $\beta_\sigma$ form a superfluid. At the mean-field level, the excitations are two flavors of logarithmically confined vortices in the phases of $\beta_{\sigma}$. In the presence of the dynamical gauge fields $\vec c_\sigma$, these turn into finite-energy excitations subject to the constraint that $\left(-2\pi\right)$ flux of $\vec c_\sigma$ must be accompanied by $\sigma/2n$ charge of \textit{each} boson, i.e.,
\begin{align}
\label{eq.BP_phases_QH-CSL_CS-constraint}
 \rho^\beta_\sigma= -\frac{1}{4\pi n}\vec{\nabla}\times\vec{c}_{\sigma} + \frac{1}{4\pi n}\vec{\nabla}\times\vec{c}_{\bar{\sigma}}~.
\end{align}
Since $\rho^\beta_\sigma$ is related to the physical spin [charge under $\vec A$, see Eq.~\eqref{eq.BP_phases_QH-CSL_L-eta}] via $S^z=(\rho^\beta_\uparrow-\rho^\beta_\downarrow )/2$,
the composites carry a total spin of $1/2n$. Moreover, a clockwise exchange results in a statistical phase $\pi/2n$.

\textit{Spin model}---The response to the external probing field, Eq.~\eqref{eq.BP_phases_QH-CSL_L-A-CS}, implies that the microscopic phase is a chiral QSL with topological edge states and fractionalized quasiparticles in the bulk.
Translating the interwire coupling in Eq.~\eqref{eq.BP_phases_QH-CSL_L} to microscopic variables we find
\begin{align}
\label{eq.BP_phases_QH-CSL_microscopic-L}
 \delta \mathcal{L}_{\text{CSL}} = g_{\text{QH}} \cos{\bigl(\left[\Delta\Phi\right]_{\tilde{y}} + 2n \left[S\Theta\right]_{\tilde{y}}\bigr)}~.
\end{align}
Precisely this coupling, with $n=1$, was proposed in Refs.~\onlinecite{Gorohovsky2015,Meng2015a}, where it was shown to realize the Kalmeyer-Laughlin chiral spin liquid; the generalization to arbitrary integers $n$ is straightforward. (The same coupling term also describes a bosonic Laughlin state at filling factor $\nu = 1/2n$, see Ref.~\onlinecite{TeoKane2014}.)
In particular, bulk quasiparticles carry spin $1/2n$ and acquire phases $\pi/2n$ upon (clockwise) exchange. 

\subsubsection{Pair condensate / \texorpdfstring{$\mathbb{Z}_2$}{} spin liquid}
\label{subsubsec.BP_phases_PairCond-Z2}
As the final example with bosonic partons, we construct a time-reversal-invariant gapped QSL. Here, the emergent gauge field must acquire a Higgs mass without condensation of either of the two species (which would lead to a symmetry-broken phase as discussed in Secs. \ref{subsubsec.BP_phases_SF+SF-AFM} and \ref{subsubsec.BP_phases_SF+Mott_VBSy}). These requirements are satisfied when composites with higher emergent gauge charges, such as parton pairs, condense. 

\textit{Mean field}---In the coupled-wire framework, realizing such a phase is straightforward. To form a superfluid of parton pairs $\Psi_{2y}\equiv b_{2\tilde{y},\downarrow}b_{2\tilde{y}-1,\uparrow} \equiv e^{ i 2\varphi_{2y}^+}$, we introduce the interwire coupling
\small
\begin{align}
\label{eq.BP_phases_PairCond-Z2_Lp}
 \delta \mathcal{L}_{\text{pair}} & = g_p \Psi^\dagger_{2y+2}\Psi_{2y} + \text{H.c.} = g_p \cos{\bigl(2\left[\Delta_2\varphi^+\right]_{2y+1}\bigr)}~.
\end{align}
\normalsize
Once this term flows to strong coupling, $\Psi$ spontaneously acquires an expectation value (cf.~Sec.~\ref{subsubsec.BP_phases_SF+SF-AFM}). Individual partons, however, are not condensed. To describe their properties, we introduce $\varepsilon_{2y} = e^{-i\varphi^-_{2y}}$ with $\varphi_{2y}^- \equiv \left( \varphi_{2\tilde{y},\downarrow}-\varphi_{2\tilde{y}-1,\uparrow}\right)/2$. The expectation value acquired by the pair field relates $ \varepsilon_{2y} \propto b_{2\tilde{y},\downarrow} \propto b_{2\tilde{y}-1,\uparrow}^\dagger $. A gapped phase where neither $b_\sigma$ condense is realized by including the interaction
\small
\begin{align}
\label{eq.BP_phases_PairCond-Z2_Lq}
 \delta\mathcal{L}_{\text{Mott}-\varepsilon} = g_q \cos \left(2\theta_{2y}^-\right) = g_q \cos \left(2\theta_{2\tilde{y},\down}-2\theta_{2\tilde{y}-1,\up}\right)~.
\end{align}
\normalsize
Elementary domain walls in this cosine are created by individual partons, which constitute the fundamental quasiparticles at the mean-field level. Vortices in the pair condensate, created by $m^\dagger_{2y+1}\equiv e^{i\left(\tilde{\varphi}_{\up,2\tilde{y}+1} +\tilde{\varphi}_{\down,2\tilde{y}}\right)/2}$, are logarithmically confined.

\textit{Gauge fluctuations}---We reinstate the gauge-field dynamics and integrate out the matter fields. The pair condensate leads to a Higgs mass for the emergent gauge field, $\vec{a}$, as in Sec.~\ref{subsubsec.BP_phases_SF+SF-AFM}. Consequently, monopoles are again strongly irrelevant and can be safely discarded.

\textit{Quasiparticles}---The analysis of quasiparticle excitations closely mirrors the one in Sec.~\ref{subsubsec.BP_phases_SF+Mott_VBSy}. Integrating out both $\theta^+,\varphi^+$ and the gauge field $\vec{a}$ results, to lowest order in $\Delta$, in 
\small
\begin{align}
\label{eqnphiminus}
 \nonumber \mathcal{L}^{\text{eff}}_{\varepsilon} = & \frac{i}{\pi} \partial_x \theta_{2y}^- \left( \partial_\tau \varphi^-_{2y}+\textstyle\frac{1}{2}A_{0,2y}\right) + \frac{\bar{v}}{2\pi}\left(\partial_x \varphi_{2y}^- + \textstyle\frac{1}{2}A_{1,2y}\right)^2\\
 & + \frac{\bar{u}}{2\pi}\left(\partial_x \theta_{2y}^-\right)^2 + g_q \cos\left(2\theta_{2y}^-\right)~,
\end{align}
\normalsize
with renormalized parameters $\bar{v}$ and $\bar{u}$. Consequently, $\varepsilon \sim \langle b_\up^\dagger\rangle_{\varphi^+,\vec{a}} \sim \langle b_\down\rangle_{\varphi^+,\vec{a}}$ creates a deconfined bosonic spin-$1/2$ excitation, as in the mean-field discussion. In addition, the dynamical gauge field liberates vortices, $m$, from their logarithmic confinement as in Sec.~\ref{subsubsec.BP_phases_SF+Mott_VBSy}; they become bona fide spin-0 bosonic quasiparticles. 

\begin{figure} [htbp]
\centering
 \includegraphics[width=\columnwidth]{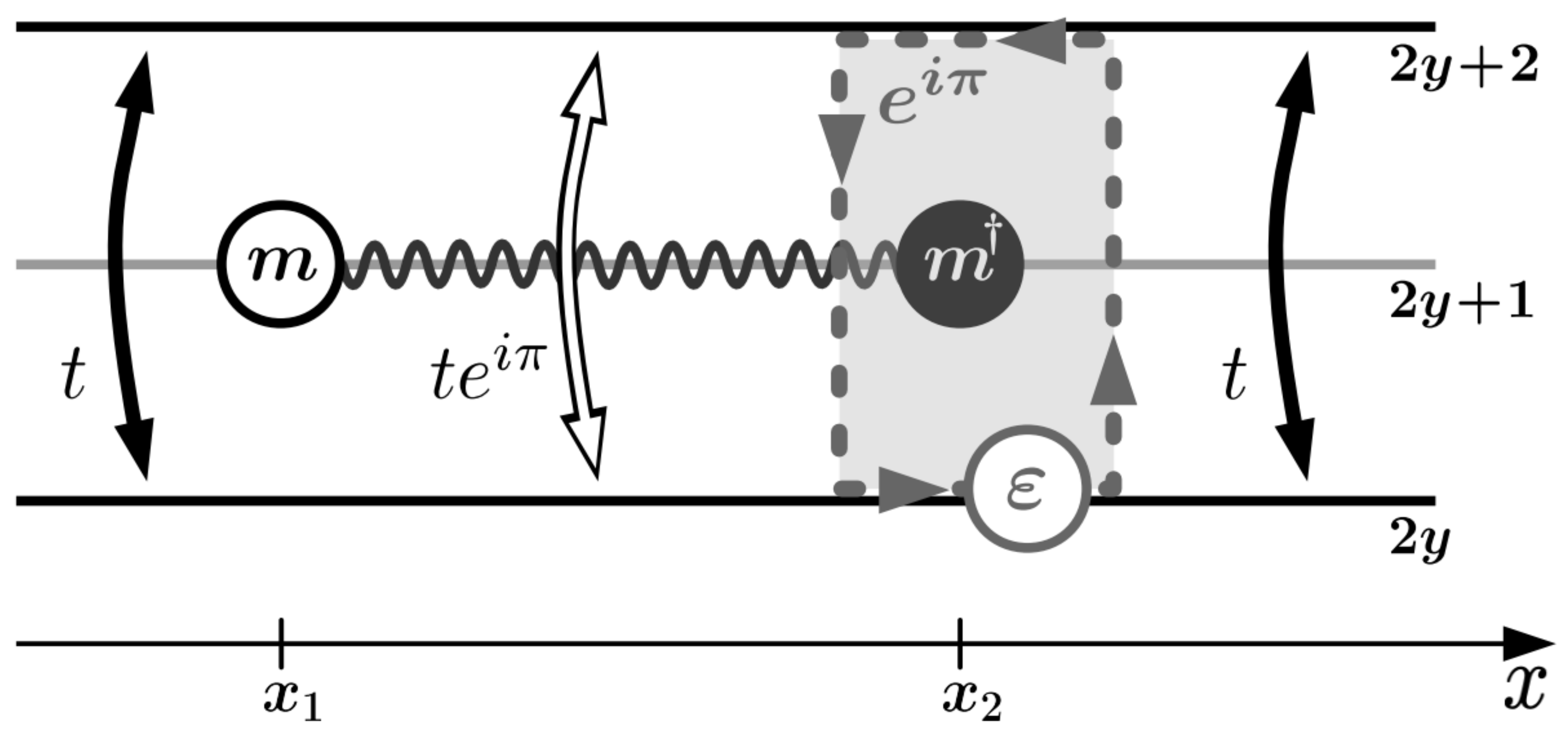}
 \caption{In the parton-pair condensate, a static configuration of $m$ quasiparticles modifies the hopping amplitudes of $\varepsilon$ excitations. An $m$
 quasiparticle-quasihole pair at $x_1$ and $x_2$ is connected by a branch cut (wavy line), across which the phase of $\varepsilon$-hopping changes by $\pi$. Consequently, the $\varepsilon$ quasiparticle acquires a minus sign upon encircling $m$, i.e., the two are mutual semions. 
 \label{fig:modified-hopping}
 }
\end{figure}

To infer the mutual statistics between the two quasiparticles, consider the hopping of $\varepsilon$. It stems from the microscopic term
\small
\begin{align}
 N^+_{2y+1}N^-_{2y} = e^{i \varphi^+_{2y+2}-i\varphi^+_{2y}}\varepsilon^\dagger_{2y}\varepsilon_{2y+2} \equiv t_{2y,2y+2}\varepsilon^\dagger_{2y}\varepsilon_{2y+2}~,
\end{align}
\normalsize
i.e., the hopping amplitude is set by the pair condensate. In the absence of vortices, $t$ is uniform. For a static vortex---anti-vortex pair, there is instead a branch cut connecting the two, across which the phase of $t$ jumps by $\pi$ (see Fig.~\ref{fig:modified-hopping}). For dynamical quasiparticles $m$ and $\varepsilon$, this property implies mutual semionic statistics.

\textit{Spin model}---The quantum numbers and braiding properties of quasiparticles are characteristic of a $\mathbb{Z}_2$ QSL. Time-reversal symmetry is preserved in this phase, but translation symmetry in the $\hat y$ direction is reduced to translations by two wires. To analyze this phase in terms of microscopic spin variables, we translate the intrawire couplings of Eqs.~\eqref{eq.BP_phases_PairCond-Z2_Lp} and \eqref{eq.BP_phases_PairCond-Z2_Lq}, finding
\begin{align}
\label{eq.BP_phases_PairCond-Za_L-micro}
 \delta{\cal L}_{\mathbb{Z}_2} =& g_p \cos\left(\Phi_{2y+2} - 2 \Phi_{2y+1} + \Phi_{2y}\right) \\
 \nonumber & + g_q \cos\left(2 \Theta_{2y+1} + 4 \Theta_{2y} + 2 \Theta_{2y-1}\right)~.
\end{align}
The two cosines do not compete, and their arguments can thus be pinned simultaneously.
For an even number of wires with periodic boundary conditions in the $\hat{y}$ direction, there are as many linearly independent pinned fields as there are wires. Consequently, a fully gapped phase can form. 

The microscopically allowed operator $e^{i 2\Theta_{2y}}$ increments the arguments of two adjacent $g_p$-cosines by $2\pi$, i.e., it creates a pair of fundamental domain walls. Similarly, $e^{-i 2\Theta_{2y+1}}$ creates a strength-2 domain wall in a single $g_p$-cosine. Consequently, a domain wall and an anti-domain wall on the wires $2 y_2+1$ and $2 y_1-1$ are created by
\begin{align}
\label{eq.BP_phases_PairCond-Z2_m-pair}
 {\cal O}^{m}_{2y_1-1,2y_2+1} & \equiv \exp\left[-2i \sum\nolimits_{2y_1 \leq y \leq 2y_2} \Theta_{y}\right]\\
 \nonumber & = \exp\left[i\Theta_{2y_1-1}\right] \exp\left[-i\Theta_{2y_2+1}\right] s^{m}_{2y_1,2y_2}~.
\end{align}
The string operator $s^{m}_{2y_1,2y_2}$ is comprised solely of the pinned combinations $\Theta_{2y+1}+2\Theta_{2y}+\Theta_{2y-1}$ and acquires a non-zero expectation value. Domain walls are thus deconfined in the $\hat{y}$ direction. 
Additionally, one readily verifies that ${\cal O}^{m}_{2y_1-1,2y_2+1} \propto m^\dagger_{2y_2-1} m_{2y_1+1}$ in the ground state. To move the domain wall along the wire direction, one need only apply ${\cal O}^{m}_{2y_0-1}(x_1,x_2) = e^{-i\intop_{x_1}^{x_2} dx \partial_x \Theta_{2y_0-1}}$. These domain-wall excitations are thus precisely the $m$ quasiparticles discussed in the gauge-theory analysis.

Similarly, we construct a second species of deconfined excitations via the operator $e^{i\left(-1\right)^y\Phi_y}$. A quasiparticle-quasihole pair on wires $2 y_2$ and $2 y_1$ is created by
\begin{align}
\label{eq.BP_phases_PairCond-Z2_e-pair}
 {\cal O}^{\varepsilon}_{2y_1,2y_2}&\equiv \exp\left[i \sum\nolimits_{2y_1 < y \leq 2y_2} \left(-1\right)^y \Phi_{y}\right]\\
 \nonumber & = \exp\left[-\textstyle\frac{i}{2}\Phi_{2y_1}\right] \exp\left[\textstyle\frac{i}{2}\Phi_{2y_2}\right] s^{\varepsilon}_{2y_1+1,2y_2-1}~.
\end{align}
The string operator $s^{\varepsilon}_{2y_1+1,2y_2-1}$ is also expressible in terms of pinned fields only, specifically the combination $\Phi_{2y+2}-2\Phi_{2y+1}+\Phi_{2y}$. Consequently, ${\cal O}^{\varepsilon}_{2y_1,2y_2} \propto \varepsilon^\dagger_{2y_2} \varepsilon_{2y_1}$ in the ground state and $\varepsilon$ is deconfined in the $\hat y$ direction. Finally, ${\cal O}^{\varepsilon}_{2y_0}(x_1,x_2) = e^{\frac{i}{2}\intop_{x_1}^{x_2} dx \partial_x \Phi_{2y_0}}$ moves $\varepsilon$ along $\hat x$.

The spin of $m$ and $\varepsilon$ can be inferred from the operators that terminate the strings in ${\cal O}^{m}$ and ${\cal O}^{\varepsilon}$ using Eq.~\eqref{eq.BP_phases_SF+Mott_VBSy_CalcSz-excitation}. The $m$ quasiparticle is spinless while $\varepsilon$ carries spin 1/2.
Finally, we compute the exchange statistics of the quasiparticles. Since all terms in Eq.~\eqref{eq.BP_phases_PairCond-Z2_m-pair} commute, $m$ has trivial self-statistics, i.e., it is a boson. The same holds for $\varepsilon$. Their mutual statistics can be read off from 
\begin{align}
\label{eq.BP_phases_PairCond-Z2_braid}
 e^{i 2 \alpha} = U_1 U_2 U_1^\dagger U_2^\dagger~, 
\end{align}
where $U_1={\cal O}^{m}_{2y_0-1}\left(x_1,x_2\right)$ and $U_2={\cal O}^{\varepsilon}_{2y_1,2y_2}\left(x_0\right)$. Braiding occurs when the paths of $\varepsilon$ and $m$ interlink, i.e., for $y_0\in[y_1,y_2]$ and $x_0\in[x_1,x_2]$. In that case, we find $e^{i 2 \alpha}=-1$, which implies that the two quasiparticles are mutual semions. Consequently, they can combine to form a fermion, schematically ${\cal O}^{\psi} \sim {\cal O}^{\varepsilon}{\cal O}^{m}$. We will see that these fermions exactly coincide with the partons that are the focus of the next section.

\section{Fermionic partons from coupled wires}
\label{sec.FP}
\begin{figure}[tb]
\centering
 \includegraphics[width=\columnwidth]{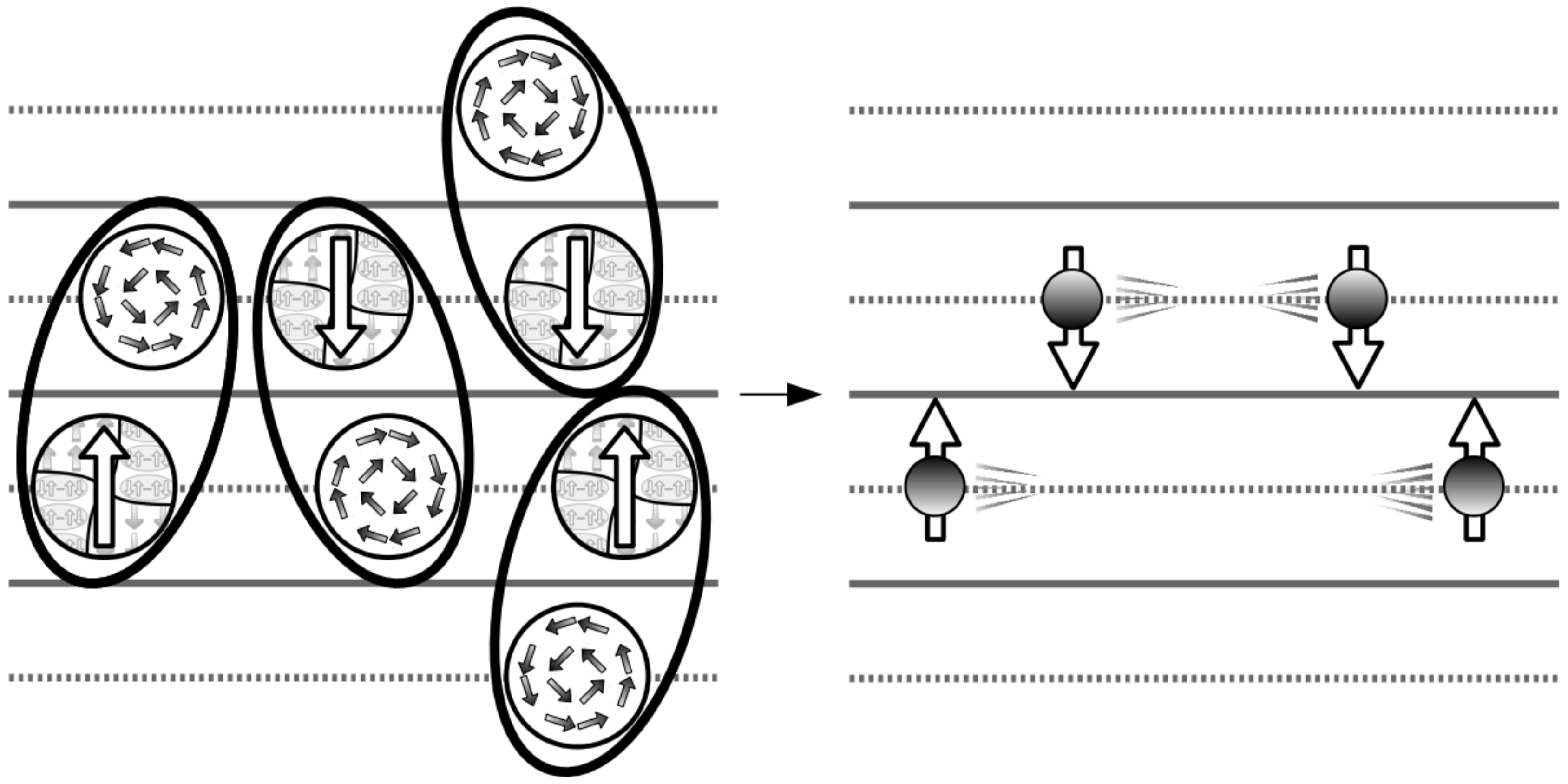}
 \caption{Magnetic vortices and dislocations are dual to each other, similar to bosons and vortices in a superfluid. Combinations of the two defects, therefore, exhibit fermionic statistics. These composite fermions inherit their species from the dislocation, i.e., $\sigma=\uparrow$ or $\sigma=\downarrow$ depending on the dual-wire parity. The relative orientation of magnetic vortex and dislocation determines the chirality of the fermion, as in the coupled-wire implementation of conventional flux attachment~\cite{Mross2017}.
 \label{fig:fermionic-partons}}
\end{figure}

Above, we have seen that the topological defects in a VBS form a bosonic-parton representation of spins. We now show that fermionic partons can also be constructed from topological defects, specifically as composites of magnetic vortices and dislocations. This route to fermionic partons is closely related to the well-known flux attachment~\cite{Wilczek1982,Fradkin1989} (see also Refs.~\onlinecite{Seiberg2016,Karch2016,Murugan2017,Senthil+Son+Wang+Xu2019} for recent refinements). Recall that $b^\dagger_{\uparrow}$ is dual to one flavor of magnetic vortices, $V_{\uparrow}=e^{i \tilde \varphi_\uparrow}$, while $b_{\downarrow}$ is dual to the other, $V_\downarrow=e^{i \tilde \varphi_\downarrow}$ [see discussion near Eq.~\eqref{eq.BP_bosonic-parton-vortices}]. Therefore, fermions can be constructed as $f_\sigma = b_\sigma V_\sigma$, i.e., by attaching $2\pi\sigma$ flux to the bosonic partons (see Fig.~\ref{fig:fermionic-partons}). Schematically this transformation can be expressed as
\begin{align} 
\label{eq.FP_field-theory-flux-attachment}
 i\sum\nolimits_{\sigma} \vec j_{b,\sigma}\cdot \vec a \rightarrow i\sum\nolimits_\sigma \vec j_{f,\sigma}\cdot (\vec c_\sigma+\vec a) + \frac{\sigma}{4\pi}c_\sigma d c_\sigma~.
\end{align}
Performing the shift $ c_{\mu,\sigma} \rightarrow c_{\mu,\sigma} - a_\mu$ and integrating out $ a_\mu$ yields the constraint $\vec{c}_\uparrow = \vec{c}_\downarrow \equiv \vec c$. The resulting theory has the same structure as the bosonic one: two species of fermions that are minimally coupled to an emergent gauge field $\vec{c}$, which obeys Maxwell dynamics.

To implement these manipulations in the wire array, we introduce new variables 
\begin{subequations} 
\label{eq.FP_fermionic-partons}
 \begin{align}
 \label{eq.FP_fermionic-partons-phi}
 \varphi_{f,\tilde y,\sigma}&=\varphi_{\tilde y,\sigma}+ \textstyle\frac{\sigma}{2}\left(\tilde \varphi_{\sigma,\tilde y +1}+\tilde \varphi_{\sigma,\tilde y -1}\right)~,\\
 \label{eq.FP_fermionic-partons-theta}
 \theta_{f,\tilde y,\sigma}&=\theta_{\tilde y,\sigma}~,
 \end{align}
\end{subequations}
where $\tilde \varphi_{\sigma}$ was defined in Eq.~\eqref{eq.BP_bosonic-parton-vortices}. The linear combinations $\phi^{\chi}_\sigma =\varphi_\sigma + \chi \theta_\sigma$, with $\chi = R/L = +/-$, satisfy
\begin{align}
\label{eq.FP_commutator}
 \left[\phi^{\chi}_{\tilde y,\sigma}(x),\phi^{\chi'}_{\tilde y',\sigma'}(x')\right] &=i\pi\delta_{\sigma\sigma'}\big\{ \chi \delta_{\chi\chi'}\delta_{\tilde y\tilde y'} \text{sgn}(x-x') \nonumber\\
 & + \sigma \text{sgn}(\tilde{y}'-\tilde{y})+\delta_{\tilde y\tilde y'}\epsilon_{\chi\chi'}\big\}~,
\end{align}
with the convention that $\text{sgn}(0)=0$. These commutators imply that the operators $f_{\tilde{y},\sigma,\chi}^\dagger \sim e^{i\phi^{\chi}_{\tilde{y},\sigma}}$ anti-commute for equal $\sigma$. The associated densities $\rho_{\tilde y ,\sigma,\chi}\equiv \frac{\chi }{2\pi}\partial_x \phi_{\tilde y ,\sigma,\chi} $ are chiral; they describe right- and left-movers for $\chi=R$ and $\chi=L$, respectively. Moreover, the \textit{total} density on the $\tilde y$th dual wire, $\rho_{\tilde{y},\sigma} =\rho_{\tilde{y},\sigma,R}+\rho_{\tilde{y},\sigma,L} = \frac{1}{\pi}\partial_x \theta_{f,\tilde{y},\sigma}$, is identical to that of bosonic partons. Consequently, the particle number of each species, $Q_\sigma$, is separately conserved [cf.~Eq.~\eqref{eq.CW_coupled_VBS_partons-number}] and particles created by $f_{\uparrow}$ and $f_{\downarrow}$ are distinguishable. Their exchange phase, which is trivial according to Eq.~\eqref{eq.FP_commutator}, is merely a gauge choice.\footnote{A redefinition $f_{\uparrow} \rightarrow f_{\uparrow} e^{i \pi Q_{\downarrow}}$ results in the anti-commutation of $f_{\uparrow}$ and $f_{\downarrow}$ without affecting the action or the full-braiding phase.} Only phases acquired during full braiding processes carry significance; in the present case, they are also trivial. These properties identify $f_{\sigma}$ with the fermions obtained via the schematic flux attachment described by Eq.~\eqref{eq.FP_field-theory-flux-attachment}.

To translate generic interwire couplings, the following identities are useful:
\begin{subequations}
\label{eq.FP_coupling-identities}
\begin{align}
 &e^{i \left(2\Theta_{y+1} + 2\Theta_{y}\right)}=
 \begin{cases}
 f_{\tilde{y},\downarrow,L}^\dagger f_{\tilde{y},\downarrow,R}
 & \tilde{y} \text{ even}~, \\
 f_{\tilde{y},\up,R}^\dagger f_{\tilde{y},\up,L}
 \quad & \tilde{y} \text{ odd}~,
 \end{cases}\\
 &e^{i \left(\Phi_{y+1} - \Phi_{y}\right)}=
 \begin{cases}
 f_{\tilde{y}+1,\up,R}^\dagger f_{\tilde{y}-1,\uparrow,L}
 \quad & \tilde{y} \text{ even}~, \\
 f_{\tilde{y}-1,\downarrow,R}^\dagger f_{\tilde{y}+1,\downarrow,L}
 &\tilde{y} \text{ odd}~.
 \end{cases}
\end{align}
\end{subequations}
In particular, the interwire couplings in Eq.~\eqref{eq.CW_coupled_couplings}, which generate the AFM and VBS phases, become simple hopping terms for the fermions
\begin{subequations}
\label{eq.FP_couplings}
\begin{align}
\label{eq.FP_couplings-tunneling}
 \sum\nolimits_y \mathcal{L}_{t} = & g_t \sum\nolimits_{\tilde {y}} f^\dagger_{\tilde{y}+\sigma,\sigma,R} f_{\tilde{y}-\sigma,\sigma,L} + \text{H.c.}~, \\ 
\label{eq.FP_couplings-backscatter}
 \sum\nolimits_y \mathcal{L}_{u} = & g_u \sum\nolimits_{\tilde y}f^\dagger_{\tilde{y},\sigma,R} f_{\tilde{y},\sigma,L}+\text{H.c.}~
\end{align}
\end{subequations}
As in the case of bosonic partons, trivial umklapp processes are allowed, which implies unit filling. When it is the most relevant term, $\mathcal{L}_{u}$ opens a band gap, as would be the case for weakly interacting fermions. In the present case, such a trivial phase can be avoided by several mechanisms: Firstly,
it stands in competition with a quantum Hall insulator generated by $\mathcal{L}_u$. Secondly, interactions can render correlated processes strongly relevant and drive the partons to a new fixed point where umklapp processes are irrelevant. Lastly, in certain microscopic spin models ${\cal L}_u$, is altogether absent. This is the case, e.g., on a triangular lattice due to geometric frustration~\cite{Nersesyan98}.

Finally, destroying a parton of spin $\downarrow$ and creating one with spin $\uparrow$ yields the smooth component of the microscopic spin-raising operators
\begin{subequations}
\label{eqn.jinfermions}
\begin{align}
 f_{2\tilde{y}+1,\uparrow,R}^\dagger f_{2\tilde{y}\downarrow,R} &= J^+_{2y+1,R}\equiv e^{ i \Phi_{2y+1}+2 i \Theta_{2y+1}}~,\\
 f_{2\tilde{y}-1,\uparrow,L}^\dagger f_{2\tilde{y},\downarrow,L} &= J^+_{2y,L}\equiv e^{ i \Phi_{2y}-2 i \Theta_{2y}}~,
\end{align}
\end{subequations}
similar to the parton decomposition of Eq.~\eqref{eq.PC_definitions}.
The analogous expression for bosonic partons instead gives the staggered component of the spin. Of course, both contributions must be encoded in either parton representation. The respective missing ones are encoded nonlocally in monopole operators, as we will see below.

\subsection{Gauge theory}
\label{subsec.FP_symm+theory}
To derive the action for fermionic partons, we proceed as we did for the bosons in Sec.~\ref{subsec.BP_symm+theory}. We find $\mathcal{S}_f = \int_{x,\tau} \sum\nolimits_{\tilde{y}} \left[ \mathcal{L}_{f} + \mathcal{L}_{\text{Maxwell}}+\mathcal{L}_{\text{int}}\right]$ with
\begin{align}
\label{eq.FP_symm+theory_L-feilds}
 \nonumber &\mathcal{L}_{f} = 
 \frac{i}{\pi}\partial_x\theta_{f,\tilde{y},\sigma}(\partial_\tau \varphi_{f,\tilde{y},\sigma} -a_{0,\tilde{y}}) + \frac{v_F}{2\pi} \left(\partial_x \varphi_{f,\tilde{y},\sigma} - a_{1,\tilde{y}}\right)^2 \\
 &\qquad + \frac{u_F}{2\pi}\left(\partial_x \theta_{f,\tilde{y},\sigma}\right)^2~,\\
\label{eq.FP_symm+theory_L-maxwell}
 &\mathcal{L}_{\text{Maxwell}}=\frac{\kappa}{8\pi \tilde{v}} \left[\Delta a_{0}\right]_{y}^2 + \frac{\kappa\tilde{v}}{8\pi} \left[\Delta a_{1}\right]_{y}^2~.
\end{align}
The final term, $\mathcal{L}_\text{int}$, contains exponentially decaying interwire terms (see Appendix \ref{app.gauge_fermions} for a detailed derivation and expressions for $v_F,$ $\kappa,$ and $\tilde v$). It is instructive to express ${\cal L}_f$ in terms of the non-chiral fermions $f_{\tilde{y},\sigma}=f_{\tilde{y},\sigma,R}e^{i k_F x}+f_{\tilde{y},\sigma,L}e^{-i k_F x}$. We find
\begin{align}
\label{eq.FP_symm+theory_L}
 \mathcal{L}_f^\prime = & f^\dagger_{\tilde{y},\sigma} \left(\partial_\tau - ia_0 - \mu \right)f_{\tilde{y},\sigma}+ \frac{v_F}{2k_F}\left| \left(\partial_x - i a_1\right) f_{\tilde{y},\sigma}\right|^2 ~.
\end{align}
To determine the chemical potential $\mu$, notice that the value of $k_F$ only carries significance relative to another length. In the present case, this scale is given by the lattice spacing of the underlying spin-chain, which enters the fermionic theory through Eq.~\eqref{eq.FP_couplings}. 

\subsubsection*{Monopoles}
The $4\pi$ phase-slip term of Eq.~\eqref{eq.CW_4pi-phase-slip} is nonlocal in terms of fermionic partons. In the discussion of bosonic partons, we expressed phase slips through the operator $\widehat{\mathcal{M}}$ [cf.~Eq.~\eqref{eqn.monopoleoperator}]. Since $\theta_{\sigma}=\theta_{f,\sigma}$, it again acts as the insertion of a fundamental ($2\pi$) monopole in the emergent gauge field $\vec{a}$. Microscopically, monopoles encode the N\'eel vector through $
N^+_{y} = \widehat{\mathcal{M}}_{y} J^+_{y,\chi}$,
with $\chi = (-1)^y$, and where $J^+$ is expressed using fermion operators in Eq.~\eqref{eqn.jinfermions}. The same procedure as for bosonic partons (cf.~Sec.~\ref{subsubsec.BP_symm+theory_monopole}) leads to the monopole Lagrangian
\begin{align} 
\label{eq.FP_symm+theory_monopoles2}
 \mathcal{L}_\mathcal{M} = \frac{g_{\mathcal{M}}}{2} \left[\mathcal{M}_{\vect{r}}^2 + \mathcal{M}^{2\dagger}_{\vect{r}}\right]-\frac{\kappa}{8\pi \tilde{v}} \left[\Delta a_0 - \frac{i 2\tilde{v}}{\kappa} \partial_x \phi_{\mathcal{M}}\right]_{y}^2
 ~,
\end{align} 
with parameters $\kappa,\tilde v$ as in Eq.~\eqref{eq.FP_symm+theory_L-maxwell}. As before, when $g_{\mathcal{M}}=0$, the monopole field $\phi_{\mathcal{M}}$ does not affect any gauge-field or matter correlation function. 

\subsubsection*{Symmetries}
We conclude the description of the parton gauge theory by discussing how microscopic symmetries are encoded. One significant difference from the case of bosonic partons is, that certain microscopic symmetries are realized nonlocally. This property may be readily understood from the flux-attachment interpretation of fermionic partons: Time-reversal flips the winding of dislocations (cf.~Fig.~\ref{fig:dislocations}), but not of magnetic vortices. Consequently, it transforms $f_\sigma=b_\sigma V_\sigma$ onto a dual set of fermions $d_\sigma \equiv b_\sigma^\dagger V_\sigma$. The same dual fermions also arise under translation along $\hat y$, which takes $b_{\sigma} \rightarrow \sigma b^\dagger_{\bar \sigma}$ (cf.~Table \ref{tab.BP_symm+theory_symmetries}). 

We summarize the actions of all previously discussed microscopic symmetries on the fermionic partons in Table \ref{tab.FP_symm_fermionic-parton-symmetries}. As in the case of bosonic partons, it is convenient to keep track of the $U(1)$ spin-rotation symmetry by introducing the appropriate external probing field $\vec{A}$ [see Eq.~\eqref{eq.BP_symm+theory_A-microscopic}]. To lowest order in $\Delta$, it enters $\mathcal{L}_{f}$ and $\mathcal{L}_{f}'$ by replacing $a_{\mu,\tilde y} \rightarrow a_{\mu,\tilde y} - \frac{1}{4} (-1)^{\tilde{y}}\left[S A_\mu\right]_{\tilde y}$. 

\begin{center}
\begin{table}[ht]
\caption{The action of microscopic symmetries on the fermionic partons, and their gauge-theory interpretation. Certain symmetries such as time reversal act nonlocally as dualities, i.e., transform $f_\sigma$ into $d_\sigma$. As before, asterisks denote anti-unitary symmetries.}
\begin{tabularx}{\linewidth}{l r@{}l l l }
\hline\hline
\textbf{Microscopic} & & \multirow{2}{*}{$f^\dagger_{\up,R}f_{\down,L}$\quad} & \multirow{2}{*}{$f^\dagger_{\sigma,R}f_{\sigma,L}$\qquad\hspace{-2pt}} & \textbf{Parton} \\
\textbf{symmetry} & & & & \textbf{interpretation} \tabularnewline
\hline
$U\left(\alpha\right)$ & $e^{i\alpha}$ & $f^\dagger_{\up,R}f_{\down,L}$ & $f^\dagger_{\sigma,R}f_{\sigma,L}$ & Global $Q_\uparrow - Q_\downarrow$ \tabularnewline
& & & & gauge transform. \tabularnewline
$\mathcal{T}^*$ & $-$ & $d^\dagger_{\up,L}d_{\down,R}$ & $d^\dagger_{\sigma,L}d_{\sigma,R}$ & Duality \tabularnewline
$\Pi_y$ & $-$ & $f^\dagger_{\down,L}f_{\up,R}$ & $f^\dagger_{\sigma,L}f_{\sigma,R}$ & PH \tabularnewline
$T_x$& & $f^\dagger_{\up,R}f_{\down,L}$ & $f^\dagger_{\sigma,R}f_{\sigma,L}$ & $x$-translation \tabularnewline
$T_y$ & & $d^\dagger_{\down,R}d_{\up,L}$ & $d^\dagger_{\bar{\sigma},R}d_{\bar{\sigma},L}$& TR + Duality \tabularnewline
$\mathcal{T}^{\prime *}$ & & $f^\dagger_{\down,L}f_{\up,R}$ & $f^\dagger_{\bar{\sigma},L}f_{\bar{\sigma},R}$ & TR \tabularnewline
$\Pi_y \cdot T_y \cdot \mathcal{I}_x $ & & $f^\dagger_{\down,L}f_{\up,R}$ & $f^\dagger_{\bar{\sigma},L}f_{\bar{\sigma},R}$ & $x$-inv. + spin-flip
\tabularnewline
$\Pi_y \cdot T_y \cdot \mathcal{I}_{y} $ & $-$ & $f^\dagger_{\down,R}f_{\up,L}$ & $f^\dagger_{\bar{\sigma},R}f_{\bar{\sigma},L}$ & $\tilde{y}$-inv. + spin-flip
\tabularnewline
\hline\hline
\end{tabularx}\\ 
\label{tab.FP_symm_fermionic-parton-symmetries}
\end{table}
\end{center}

\subsection{Phases of fermionic partons / spins}
\label{subsec.FP_phases}
We now apply the formalism developed above to several specific phases of the fermionic-parton gauge theory. Following the same steps as for bosonic partons in Sec.~\ref{subsec.BP_phases}, we first study mean-field states without gauge-field dynamics. We then include gauge fluctuations and determine the quasiparticle content. Finally, we analyze the corresponding microscopic model. 

\subsubsection{Trivial band insulator / intrawire VBS}
\label{subsubsec.FP_phases_BI+BI-VBSx}
The fermionic partons are at unit filling and can form a trivial band insulator. To generate it, we perturb the parton gauge theory with the umklapp term of Eq.~\eqref{eq.FP_couplings-backscatter}. The resulting theory has the same form as the one describing Mott-gapped bosonic partons (see Sec.~\ref{subsubsec.BP_phases_Mott+Mott-VBSx}), and its analysis is identical. In particular, we obtain the same microscopic model.

\subsubsection{Spin Hall insulator / Easy-plane AFM}
\label{subsubsec.FP_phases_QSH-AFM}
\textit{Mean field}---Consider a quantum spin Hall insulator of partons realized by the interwire couplings
\begin{align}
\label{eq.FP_phases_QSH-AFM_coupling}
 \delta \mathcal{L}_\text{QSH} =g_\text{QSH} \sum\nolimits_\sigma f^\dagger_{R,\tilde{y}+\sigma,\sigma} f_{L,\tilde{y}-\sigma,\sigma} +\text{H.c.}~,
\end{align}
where spin $\sigma=\up(\down)$ fermions reside on odd (even) dual wires. This model describes a fermionic band structure with gap $\propto g_\text{QSH}$. The induced Lagrangian for $\vec a$, to lowest order in $\Delta_2$, is
\begin{align}
\label{eq.FP_phases_QSH-AFM_CS}
 {\cal L}_\text{ind} = \frac{i}{4\pi} \epsilon_{\mu\nu} \left[S_2 a^s_\mu\right]_{2y} \left[\Delta_2 a^c_\nu\right]_{2y} \sim \frac{i}{2\pi} \epsilon_{\mu\nu} a^s_\mu \partial_y a^c_\nu~,
\end{align}
\normalsize
where $(S_2)_{y,y'} = \delta_{y+2,y'}+\delta_{y,y'}$, and we have introduced the `charge' and `spin' gauge fields $\vec a^{c,s}$, which couple to $\rho^c_{2y+1}=(\rho_{2\tilde y +1,\uparrow }+ \rho_{2 \tilde y,\downarrow})$ and
$\rho^s_{2y+1}=(\rho_{2\tilde y +1,\uparrow }- \rho_{2 \tilde y,\downarrow})/2$, respectively.

\begin{figure} [tb]
\centering
 \includegraphics[width=\columnwidth]{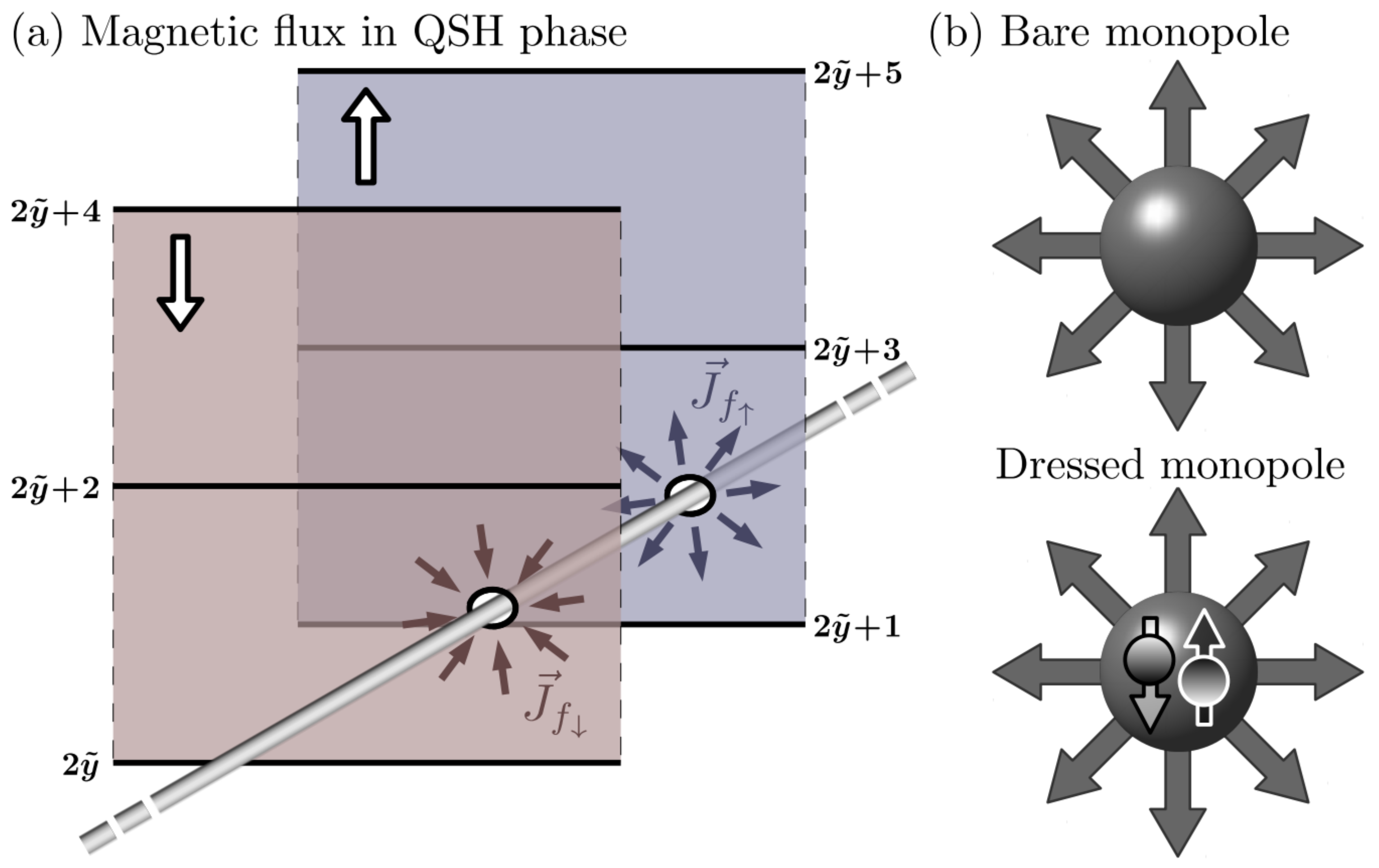}
 \caption{(a) Adiabatically threading magnetic flux generates an electromotive force that creates an inwards current of $\downarrow$ and an outwards current of $\uparrow$ partons. When $2\pi$ flux is introduced in this manner, a single $\downarrow$ particle and $\uparrow$ hole are pulled in. (b) Accordingly, fundamental monopoles become dressed by $f^\dagger_\down f_\up$ and carry spin 1. Their proliferation thus results in a phase with (spontaneously) broken $U(1)$ spin-rotation symmetry, i.e., magnetic order.
 \label{fig:QSH}}
\end{figure}

This response reflects the well-known property of quantum spin Hall systems that a $2\pi$ flux is accompanied by a spin $1$~\cite{Qi+Zhang2011}. To see this explicitly, consider a configuration of $\vec{a}$ that includes a single $2\pi$ flux tube penetrating through the plaquette delimited by $\tilde{y}_0$ and $\tilde{y}_0+1$ [see Fig.~\ref{fig:QSH}(a)]. In the $a_2=0$ gauge, such a configuration satisfies
\begin{align}
\label{eq.FP_phases_QSH-AFM_flux-tube}
 \varointctrclockwise_\text{plaquette} d\vect {l} \cdot \vect{\nabla}\times\vect{a}= \int dx \left[\Delta a_1\right]_{y_0+1} = -2\pi~.
\end{align}
Shifting $\varphi_{f,\tilde{y}} \rightarrow \varphi_{f,\tilde{y}} + \int^x dx' a_{1,\tilde{y}}(x')$ transfers $a_1$ from the kinetic term in Eq.~\eqref{eq.FP_symm+theory_L} to Eq.~\eqref{eq.FP_phases_QSH-AFM_coupling}. In bosonized form, the latter becomes
\begin{align}
 \delta \mathcal{L}_{\text{QSH}} = \sum_\sigma \cos\left(2\bar{\theta}_{\tilde{y},\sigma} + \sigma \int^x dx' \Delta_2 a_{1,\tilde{y}}\right)~,
\end{align}
with $\bar{\theta}_{\tilde{y},\sigma} = \left(\phi^R_{\tilde{y}+\sigma,\sigma} - \phi^L_{\tilde{y}-\sigma,\sigma} \right)/2$. During an adiabatic flux insertion, the argument of the cosine remains locked to its
minimum and thus $\bar{\theta}_{\tilde{y},\sigma}\rightarrow \bar{\theta}_{\tilde{y},\sigma}-\sigma \int^x dx' \Delta_2 a_{1,\tilde{y}}/2$. The resulting change in the parton numbers is $\delta Q_{\sigma}=\sigma$; recall $ Q_{\sigma}$ are the same as for bosonic partons and given by Eq.~\eqref{eq.CW_coupled_VBS_partons-number}. 
Inserting $2\pi$ flux thus pulls in an $\uparrow$ parton and a $\downarrow$ hole, which together carry physical spin $1$. 

\textit{Gauge fluctuations}---Upon reinstating the status of $\vec{a}$ as a dynamical gauge field, we find its universal long-distance behavior to be unaffected by ${\cal L}_\text{ind}$. The field $\vec a^s$ describes modes near momentum $q=\pi$, which are massive according to the bare Maxwell term, ${\cal L}_\text{Maxwell}$, and thus do not affect long-wavelength fluctuations. To lowest order in $\Delta$ and for frequencies and $x$-momenta small compared to the QSH gap, $\vec a$ is governed by 
\begin{align}
\label{eq.FP_phases_QSH-AFM_L-gauge}
 \nonumber \mathcal{L}_{\text{MW}} = & \mathcal{L}_{\text{Maxwell}} + \frac{1}{2\pi v_F} \left[\Delta a_0 \right]^2_{\tilde{y}} + \frac{u_F}{2\pi} \left[ \Delta a_1 \right]^2_{\tilde{y}} \\
 & + \frac{\kappa d_*^2}{8\pi \tilde{v}}\left(\partial_x a_{0,\tilde{y}} - \partial_\tau a_{1,\tilde{y}} \right)^2~.
\end{align}
Here, ${\cal L}_\text{Maxwell}$ is the bare gauge-field action of Eq.~\eqref{eq.FP_symm+theory_L} and $d_*$ is a nonuniversal length scale proportional to the inverse QSH gap. While this effective action describes a propagating photon, the monopole operator $\mathcal{M}$ is strongly irrelevant. Its correlation function, according to Eq.~\eqref{eq.BP_phases_Mott+Mott-VBSx_Monopole-corr-def}, is
\small
\begin{align}
\label{eq.FP_phases_QSH-AFM_naive-monopole-corr}
 C^{\mathcal{M}}_{\vect{R}} = \exp\left\{ \frac{\kappa d_*}{8\pi^2\tilde{v}} \int_{\vect{k},\omega} \frac{\cos\left(\vect{k} \cdot \vect{R} \right)-1}{k_x^2}\left[1-\frac{\kappa d_*^2}{4\pi \tilde{v}} \left< \left|\varepsilon_2 \right|^{2} \right>_0 \right] \right\}~,
\end{align}
\normalsize
with $\varepsilon_2$ as in Sec.~\ref{subsubsec.BP_phases_Mott+Mott-VBSx} and $\vect{R}=\left(x,~d_*y\right)$. The Gaussian average $\left<\ldots\right>_0$, with respect to $\mathcal{L}_{\text{MW}}$, is readily evaluated; we find the asymptotic behavior of $C^{\mathcal{M}}_{\vect{R}}$ is as given in Eq.~\eqref{eq.BP_phases_SF+SF-AFM_monopole-corr-result} (see Appendix \ref{app.gauge-calc_QSH} for details).

The reason for the rapid decay is, that $\mathcal{M}$ attempts to introduce a gauge flux without the accompanying spin discussed above. Consider instead the `dressed' monopole $\mathcal{M}_\text{dressed} \sim f_\downarrow^\dagger f_\uparrow \mathcal{M}$ [see Fig.~\ref{fig:QSH}(b)]. 
Its correlation function reproduces Eq.~\eqref{eq.BP_phases_Mott+Mott-VBSx_Monopole-corr}, i.e., approaches a non-zero constant at long distances, and $\mathcal{M}_\text{dressed}$ \textit{spontaneously} acquires an expectation value. Notice, however, that a term of the form $\delta{\cal L} \sim \left( \mathcal{M}^n_{\text{dressed}}+\text{H.c.}\right)$ would explicitly break $U(1)$ spin-rotation symmetry for any $n \neq 0$ and is therefore disallowed. The gauge field thus remains gapless in this phase, unlike in the trivial parton Mott insulator. For the details of these calculations, see Appendix \ref{app.gauge-calc_QSH}. In particular, the dressed monopole correlation function coincides with the one obtained by evaluating Eq.~\eqref{eq.PC_gauge_monopole-corr} using the singular configuration introduced in Sec.~\ref{subsubsec.BP_phases_Mott+Mott-VBSx}.

\textit{Quasiparticles}---The above analysis implies that, conversely, spin-$s$ operators must be accompanied by $2\pi s$ gauge flux. Since $2\pi$ is the fundamental monopole, there are no low-energy excitations with half-odd integer spin. Spinless excitations, such as $f_\uparrow f_\downarrow$, are charged under the emergent gauge field. They are thus subject to logarithmic confinement (cf.~Sec.~\ref{subsubsec.BP_phases_Mott+Mott-VBSx}).

\textit{Spin model}---The parton QSH breaks the microscopic time-reversal symmetry as well as translations by a single site in the $\hat x$ or $\hat y$ direction. It also exhibits spontaneous breaking of $U(1)$ spin-rotation symmetry and an associated linear spectrum, as well as logarithmically confined neutral excitations. These properties exactly match those of an easy-plane AFM. Indeed, $\sum\nolimits_{\tilde{y}}\delta {\cal L}_{\text{QSH}}$ maps onto $\sum\nolimits_y\mathcal{L}_t$, which generates the easy-plane AFM described in Sec.~\ref{subsubsec.CW_coupled_AFM}.

\subsubsection{Mixed insulators / interwire VBS}
\label{subsubsec.FP_phases_QH+BI-VBSy}
\textit{Mean field}---Consider now an integer quantum Hall state for one parton while the second forms a trivial band insulator. This is achieved, e.g., by introducing 
\begin{align}
\label{eq.FP_phases_QH+BI-VBSy_L}
 \delta \mathcal{L}_{\text{MI}} = g_t f^\dagger_{2\tilde{y}+2,\down,L}f_{2\tilde{y},\down,R} + g_u f^\dagger_{2\tilde{y}+1,\up,R} f_{2\tilde{y}+1,\up,L} +\text{H.c.}~
\end{align}
At the mean-field level, the fermionic partons constitute gapped spin-$1/2$ quasiparticles. The induced action for the gauge field $\vec{a}$ is 
\begin{align}
\label{eq.FP_phases_QH+BI-VBSy_L-ind}
 \mathcal{L}_{\text{CS-odd}} = \frac{i}{4\pi} \left[S_2 a_{1}\right]_{2\tilde{y}+1}\left[\Delta_2 a_{0}\right]_{2\tilde{y}+1}~,
\end{align}
as expected for a quantum Hall state at unit filling.

\textit{Gauge fluctuations}---When the gauge field $\vec{a}$ is promoted to a dynamical variable, it acquires a mass through the Chern-Simons term. Monopoles are thus strongly irrelevant and can be safely discarded. The external probing field $\vec{A}$ can be included in Eq.~\eqref{eq.FP_phases_QH+BI-VBSy_L-ind} by replacing $a_{\mu,2\tilde y} \rightarrow a_{\mu,2\tilde y} + \frac{1}{4} \left[S A_\mu\right]_{2\tilde y}$ (without modifying the bare Maxwell term). Integrating out $a_\mu$ does not result in a Chern-Simons term for $\vec{A}$, which raises the possibility that the microscopic time-reversal symmetry is preserved. Indeed, it translates into the combination of fermionic time-reversal and $y$-translation symmetries (cf.~Table \ref{tab.FP_symm_fermionic-parton-symmetries}), which is preserved by the parton band structure in Eq.~\eqref{eq.FP_phases_QH+BI-VBSy_L}.

\textit{Quasiparticles}---The Chern-Simons term in Eq.~\eqref{eq.FP_phases_QH+BI-VBSy_L-ind} attaches $2\pi$ emergent gauge flux to the fermionic mean-field excitations, converting them into bosonic quasiparticles. The way that $\vec{A}$ enters in Eq.~\eqref{eq.FP_phases_QH+BI-VBSy_L-ind} (see above) implies that the flux of $\vec{a}$ carries physical spin $1/2$. There are, thus, two types of bosonic quasiparticles, one with spin $1$ and one with spin $0$.

\textit{Spin model}---The $y$-translation symmetry is broken, while $\tilde{y}$-inversion is preserved (cf.~Table \ref{tab.BP_symm+theory_symmetries}). Moreover, $U(1)$ spin-rotation, time-reversal, and $x$-translation symmetries all remain intact. These symmetry properties, along with the integer-spin quasiparticles, identify the phase as an interwire VBS. Indeed, translating $\delta \mathcal{L}_\text{MI}$ to the microscopic spin variables, we find the wire construction of Eq.~\eqref{eq.BP_phases_SF+Mott_VBSy_L-microscopic}, which realizes an interwire VBS. 

\subsubsection{Generic \texorpdfstring{$K$}{}-matrix / Chiral spin liquid}
\label{subsubsec.FP_phases_QH-CSL}
\textit{Mean field}---Consider now quantum Hall states characterized by a non-singular $2\times 2$ $K$-matrix 
\begin{align}
 \label{eq.FP_phases_QH-CSL_Kmat}
 K = \begin{pmatrix}
 m_\up & m_0 \\
 m_0 & m_\down
 \end{pmatrix}~,
\end{align}
where $m_\sigma$ are odd integers and $\det\left[K\right] \neq 0$. The corresponding wire construction was worked out in Ref.~\onlinecite{TeoKane2014} and is given by
\small
\begin{align}
\label{eq.FP_phases_QH-CSL_L}
 \delta \mathcal{L}_K = \tilde{g} \cos{\bigl(\left[\Delta_2 \varphi_{f,\sigma}\right]_{\tilde{y}} - m_{\sigma} \left[S_2 \theta_{f,\sigma}\right]_{\tilde{y}} - 2m_0\theta_{f,\tilde{y},\bar{\sigma}}\bigr)}~.
\end{align}
\normalsize
For generic $m_\sigma,m_0$ such a state exhibits a quantum Hall effect (associated with the total charge), a spin quantum Hall effect (associated with the relative charge), and a quantum spin Hall effect that connects the total and relative charges. In terms of $q^T=(1,\ 1)$ and $s^T=(1,\ -1)$m, these are given by $\nu_{cc}=q^T K^{-1}q$, $\nu_{ss}=s^TK^{-1}s$, and $\nu_{cs}=q^T K^{-1}s$, respectively. Integrating out the matter field we find, at leading order in $\Delta_2$,
\begin{align}
\label{eq.FP_phases_QH-CSL_L-ind}
 \mathcal{L}_{\text{CS}-K} = & \frac{i }{4\pi} \sum_{i,j=c/s}\nu_{ij}\bigl[S_2 a^i_0\bigr]_{2y+2} \bigl[\Delta_2 a^j_1\bigr]_{2y+2}~,
\end{align}
where we have introduced `charge' and `spin' gauge fields $\vec a^{c/s}_{2y+1}=\left(\vec a_{2\tilde{y}+1}\pm \vec a_{2\tilde{y}}\right)/2$. The quasiparticles, at the mean-field level, are anyons and carry fractional charges under $\vec a$. They can be determined by a standard $K$-matrix analysis (see, e.g., Ref.~\onlinecite{WenBook2004}). 

\textit{Gauge fluctuations}---Upon reinstating the dynamics of $\vec{a}$, governed by $\mathcal{L}_{\text{MW}} + \mathcal{L}_{\text{CS}-K}$, we find two distinct cases. For $\nu_{cc}=0$, the gauge field remains gapless, and monopoles are important. The assumption of non-singular $K$ implies a non-zero spin Hall response. Therefore, $U(1)$ spin-rotation symmetry is spontaneously broken, as in the special case $m_0=0$ and $m_\sigma =\sigma$ (cf.~Sec.~\ref{subsubsec.FP_phases_QSH-AFM}). By contrast, for non-zero $\nu_{cc}$, the gauge field is massive, and monopoles can be safely discarded. 

Recall, that the microscopic time-reversal symmetry acts as a duality transformation on the partons. Specifically, it attaches $- 2\pi \sigma$ flux to the fermions (followed by particle-hole transformation, see Sec.~\ref{sec.FP}). Therefore, $K$-matrix states for the $f_\sigma$ fermions and for the dual $d_\sigma$ fermions are related by $-K_f = K_d - 2\sigma^z$. For non-zero $\nu_{cc}$, the two $K$-matrices cannot coincide, and time-reversal symmetry is broken explicitly. To obtain the physical response, we include the external probing field $\vec{A}$ in Eq.~\eqref{eq.FP_phases_QH-CSL_L-ind}, according to $a_{\mu,\tilde y} \rightarrow a_{\mu,\tilde y} - \frac{1}{4} \left(-1\right)^{\tilde{y}} \left[S A_\mu\right]_{\tilde y}$. Integrating out the emergent gauge field we obtain
\small
\begin{align}
\label{eq.FP_phases_QH-CSL_L-A}
 \mathcal{L}_{A-\text{CS}} = -\frac{i}{8\pi n} \left[SA_1\right]_{\tilde{y}} \left[\Delta A_0\right]_{\tilde{y}} \sim -\frac{i}{8\pi n} \epsilon_{\mu\nu} A_{\mu}\partial_y A_\nu~,
\end{align} 
\normalsize
where $n \equiv\nu_{cc}\det\left[K\right]/2$. Consequently, the phase is chiral with topologically protected edge states. 

\textit{Quasiparticles}---To characterize the quasiparticles for $n \neq 0$, we adopt the strategy employed in Sec.~\ref{subsubsec.BP_phases_QH-CSL}. To each fermion, we attach $m_\sigma$ fluxes of their own species and $m_0$ fluxes of the opposite one; on an operator level, we define $\eta_{\tilde{y},\sigma} = \varphi_{f,\tilde{y},\sigma} - \frac{m_\sigma}{2}\left[S_2 \tilde{\varphi}_{f,\sigma}\right]_{\tilde{y}} - m_0\tilde{\varphi}_{f,\bar{\sigma},\tilde{y}}$. The corresponding $\beta_{\sigma}=e^{-i\eta_\sigma}$ particles are bosons; they are governed by the schematic Lagrangian 
\small
\begin{align}
\label{eq.FP_phases_QH-CSL_L-chi}
 \mathcal{L}_\beta &= i\sum\nolimits_\sigma \vec{j}^{\beta}_\sigma \cdot \left(\vec{c}_\sigma +\textstyle\frac{\sigma}{2}\vec{A}\right) - i\frac{(c_\up-c_\down)d(c_\up-c_\down)}{8\pi n}~.
\end{align}
\normalsize
Notice that this Lagrangian is the same as Eq.~\eqref{eq.BP_phases_QH-CSL_L-eta}. The quasiparticles thus carry spin $1/2n$ and acquire statistical phases of $\pi/2n$ upon clockwise exchange. The case $n=0$ can be analyzed as in Sec.~\ref{subsubsec.FP_phases_QSH-AFM}. In particular, these phases feature linearly dispersing Goldstone modes associated with the broken $U(1)$ spin-rotation symmetry and logarithmically confined topological excitations.

\textit{Spin model}---The response to the external probing field, Eq.~\eqref{eq.FP_phases_QH-CSL_L-A}, implies that the microscopic phase is a chiral QSL. Translating the interwire coupling in Eq.~\eqref{eq.FP_phases_QH-CSL_L} to microscopic variables, we find 
\small
\begin{align}
\label{eq.FP_phases_QH-CSL_L-micro}
 \delta \mathcal{L}_{\text{CSL}}' = \tilde{g}\cos{(\left[\Delta\Phi\right]_{\tilde{y}} - 2m_0\left[S\Theta\right]_{\tilde{y}}+m_{\tilde{y}}\left[S_2 S \Theta\right]_{\tilde{y}} )}~,
\end{align}
\normalsize
with $m_{2y \pm 1/2} =m_{\nicefrac{\up}{\down}} \mp 1$. 
For non-singular $K\neq \sigma^z$, these interwire couplings explicitly break time-reversal symmetry. 
The arguments of all cosines in $ \delta \mathcal{L}_{\text{CSL}}'$ commute and can flow to strong coupling simultaneously. They are, moreover, linearly independent and can thus generate a gapped phase for non-zero $n$, i.e., $m_{\tilde y}+m_{\tilde y+1} \neq 2 m_0$. To identify its quasiparticles and edge structure, we introduce chiral modes
\begin{align}
 \tilde \phi_{\chi,y} = \Phi_y + \chi 2 n \Theta_y +2m_{y+1}\Theta_{y+1}-2m_{y}\Theta_{y-1}~,
\end{align}
which satisfy 
\begin{align}
 \left[\partial_x \tilde\phi_{\chi,y},\tilde\phi_{\chi',y'}\right]= i \chi 4\pi n \delta_{y,y'}\delta_{\chi,\chi'}\delta\left(x-x'\right)~.
\end{align}
Crucially, this change of variables preserves the locality of both the intra- and interwire terms; the latter take the form $ \delta \mathcal{L}_{\text{CSL}}' = \cos{(\tilde \phi_{R,y+1}-\tilde \phi_{L,y})}$. Domain walls in these cosines carry spin $1/2n$ and acquire exchange phases of $\pi/2n$. 

Finally, for $n=0$ the system is gapless, which can be seen by summing the arguments of all cosines in Eq.~\eqref{eq.FP_phases_QH-CSL_L-micro}, i.e., $\sum\nolimits_y [\tilde \phi_{R,y+1}-\tilde \phi_{L,y}]=4n \sum\nolimits_y \Theta_y$. This particular linear combination thus remains unpinned for $n=0$. Its conjugate describes the Goldstone mode associated with the spontaneously broken $U(1)$ spin-rotation symmetry, precisely as in Sec.~\ref{subsubsec.CW_coupled_AFM}.

\subsubsection{BCS superconductor / \texorpdfstring{$\mathbb{Z}_2$}{} spin liquid}
\label{subsubsec.FP_phases_BCS-Z2}
\textit{Mean field}---As a final example, consider now a BCS superconductor of fermionic partons. To generate pairing, we introduce interwire hopping for the Cooper-pair operator $\Psi_{\text{$f$-pair},2y+1} \equiv f_{2\tilde{y}+1\up,R} f_{2\tilde{y},\down,L}$, i.e.,
\begin{align}
\label{eq.FP_phases_BCS-Z2_Lpair}
 \delta \mathcal{L}_{\text{$f$-pair}} = & g_\text{$f$-pair} \Psi^\dagger_{\text{$f$-pair},2y+1} \Psi_{\text{$f$-pair},2y-1} + \text{H.c.} 
\end{align}
When $g_\text{$f$-pair}$ flows to strong coupling, $\Psi_{\text{$f$-pair}}$ spontaneously acquires an expectation value (cf.~Sec.~\ref{subsubsec.CW_coupled_AFM}). Vortices in the phase of this condensate, created by
$m_{2y} = e^{-i\left(\tilde{\varphi}_{f,\up,2\tilde{y}}+\tilde{\varphi}_{f,\down,2\tilde{y}-1}\right)/2}$, are logarithmically confined. While $\delta \mathcal{L}_{\text{$f$-pair}}$ renders the umklapp term in Eq.~\eqref{eq.FP_couplings-backscatter} irrelevant, the back-scattering term
\small
\begin{align}
 \delta \mathcal{L}_{\text{bs}} = g_\text{bs} f^\dagger_{2\tilde{y}+1,\up,R} f_{2\tilde{y}+1,\up,L} f^\dagger_{2\tilde{y},\down,L} f_{2\tilde{y},\down,R} + \text{H.c.}~,
\end{align}
\normalsize
can flow to strong coupling. When it does, a fully gapped phase with fermionic spin-1/2 quasiparticles, $f_\sigma$, obtains. 

\textit{Gauge fluctuations}---Since $\Psi_{\text{$f$-pair}}$ carries emergent gauge charge, its condensation leads to a Higgs mass. Monopoles are, therefore, strongly suppressed and can be safely discarded. Importantly, $\Psi_{\text{$f$-pair}}$ is neutral under the external probing field $\vec{A}$, so the microscopic $U(1)$ spin-rotation symmetry is preserved. Indeed, the response to $\vec A$ is, at leading order in derivatives, described by a Maxwell action.

\textit{Quasiparticles}---Gauge field fluctuations that are rendered massive by a Higgs term do not affect the status of $f_\sigma$ as fermionic quasiparticles. They do, however, promote vortices $m$ to deconfined bosonic spin-0 quasiparticles. Being superconducting vortices, they are experienced as $\pi$ flux by the fermions, i.e., the two are mutual semions. Consequently, the two can combine into $\varepsilon = f_{\chi,\sigma} m^\dagger$, a spin-1/2 quasiparticle with bosonic self-statistics.

\textit{Spin model}---The quasiparticle content characterizes a $\mathbb{Z}_2$ spin liquid that is, moreover, non-chiral and spin-rotation symmetric. Translating $ \delta \mathcal{L}_{\text{$f$-pair}}$ and $\delta \mathcal{L}_{\text{bs}}$ to microscopic spin variables, we find 
\begin{align}
  \delta \mathcal{L}_{\mathbb{Z}_2}' = & g_\text{bs} \cos\bigl(2 \left[S^T S \Theta\right]_{2y-1} \bigr) \\
 \nonumber & + g_\text{$f$-pair} \cos\bigl( \left[\Delta^T \Delta \Phi\right]_{2y} + 2 \left[S^T S \Theta\right]_{2y-1}\bigr)~.
\end{align}
The arguments of these cosines are linear combinations of those in Eq.~\eqref{eq.BP_phases_PairCond-Za_L-micro}. Consequently, they lead to the same $\mathbb{Z}_2$ spin liquid phase (see full analysis in Sec.~\ref{subsubsec.BP_phases_PairCond-Z2}). 

\section{Summary and discussion}
\label{sec.discussion}
We have introduced exact, nonlocal mappings between arrays of spin-$1/2$ chains and parton gauge theories. Any parton model that separately conserves both species maps onto a local spin Hamiltonian. The challenge of deriving spin models that realize exotic ground states is thereby reduced to constructing parent Hamiltonians for simple phases of bosons or fermions. Conversely, any $S^z$-conserving coupling between spins transforms into a distinct interaction or hopping term for partons. The latter are obtained without reference to a specific mean-field ansatz. They, therefore, retain not only information about the symmetries of the underlying spin model, but also about more subtle aspects, such as geometric frustration. In its presence, some symmetry-allowed terms in the dual parton description are absent. Geometric frustration may thus take the form of an \textit{emergent} symmetry and, thereby, stabilize phases that would not readily form in more generic situations.

To demonstrate the versatility of this method, we showed how to recover trivial states and access topologically ordered ones. Relatively simple phases of partons already correspond to fractionalized ground states. As a first example, we derived parent Hamiltonians for Abelian chiral spin liquids. Here, knowing wire models of bosonic integer quantum Hall states was sufficient to immediately generate a parent Hamiltonian of the Kalmeyer-Laughlin chiral spin liquid, previously constructing by different means in Refs.~\onlinecite{Meng2015a,Gorohovsky2015}. Similarly, we obtained the wire model of a time-reversal-invariant $\mathbb{Z}_2$ spin liquid as a simple $s$-wave BCS superconductor of fermionic partons. If they instead form a topological superconductor, such as $p_x\pm ip_y$, the resulting QSL will be non-Abelian.

When the partons themselves form non-trivial phases, an even wider range of exotic microscopic ground states is realized. The framework introduced here applies to such cases with no additional difficulties, once a (coupled-wire) parent Hamiltonian of the parton phase is known. We have illustrated this capability by the example of a general $2\times 2$ $K$-matrix state of fermionic partons. Based on the known fermionic wire constructions, we easily generated wire models for a range of chiral spin liquid that were not previously available. Explicit parent Hamiltonians for even more exotic states, such as the non-Abelian Read-Rezayi sequence of fractional quantum Hall states, are also known and can likewise be used to generate concrete spin models.

We primarily focused on fully gapped states. However, the dual description of spin-chain arrays in terms of fermions may be able to capture exotic gapless phases and unconventional quantum phase transitions as well. One example of the latter arises at the transition between the easy-plane AFM and the intrawire VBS. It maps onto a coupled-wire model of compact QED$_3$ with two boson or fermion species. This is precisely the effective field theory that was derived using different methods in Ref.~\onlinecite{Senthil+Vishwanath+Balents+Sachdev+Fisher2004}. Its fate in the infrared is thought to be confining (and consequently the transition to be first order). A stable gapless theory may instead arise in various ways: (i) at a transition between different phases; (ii) in the presence of emergent symmetries of the parton field theory that may arise due to geometric frustration; (iii) when the emergent fermions are doped to form a Fermi surface that suppresses monopole events. All three scenarios feature non-trivial gauge-field dynamics, which places them beyond the capability of conventional wire constructions. They should, however, be amenable to exploration within the formalism developed here.

Finally, we mention two possible generalizations of the methods developed here. The first is to itinerant electron systems. There, decomposing microscopic electron operators as $c_\sigma = b f_{\sigma}$ allows exploration of many exotic ground states. Extending our approach to wire arrays with both spin and charge modes may allow well-controlled access to those phases, and provide concrete model systems where they arise. A second interesting direction is given by spin models that do not conserve $S^z$, such as the celebrated Kitaev honeycomb model~\cite{Kitaev2006}. Systems without $U(1)$ symmetries are not readily describable within Abelian bosonization. Instead, coupled-wire techniques based on non-Abelian bosonization have been used successfully in similar contexts~\cite{Sahoo2016,Huang2016}. Generalizing our methods to these systems could provide a much-desired bridge between fine-tuned solvable models and mean-field studies of generic ones. 

\begin{acknowledgments}
It is a pleasure to thank Jason Alicea, Olexei Motrunich, and Chong Wang for many illuminating discussions. This work was supported by the Israel Science Foundation and by the Minerva foundation with funding from the Federal German Ministry for Education and Research. Part of this research was performed at the Aspen Center for Physics, which is supported by National Science Foundation grant PHY-1607611.
\end{acknowledgments}

\bibliography{partonwires.bib}

\begin{widetext}

\appendix

\section{Energy cost of topological defects}
\label{app.defect-energies}

\subsection*{Energy cost of domain walls in the VBS phase of a one-dimensional spin-chain}
\label{app.defect-energies_1d-VBS-defect}
To describe the VBS phase, consider the bosonized action of a $U(1)$ spin chain given by
\begin{align}
\label{eq.defect-energies_1d-VBS-defect_action}
 \mathcal{S}_\text{VBS} = \int_{x,\tau} \left[\frac{1}{2\pi v K}\left(\partial_\tau \Theta\right)^2 + \frac{v}{2\pi K} \left(\partial_x \Theta\right)^2 + \frac{v\tilde{g}_{4\pi}}{16\pi K d_0^2} \cos\left(4\Theta \right) \right]~,
\end{align}
with $K$ such that the dimensionless coupling constant $\tilde{g}_{4\pi}<0$ flows to strong coupling. The minima of the cosine potential, $\Theta_\text{min} = \pi n /2$, correspond to different topological sectors, labeled by the integer $n$. On length scales larger than $d_*$, where $\tilde{g}_{4\pi}$ has become of order unity, it is appropriate to expand the cosine in a single topological sector, i.e., replace $\frac{v\tilde{g}_{4\pi}}{16\pi K d_0^2} \cos\left(4\Theta \right) \rightarrow \frac{v}{2\pi K d_*^{2}} (\Theta - \pi n/2)^2$.
To find the energy cost of domain walls, we allow the system to transition between different topological sectors as a function of space, i.e., $n \rightarrow n_x$ and
\begin{align}
\label{eq.defect-energies_1d-VBS-defect_action-expanded-w/topo-sectors}
 \mathcal{S}_\text{eff}\left[n\right] =\frac{1}{2\pi v K} \int_{x,\tau} \left[\left(\partial_\tau \Theta\right)^2 + v^2 \left(\partial_x \Theta\right)^2 + \frac{v^2}{d_*^{2}} \Theta^2 - \frac{\pi v^2}{d_*^{2}} \Theta n_x + \frac{\pi^2 v^2}{4d_*^2} n_x^2 \right]~.
\end{align}
The energy cost of forcing the system into different topological sectors, relative to the uniform $n=0$ vacuum, is given by 
\begin{align}
\label{eq.defect-energies_1d-VBS-defect_energy_n}
 \nonumber\Delta E\left[n\right] = & -\lim \nolimits_{T \rightarrow 0} T \log\left(\frac{\mathcal{Z}[n]}{\mathcal{Z}[0]}\right) = -\lim \nolimits_{T \rightarrow 0} T\log\left<e^{\mathcal{S}[0]- \mathcal{S}[n]}\right>_0 \\
 = & \frac{\pi v}{8 K d_*^2} \int_x n_x^2 - \frac{v^2}{8 K^2 d_*^4} \int_{x,x'} n_x n_{x'}\left<\Theta_x\Theta_{x'}\right>_{0}\big|_{\omega=0} = \frac{v^2}{8 K^2 d_*^2} \int_{x,x'} n_x n_{x'}\left<\partial_x\Theta_x\partial_{x'}\Theta_{x'}\right>_{0}\big|_{\omega=0}~,
\end{align}
where the last equality uses the specific form of the correlation function $\left<\Theta_x\Theta_{x'}\right>$ according to Eq.~\eqref{eq.defect-energies_1d-VBS-defect_action-expanded-w/topo-sectors}. For a generic configuration of domain walls parametrized by $n(x) = \sum_i \alpha_i H(x-x_i)$ with $\alpha_i\in \mathbb{Z}$, the energy is given by
\begin{align}
\label{eq.defect-energies_1d-VBS-defect_energy_sigma}
 \Delta E\left[\{\alpha\}\right] = \frac{ v^2}{8 K^2 d_*^2} \sum\nolimits_{i,j}\alpha_i \alpha_{j}\left<\Theta_{x_i} \Theta_{x_j}\right>_{0}\big|_{\omega=0}= \frac{\pi v}{16 K d_*} \sum\nolimits_{i,j}\alpha_i \alpha_{j} e^{-\left|x_i - x_j\right|/d_*}~.
\end{align}
In particular, the energy cost of a single domain wall is given by the prefactor $\Delta E_{\text{DW}}=\frac{\pi v}{16 K d_*}$. 

The trial function provided in the main text, $\Theta_{\text{DW}}(x) = \tan^{-1}\left[e^{(x-x_0)/\xi}\right]$, produces a variational energy cost of $\Delta E_{\text{DW}}\left[\xi\right] = \frac{v}{4 \pi K}\left(1/\xi + \xi/ d_*^2\right)$, for the renormalized action, i.e., Eq.~\eqref{eq.defect-energies_1d-VBS-defect_action} with $\tilde{g}_{4\pi} \rightarrow -d_0^2/d_*^2$. Its minimal value, attained for $\xi=d_*$, is given by $\Delta E_{\text{DW}} = \frac{v}{2 \pi K d_*}$ and is parametrically the same as the result of Eq.~\eqref{eq.defect-energies_1d-VBS-defect_energy_sigma}. The somewhat smaller numerical value relative to the previous calculation arises because there, the cosine was replaced by a parabolic potential centered around the nearest minimum.

\subsection*{Energy cost of magnetic vortices in the easy-plane AFM}
\label{app.defect-energies_AFM-vortices}
We follow the same strategy as for the one-dimensional VBS domain walls. Expanding the cosine of Eq.~\eqref{eq.CW_coupled_couplings-tunnel} in topological sectors, denoted by $n_{x,y+1/2}$, we obtain
\begin{align}
\label{eq.defect-energies_AFM-vortices_action-expanded-w/topo-sectors}
 \mathcal{S}_\text{eff}\left[n\right] = \frac{ K}{2\pi v} \int_{x,\tau} \sum\nolimits_y \left[\left(\partial_\tau \Phi_y\right)^2 + v^2 \left(\partial_x \Phi_y\right)^2 + \frac{v^2}{d_*^{2}} \left(\left[\Delta\Phi\right]_{y+\frac{1}{2}} - 2\pi n_{x,y+\frac{1}{2}} \right)^2 \right]~,
\end{align}
where $d_*$ is the length scale at which $\tilde{g}_t = 2\pi d_0^2 g_t / v K$ reaches order unity. In this case, $n$ can be interpreted as counting the magnetic flux tubes in an external probing field. Consider a magnetic field $B(x)$ in the gauge $A_1=0$, i.e., $A_2 = \int_x B(x) $. Incorporating the probing field via minimal coupling amounts to replacing $\Delta \Phi \rightarrow \Delta \Phi - A_2$ above, which identifies $n =\frac{1}{2\pi}\int_x B(x)$.
The energy cost for a given configuration $n$ can be computed as in the one-dimensional case, i.e.,
\begin{align}
\label{eq.defect-energies_AFM-vortices_energy}
 \Delta E\left[n\right] = 2 v^2 K^2 d_*^{-2}\int_{x,x'} \sum\nolimits_{y,y'} n_{x,y+\frac{1}{2}}n_{x',y'+\frac{1}{2}}\left<\partial_x\Phi_{x,y}\partial_{x'}\Phi_{x',y'}\right>_{0}\Big|_{\omega=0}
 ~.
\end{align}
Parametrizing $n$ in terms of strength $\alpha_i$ vortices at positions $(x_i,\tilde y_i)$, i.e., $n_{x,y+\frac{1}{2}} = \sum_{i} \alpha_{i}\delta_{y,y_i}H\left(x-x_i\right)$, we arrive at the final expression
\begin{align}
\label{eq.defect-energies_AFM-vortices_energy-generic-n}
 \Delta E\left[\left\{\alpha\right\}\right] = 2 v^2 K^2 d_*^{-2} \left< \left(\sum\nolimits_i\alpha_i\Phi_{x_i,y_i}\right)^2 \right>_{0}\Big|_{\omega=0}
 =\frac{v K }{2\pi d_*} \sum\nolimits_{i,j}\alpha_i\alpha_j
 \int_{k_x,k_y} \frac{\cos\left[\vect k\cdot(\vect R_i -\vect R_j) \right]}{ k_x^2+ \left|\Delta_{k_y}\right|^2}
 ~,
\end{align}
with $\vect{R}_i=\left(x_i,d_*\tilde y_i\right)$ and $\vect{k} = \left(k_x, k_y\right)$. The $y$-momentum $k_y$ is measured in units of $d_*^{-1}$ and $\Delta_{k_y} \equiv \left(e^{ i d_* k_y} - 1\right)/d_*$. If the total number of vortices $N_\text{v}\equiv\sum_i \alpha_i \neq 0$, the energy diverges logarithmically with the size of the system. When $N_\text{v}=0$, the energy cost is finite; for a single vortex---anti-vortex pair we find
\begin{align}
\label{eq.defect-energies_AFM-vortices_energy-vortex-antivortex}
 \lim\limits_{|\vect{R}_1 - \vect{R}_2| \rightarrow \infty} \Delta E_{\vect{R}_1,\vect{R}_2}= \frac{2 v K}{d_*} \log\left(\frac{|\vect{R}_1 - \vect{R}_2|}{2d_*}\right)~.
\end{align}

Importantly, this calculation does not make any reference to a specific form of the vortex. Instead, it fixes the topological properties and lets the functional integral over $\Phi$ find the optimal configuration. The same result can be obtained by considering a `trial' configuration of the form
\begin{align}
 \Phi_\text{trial} = \arg\left[\left(x-x_1\right)/\xi + i \left(y-\tilde y_1\right)\right]-\arg\left[\left(x-x_2\right)/\xi + i \left(y-\tilde y_2\right)\right]~,
\end{align}
where $\xi$ is a variational parameter. Computing the corresponding energy at large vortex---anti-vortex separation, one finds
\begin{align}
\label{eqntrial}
 E_\text{trial}[\xi] = \frac{K}{2\pi v} \int_{x} \sum\nolimits_y \left[v^2 \left(\partial_x \Phi_\text{trial}\right)^2 + \frac{v^2}{d_*^2} \left[\Delta\Phi_\text{trial}\right]^2 \right] = \frac{2v K}{d_*} \frac{d_*^2+\xi^2}{2\xi d_* }\log\left(\frac{|\vect{R}_i - \vect{R}_j|}{2\xi}\right)+\mathcal{O}(1)~,
\end{align}
with $\vect R_i =(x_i,\xi y_i)$. To optimize $\xi$ within logarithmic accuracy, it is sufficient to focus on the prefactor of the logarithm; it is minimized for $\xi = d_*$ where Eq.~\eqref{eqntrial} reduces to the result provided in Eq.~\eqref{eq.defect-energies_AFM-vortices_energy-vortex-antivortex}.

\subsection*{Energy cost of dislocations / bosonic partons in the intrawire VBS}
\label{app.defect-energies_VBS-dislocations/partons}
To generate the two-dimensional VBS with dimers along the wire direction, $\hat x$, we introduce two different cosines per wire pair. We introduce dimensionless coupling constants, $\tilde{g}_u>0$ and $\tilde{g}_{4\pi}<0$, and write
\begin{align}
\label{eqn.app.isingvbs}
 \delta \mathcal{L}_\text{intra$-$VBS} = -\frac{v \tilde{g}_u}{4\pi K d_0^2} \cos{(2\left[\Delta\Theta\right]_{y+1/2})} + \frac{v \tilde{g}_{4\pi}}{16\pi K d_0^2} \cos\left(4\Theta_y\right)~,
\end{align}
where, for convenience, we have redefined $\Theta_y \rightarrow \left(-1\right)^y\left(\Theta_y+\pi y/2\right)$. Beyond the length scales $d_*$ and $l_*$, where the coupling constants renormalize to order unity, we expand the action in topological sectors. Labeling said sectors by integers $n$ and $p$, we obtain
\begin{align}
 \mathcal{S}_\text{eff}\left[n,p\right] = \frac{1}{2\pi vK} \int_{x,\tau} \sum\nolimits_y \left[\left(\partial_\tau \Theta_y\right)^2 + v^2 \left(\partial_x \Theta_y\right)^2 + \frac{v^2}{d_{*}^2} \left(\left[\Delta\Theta\right]_{y+\frac{1}{2}} - \pi n_{x,y+\frac{1}{2}}\right)^2 + \frac{v^2}{l_*^2} \left(\Theta_{y} - \textstyle\frac{\pi}{2} p_{x,y}\right)^2\right]~.
\end{align}
As before, we parametrize the integer functions $n$ and $p$ by the locations and strengths of topological defects, i.e., $n = \sum_{i}\alpha_i H\left(x-x_i\right) \delta_{\tilde y,\tilde y_i}$, and $p = \sum_{i}\beta_i H\left(x-X_i\right) \delta_{y,Y_i}$. The energy cost for a general configuration naturally decomposes into a manifestly local, system-size independent contribution, and a nonlocal one that may be IR divergent, i.e., $\Delta E\left[n,p\right] = \Delta E_{\text{local}}\left[n,p\right]+\Delta E_{\text{nonlocal}}\left[n-\frac{1}{2}\Delta p\right]$ with
\begin{align}
 \Delta E_{\text{local}} = &\frac{v^2}{2K^2 d_{*}^2} \sum_{i,j} \alpha_i \alpha_j \left<\Theta_{x_i,y_i} \Theta_{x_j,y_j}\right>_{0}\big|_{\omega=0} + \frac{ v^2 }{8 K^2 l_*^2} \sum_{i,j} \beta_{i}\beta_j \left<\Theta_{x_i,y_i}\Theta_{x_j,y_j}\right>_{0}\big|_{\omega=0}~, \\
 \Delta E_{\text{nonlocal}} = & \frac{v^2 }{2K^2 d_{*}^2 l_*^2} \int_{x,x'} \sum\nolimits_{y,y'} \left(n_{x,y+\frac{1}{2}}-\textstyle\frac{1}{2}\left[\Delta p\right]_{x,y+\frac{1}{2}}\right) \left(n_{x',y'+\frac{1}{2}}-\textstyle\frac{1}{2}\left[\Delta p\right]_{x',y'+\frac{1}{2}}\right)\left<\Theta_{x,y}\Theta_{x',y'}\right>_{0}\Big|_{\omega=0}~.
\end{align}
\textit{A priori}, $n$ and $p$ are independent, and the lowest energy cost for a given $n$ must be found by optimizing over all possible $p$. Fortunately, this minimization can be avoided when the density of defects $n$ is low; there, the optimal $p$ can be inferred based on general considerations.

Consider a single dislocation---anti-dislocation pair defined by $\alpha_1=1$ and $\alpha_2=-1$, i.e., $n_{x,\tilde y}= H\left(x-x_1\right)\delta_{\tilde y,\tilde y_1}-H\left(x-x_2\right)\delta_{\tilde y,\tilde y_2}$. The corresponding $p$ can be deduced as follows: (i) To avoid energies that diverge with system size, the topological sectors must match asymptotically, i.e., $\lim\nolimits_{x \rightarrow \infty}(n_{x,\tilde{y}} - \left[\Delta p\right]_{x,\tilde{y}}/2)=0$. (ii) The minimal number of $p$ defects within this constraint has exactly one of strength $\beta_i = 2$ for each wire $Y_i \in \left[\tilde y_1,\tilde y_2\right]$. (iii) Their optimal locations are along the line connecting $\vect{\tilde{r}}_1 =\left(x_1,~\tilde{y}_1\right)$ and $\vect{\tilde{r}}_2 =\left(x_2,~\tilde{y}_2\right)$. To determine the energy for distances much larger than $l_*$, we approximate
\begin{align}
 \left<\Theta_{x,y}\Theta_{x',y'}\right>_{0}\big|_{\omega=0} = \frac{K d_{*}}{4\pi v} \int_{k_x,k_y} \frac{\cos\left[\vect{k} \cdot (\vect{R}-\vect{R}')\right]}{ k_x^2 + \left|\Delta_{k_y}\right|^2 + l_*^{-2}} \approx \frac{\pi K l_*^2}{v}\delta(x-x')\delta_{y,y'}~,
\end{align}
where $\vect{R} = \left(x,~d_{*} y\right)$. [When appearing in discrete sums as in $\Delta E_\text{\cal}$, $\delta\left(x_i-x_j\right)$ is to be understood as $d_0^{-1}\delta_{x_i,x_j}$, where the microscopic length scale $d_0$ acts as a UV cutoff.] In this limit we find simplified expressions for the local and nonlocal contributions to the energy, which in case of a single dislocation---anti-dislocation pair give
\begin{align}
 \Delta E_{\text{local}} =& \frac{\pi v l_*^2}{2 K d_0 d_{*}^2} \sum_i \alpha_i^2 + \frac{\pi v }{8 K d_0} \sum_{i} \beta_{i}^2 = \frac{\pi v l_*^2}{K d_0 d_{*}^2} + \frac{\pi v }{2 K d_0} \left(\tilde{y}_2-\tilde{y}_1\right)~, \\
 \Delta E_{\text{nonlocal}} = & \frac{\pi v}{2 K d_*^2} \int_{x} \sum\nolimits_{y} \left(n_{x,y+\frac{1}{2}}-\textstyle\frac{1}{2}\left[\Delta p\right]_{x,y+\frac{1}{2}}\right)^2 = \frac{\pi v}{2 K d_*^2} \frac{\left(x_2-x_1\right)}{1+\left(\tilde{y}_2-\tilde{y}_1\right)^{-1}}~.
\end{align}
Topological defects in the intrawire VBS phase are thus linearly confined. As for the case of magnetic vortices, the same conclusion can be reached by studying appropriate trial states. For the Ising-AFM phase, the sign of $\tilde g_{4\pi}$ in Eq.~\eqref{eqn.app.isingvbs} is reversed, but the analysis is otherwise identical, i.e., defects are also linearly confined.

\subsection*{Energy cost of bosonic parton vortices in the correlated Mott insulator}
\label{app.defect-energies_Higgs-vortices} In the correlated Mott insulator (cf.~Sec.~\ref{subsubsec.BP_phases_SF+Mott_VBSy}), there are two cosines per wire pair, i.e.,
\begin{align}
 \delta \mathcal{L}_\down + \delta \mathcal{L}_\up = \frac{v_B \tilde{g}_u}{\pi d_0^2} \cos{(\left[\Delta_2 \varphi_\down\right]_{2\tilde{y}+1})} + \frac{u_B \tilde{g}_{t}}{\pi d_0^2} \cos\left(2\theta_{2\tilde{y}+1,\up}\right)~.
\end{align} 
Beyond the length scales $d_*$ and $l_*$, where coupling constants renormalize to order unity, we expand the action in the topological sectors. Labeling said sectors by the integer functions $n$ and $p$, we write
\begin{align}
 \mathcal{S}_\text{eff}\left[n,p\right] = \mathcal{S}_0 + \int_{x,\tau} \sum\nolimits_{\tilde{y}} \left[\frac{2 v_B}{d_*^2}\left(\pi n^2_{x,2\tilde{y}+1} - \left[\Delta_2 \varphi_\down\right]_{2\tilde{y}+1} n_{x,2\tilde{y}+1}\right)
 + \frac{2 u_B}{l_*^2} \left(\pi p^2_{x,2\tilde{y}+1} -2\theta_{2\tilde{y}+1,\up}p_{x,2\tilde{y}+1}\right)\right]
 ~,
\end{align}
where $\mathcal{S}_0$ is the effective action of the trivial topological sector with $n=p=0$. Vortex---anti-vortex pairs are encoded in $n$, so we set $p$ to zero. To evaluate the energy cost for a given $n$, we need the Green function of $\Delta \varphi_\downarrow$ within the Gaussian theory $\mathcal{S}_0[\varphi_\sigma,\theta_\sigma,a]$. Its leading order in $k_x$ and $\Delta_{k_y}$ is 
\begin{align}
\langle |\Delta_{2 k_y}\varphi_\down|^2\rangle \big|_{\omega=0} =&\frac{ \pi \alpha^2}{ v_B d_*^2}\frac{1+\alpha^2 d_*^2\left|\Delta_{k_y}\right|^2}{k_x^2+\left|\Delta_{k_y}\right|^2 + \frac{\alpha^2}{d_*^2}}~,
\end{align}
where the dimensionless parameter $\alpha^2 =\frac{4v_B}{\kappa \tilde{v}}$ encodes the ratio of boson and gauge-field velocities.
For a single vortex---anti-vortex pair at positions $\vect{R}_i = \left(x_i, d_* \tilde{y}_i\right)$, with $\tilde{y}_i$ odd and $|\vect R_1 -\vect R_2|\gg d_*$ we find
\begin{align}
 \Delta E_{\vect{R}_1,\vect{R}_2} = \frac{v_B }{2 \pi d_*} \int_{k_x,k_y} \frac{2-2\cos\left[\vect{k}\cdot\left(\vect{R}_1-\vect{R}_2\right)\right]}{k_x^2+\left|\Delta_{k_y}\right|^2 + \frac{\alpha^2}{d_*^2}} =\frac{2v_B }{d_*}f\left(\alpha^2\right)- \frac{2 v_B}{d_*}
 \sqrt{\frac{\pi d_*}{2\alpha|\vect R_1 - \vect R_2|}}
 e^{-\alpha\left|\vect{R}_1-\vect{R}_2\right|/d_*}
 ~.
\end{align}
Here, the dimensionless integral $f(x) = \frac{1}{2} \int_{-\pi}^\pi d\theta \frac{1}{\sqrt{x+2- 2\cos(\theta)}}$ diverges logarithmically for $x\rightarrow 0$ but is otherwise finite. The energy required for creating an isolated vortex is thus $v_B/d_*$ times a number of order unity.

\section{Derivation of the bosonic-parton field theory}
\label{app.gauge_bosons}
The equivalence between the microscopic coupled-wire model described in Sec.~\ref{subsec.CW_coupled} and the gauge theory for bosonic partons introduced in Sec.~\ref{subsec.BP_symm+theory} can be shown using the methods of Refs.~\onlinecite{Mross2016,Mross2017}. We begin with the gauge theory $\mathcal{S}=\int_{x,\tau}\sum\nolimits_y\left[ \mathcal{L}_{b} + \mathcal{L}_{\text{Maxwell}}+ \mathcal{L}'_{b} \right]$ in the $a_2=0$ gauge, where
\begin{align}
\label{eq.gauge_bosons_L-partons}
 \mathcal{L}_{b} & = \frac{i}{\pi}\partial_{x}\theta_{\tilde{y}}\left(\partial_{\tau}\varphi_{\tilde{y}} - a_{0,\tilde{y}} \right) + \frac{v_B}{2\pi} \left(\partial_x \varphi_{\tilde{y}} - a_{1,\tilde{y}}\right)^2 + \frac{u_B}{2\pi}\left(\partial_x\theta_{\tilde{y}}\right)^2 \\
 \mathcal{L}_{\text{Maxwell}} & = \frac{\kappa}{4\pi \tilde{v}} \left[\Delta a_{0}\right]_{y}^2 + \frac{\kappa \tilde{v }}{4\pi}\left[\Delta a_{1}\right]^2_y~.
\end{align}
We here omit the redundant label $\sigma=\left(-1\right)^{y+1}$ to lighten the notation.
The final term, $\mathcal{L}'_{b}$, contains exponentially decaying interwire interactions. Specifically, density-density and current-current interactions
\begin{align}
\label{eq.gauge_bosons_L-partons-prime}
 \mathcal{L}'_{b}=\sum_{y'} \left\{\left[\Delta\partial_x\theta\right]_{y}\left(W_1\right)_{y,y'}\left[\Delta\partial_x\theta\right]_{y'} + \left[\Delta \partial_x \varphi\right]_{ y }\left(W_2\right)_{ y, y'} \left[\Delta \partial_x \varphi\right]_{ y'} \right\}~,
\end{align}
where $W_i$ decay exponentially with $\left|y-y'\right|$.
Crucially, ${\cal L}'_{b}$ is short-ranged in both parton and spin variables and thus does not affect the structure of the gauge theory.
Performing the Gaussian integrals over $a_\mu$ and expressing $\varphi, \theta$ in terms of $\Phi, \Theta$ using Eq.~\eqref{eq.CW_coupled_VBS_partons}, we find $\mathcal{S} = \int_{x,\tau} \sum\nolimits_y \left[\mathcal{L}_{\text{spins}}+ \mathcal{L}'_{\text{spins}}\right]$ with
\begin{align}
\label{eq.gauge_bosons_L-microscopic}
 \mathcal{L}_{\text{spins}}& =\frac{i}{\pi} \partial_x \Theta_y \partial_\tau \Phi_y + \frac{v K}{2\pi} \left(\partial_x \Phi_y\right)^2 +\frac{v}{2\pi K}\left(\partial_x \Theta_y\right)^2 + \frac{u_B}{2\pi}\left[S \partial_x \Theta\right]_{\tilde{y}}^2~,
\end{align}
where $K = \frac{\kappa}{2} \sqrt{\frac{v_B}{v_B + 2\kappa \tilde{v}}}$ and $v = 2 \tilde v K/\kappa$. The term ${\cal L}'_\text{spins}$ contains exponentially decaying interwire interactions, i.e.,
\begin{align}
 \nonumber \mathcal{L}_{\text{spins}}' = & 
 \frac{v^2 K^2}{2 \pi v_B }\sum_{\tilde{y}'} \left[\Delta\partial_x \Phi\right]_{\tilde y} \left[1-\frac{vK}{v_B}\Delta^T\Delta\right]^{-1}_{\tilde{y},\tilde{y}'}\left[\Delta \partial_x \Phi\right]_{\tilde y'}+ \sum_{ y'} \partial_x \Phi_{ y }\left(PW_2P\right)_{ y,y'} \partial_x \Phi_{ y'} \\
 & + \sum_{ y'} \left[S^T S \partial_x \Theta\right]_{ y }\left(PW_1P\right)_{ y, y'} \left[S^T S \partial_x \Theta\right]_{y'}~,
\end{align}
where $P_{y,y'}=\left(-1\right)^{y}\delta_{y,y'}$. Since $\frac{vK}{v_B} < \frac{1}{4}$ for any choice of parameters, all terms in $ \mathcal{L}_{\text{spins}}'$ decay exponentially with $\left|y-y'\right|$; a suitable choice of $W_i$ in Eq.~\eqref{eq.gauge_bosons_L-partons-prime} can thus be used to achieve ${\cal L}_\text{spins}'=0$. Alternatively, $W_i$ can be chosen such that ${\cal L}_\text{spins}'=\frac{u_V}{8\pi}\left[\Delta \partial_x \Phi \right]_{\tilde{y}}^2$, which is the result stated in the main text, i.e., Eqs.~\eqref{eq.CW_decoupled_LL} and \eqref{eq.BP_symm+theory_inter}.

\subsection*{External probing field}
Before concluding this appendix, we include the external probing field $\vec{A}$ that minimally couples to the conserved $S^z$ of the microscopic spins. Translating $\mathcal{L}_{A}$ of Eq.~\eqref{eq.BP_symm+theory_A-microscopic} to bosonic parton variables we find
\begin{align}
 \mathcal{L}_{A} = & \frac{i}{\pi}(-1)^{\tilde{y}} \partial_x \theta_{\tilde{y}} \left[S^{-T} A_{0}\right]_{\tilde{y}} + \frac{v K}{\pi}(-1)^{\tilde{y}}\partial_x\varphi_{\tilde{y}} \left[S A_{1}\right]_{\tilde{y}} + \frac{v K}{2\pi}A_{1,y}^2~,
\end{align} 
where $S^{-T}_{\tilde{y},y}$ is the inverse-transpose of $S_{\tilde{y},y}$. To cast the coupling into a more revealing form, we shift $a_\mu$ according to 
\begin{subequations} 
\begin{align}
 &a_{0,\tilde{y}}\rightarrow a_{0,\tilde{y}}-\frac{1}{4}(-1)^{\tilde{y}}\left[S A_{0}\right]_{\tilde{y}} +(-1)^{\tilde{y}}\left[S^{-T} A_{0}\right]_{\tilde{y}}~, \\
 &a_{1,\tilde{y}}\rightarrow a_{1,\tilde{y}}-\left(\frac{1}{4}-\frac{{v} K}{v_B}\right)(-1)^{\tilde{y}} \left[S A_{1}\right]_{\tilde{y}}~.
\end{align}
\end{subequations}
This shift completely cancels ${\cal L}_A$; the external probing field instead appears in $\mathcal{L}_{b}$ of Eq.~\eqref{eq.gauge_bosons_L-partons} through the replacement $a_{\mu,\tilde{y}} \rightarrow a_{\mu,\tilde{y}} - \frac{1}{4}\left(-1\right)^{\tilde{y}}\left[S A_{\mu}\right]_{\tilde{y}}$ and in higher-order couplings to the emergent gauge field given by
\begin{align}
 \mathcal{L}_{a-A} = \frac{K}{\pi v} \left(-1\right)^y \left[\Delta^T \Delta A_{0}\right]_y \left[\Delta a_0\right]_y + \frac{v K}{4\pi} \left(-1\right)^y \left[\Delta^T \Delta A_{1}\right]_y \left[\Delta a_1\right]_y~.
\end{align} 

\section{Derivation of the fermionic-parton field theory}
\label{app.gauge_fermions}
To derive the gauge theory for the fermionic partons, we follow the same steps as for the bosonic partons. Specifically, we consider the action $\mathcal{S}=\int_{x,\tau}\sum\nolimits_y\left[\mathcal{L}_{f}+\mathcal{L}_{\text{Maxwell}}+\mathcal{L}'_{f} + \mathcal{L}' \right]$ with
\begin{align}
\label{eq.gauge_fermions_L}
 \mathcal{L}_{f} = & \frac{i}{\pi}\partial_{x}\theta_{f,\tilde{y}}\left(\partial_{\tau}\varphi_{f,\tilde{y}} - a_{0,\tilde{y}} \right) + \frac{v_F}{2\pi} \left(\partial_x \varphi_{f,\tilde{y}} - a_{1,\tilde{y}}\right)^2 + \frac{u_F}{2\pi}\left(\partial_x\theta_{f,\tilde{y}}\right)^2~, \\
\label{eq.gauge_fermions_L-Maxwell}
 \mathcal{L}_{\text{Maxwell}} = & \frac{\kappa}{8\pi \tilde{v}} \left[\Delta a_{0}\right]_{y}^2 + \frac{\kappa \tilde{v}}{8\pi}\left[\Delta a_{1}\right]_y^2 ~.
\end{align}
 The first `primed' term, ${\cal L}'_f$, contains interwire interactions between fermion densities and currents, that decay exponentially with the wire separation, $\left|y-y'\right|$. This term is not essential for understanding the structure of the gauge theory. For completeness, we still provide an explicit treatment below. The final term is
\begin{align}
\label{eq.gauge_fermions_L-alpha}
 \mathcal{L}' = & \frac{i}{4\pi} \alpha \left(-1\right)^{y} \left[\Delta a_1\right]_y \left[\Delta a_0\right]_y~.
\end{align}
It has an alternating sign and thus does not affect the long-distance behavior of the photon propagator (see Appendix \ref{app.gauge-calc_QSH} below). However, as we will see, it is essential for the microscopic time-reversal symmetry of the spin system. To understand its origin, recall that the fermionic partons are obtained by attaching fluxes of opposite sign to the two bosonic-parton species. In the schematic continuum manipulations described by Eq.~\eqref{eq.FP_field-theory-flux-attachment}, the two Chern-Simons terms that implement this flux attachment exactly cancel. By contrast, in the wire regularization, the two species of fermionic partons reside at different locations (even vs. odd dual wires). Consequently, the cancellation is imperfect, and a residual term $\mathcal{L}_{\text{CS,even}} + \mathcal{L}_{\text{CS,odd}} \propto \mathcal{L}' +(-1)^{\tilde y}\mathcal{O}(\Delta^4)$ remains. 

To map the gauge theory to spin variables, we perform the Gaussian integrals over $a_\mu$ and express $\varphi_f$, $\theta_f$ in terms of $\Phi$, $\Theta$ using 
\begin{align}
\label{eq.gauge_fermions_Parton2Micro}
 \varphi_{f,\tilde{y}} = & - \left(-1\right)^{\tilde{y}} \left[ S^{-T}\Phi\right]_{\tilde{y}} + 2\left[\Delta^{-T}\Theta\right]_{\tilde{y}} - \left[\Delta\Theta\right]_{\tilde{y}}~,\\ \theta_{f,\tilde{y}} = & - \left(-1\right)^{\tilde{y}}\left[ S\Theta\right]_{\tilde{y}}~,
\end{align}
with $\left(\Delta^{-T}\right)_{\tilde{y},y'}$ the inverse-transpose of $\Delta_{\tilde{y},y'}$.
We find $\mathcal{S} = \int_{x,\tau} \sum\nolimits_y \left[ \mathcal{L}_{\text{spins}}+ \mathcal{L}_{\text{spins}}'+ \mathcal{L}_{\text{spins}}'\right]$, where
\begin{align}
\label{eq.gauge_fermions_L-microscopic}
 \mathcal{L}_{\text{spins}}& =\frac{i}{\pi} \partial_x \Theta_y \partial_\tau \Phi_y+ \frac{v K}{2\pi} \left(\partial_x \Phi_y\right)^2 +\frac{v}{2\pi K}\left(\partial_x \Theta_y\right)^2 + \frac{u_V}{4\pi} \left[\Delta \partial_x \Phi\right]_{\tilde{y}} + \frac{\overline{u_F}}{2\pi}\left[S \partial_x \Theta\right]_{\tilde{y}}^2~,
\end{align}
with parameters $K=\frac{\alpha}{2\kappa}\sqrt{\frac{\kappa v_F}{\kappa v_F + \left(\alpha^2+\kappa^2\right) \tilde{v}}}$, $v=2 \tilde{v} \kappa K/\alpha $, $u_V = v_F \left(\frac{\alpha \tilde{v}}{2\alpha \tilde{v}+\kappa v_F}\right)^2$, and $\overline{u_F} = u_F + 4 v K$. The term ${\cal L}'_\text{spins}$ contains exponentially decaying interwire terms whose explicit form is provided below. Crucially, a suitable choice of the ${\cal L}_\text{parton}'$ results in ${\cal L}'_\text{spins}=0$, and can additionally be used to tune the parameters $u_V$ and $\overline{u_F}$ in Eq.~\eqref{eq.gauge_fermions_L-microscopic}. The final term
\begin{align}
 \mathcal{L}_{\text{spins}}'& = \frac{\overline{v}}{2\pi} \left(-1\right)^{y} \partial_x\Theta_y \partial_x\Phi_{y}~,
\end{align}
with $\overline{v}=4v K\bigl( \frac{2\alpha}{\alpha^2+\kappa^2}-1\bigr)$, cannot be eliminated by any choice of ${\cal L}_\text{parton}'$, which contains only higher orders in $\Delta$. Moreover, it violates the microscopic time-reversal symmetry. To preserve this symmetry, we must, therefore, choose $\alpha$ in the gauge theory of Eq.~\eqref{eq.gauge_fermions_L-alpha} such that $\overline{v}=0$, namely, $\alpha=1\pm \sqrt{1-\kappa^2} \equiv \alpha_0$. Having eliminated ${\cal L}'_\text{spins}$ and ${\cal L}'_\text{spins}$, the action matches Eqs.~\eqref{eq.CW_decoupled_LL} and \eqref{eq.BP_symm+theory_inter} with the parameters specified above.

\subsection*{Explicit form of short-range interactions}
We include generic short-range density-density, current-current and density-current interactions for the fermionic parton in the form 
\begin{align}
\label{eq.gauge_fermions_L-exp-decay}
 {\cal L}'_\text{parton} = \sum_{\tilde y'} \left\{ \left[\Delta \partial_x \theta_f\right]_{y}\left(W_1\right)_{y,y'} \left[\Delta\partial_x \theta_f\right]_{y'} + \left[S^T \Delta \partial_x \varphi_f\right]_{\tilde y } \left(W_2\right)_{\tilde y,\tilde y'} \left[S^T \Delta \partial_x \varphi_f\right]_{\tilde y'} + \left[\Delta \partial_x \theta_f\right]_{y} \left(W_3 \right)_{y,y'} \left[P\Delta \partial_x \varphi_f\right]_{ y'}\right\}~,
\end{align}
where $W_i$ decay exponentially with $\left|y-y'\right|$. It is straightforward to express these in terms of $\Phi,\Theta$ and combine them with the exponentially decaying interactions stemming from the Gaussian integrals over $a_\mu$. We find \begin{align}
 {\cal L}'_\text{spin} = \frac{v_F}{2\pi}\sum_{\tilde{y}'} \left\{ \left[S \partial_x \Theta\right]_{\tilde{y}} (\widetilde W_1)_{\tilde{y},\tilde{y}'} \left[S \partial_x \Theta\right]_{\tilde{y}'} +\left[\Delta
 \partial_x\Phi\right]_{\tilde{y}} (\widetilde W_2)_{\tilde{y},\tilde{y}'} \left[\Delta \partial_x\Phi\right]_{\tilde{y}'} - \left[S \partial_x \Theta\right]_{\tilde{y}} (\widetilde W_3)_{\tilde{y},\tilde{y}'} \left[P \Delta \partial_x \Phi\right]_{\tilde{y}'} \right\}~,
\end{align}
with exponentially decaying $\widetilde W_i$. We are interested in $\overline v=0$, for which these Kernels are given by 
\begin{align}
 \widetilde W_1 = & S \left( V_2 - \Delta^T W_2\Delta + P W_1 P + P W_3 P \right) S^T + 4 W_2 - 2 PW_3P -4g~, \\
 \widetilde W_2 = & V_1 + P W_2 P ~, \\
 \widetilde{W_3} = & 2 \left[ V_2 + \left(2-\Delta^T\Delta\right) W_2 + \frac{1}{2} P W_3 P\right]~,
\end{align}
where $g=\frac{v K}{v_F}$, $V_1=g^3\Delta^T\left[1-g\Delta^T \Delta \right]^{-1}\Delta$, and $V_2 = \frac{g }{1- 4g}\left[1+\frac{g }{1-4g}\Delta^T \Delta\right]^{-1}$. A choice of exponentially decaying $W_i$ that results in $\mathcal{L}'_{\text{spins}}=0$ is given by 
\begin{align}
 W_1 = & 2 g \left(1-2g-2g^{2}+g^{2}\Delta^{T}\Delta-\frac{1}{2}g^{2}\Delta^{T}\Delta\Delta^{T}\Delta\right)\left[1-g\Delta^{T}\Delta\right]^{-1}~, \\
 W_2 = & -\frac{g^3}{1-4g} S^TS \left[1+\frac{g}{1-4g} \Delta^T \Delta \right]^{-1} ~, \\ 
 W_3 = & -2g\left[1+g^{2}\Delta^{T}\Delta\left(2-\Delta^{T}\Delta\right)\right]\left[1-g\Delta^{T}\Delta\right]^{-1}~.
\end{align}

\subsection*{External probing field}
Before concluding the derivation, we also include an external probing field $\vec{A}$ that minimally couples to the conserved $S^z$ of the microscopic spins. We thus supplement the action of the gauge theory action by $\mathcal{L}_{A}$ as given in Eq.~\eqref{eq.BP_symm+theory_A-microscopic}. In terms of the fermionic parton variables
\begin{align}
 \mathcal{L}_{A} = \frac{i}{\pi}\left(-1\right)^{\tilde{y}}\partial_x\theta_{f,\tilde{y}} \left[S^{-T} A_{0}\right]_{\tilde{y}} + \frac{v K}{\pi} \left(-1\right)^{\tilde{y}}\partial_x\varphi_{f,\tilde{y}}\left[S A_1\right]_{\tilde{y}} - \frac{v K}{\pi}\partial_x\theta_{f,\tilde{y}}\left[\left(2-\Delta^T\Delta\right)\Delta^{-T} A_1\right]_{\tilde{y}} +\frac{v K}{2\pi} A_{1,y}^2~.
\end{align} 
To cast this into a more revealing form, we shift $a_\mu$ according to 
\begin{subequations} 
\begin{align}
 &a_{0,\tilde{y}} \rightarrow a_{0,\tilde{y}} -\frac{1}{4}\left(-1\right)^{\tilde{y}}\left[S A_0\right]_{\tilde{y}} + \left(-1\right)^{\tilde{y}} \left[S^{-T}A_0\right]_{\tilde{y}} + i v K \left[\left(2-\Delta^T\Delta\right)\Delta^{-T}A_1\right]_{\tilde{y}} \\
 &a_{1,\tilde{y}} \rightarrow a_{1,\tilde{y}} - \left(\frac{1}{4}-\frac{v K}{v_F}\right)\left(-1\right)^{\tilde{y}}\left[S A_1\right]_{\tilde{y}}~.
\end{align}
\end{subequations}
This completely cancels ${\cal L}_A$; the external probing field instead appears in $\mathcal{L}_{f}$ of Eq.~\eqref{eq.gauge_fermions_L} through the replacement $a_{\mu,\tilde{y}} \rightarrow a_{\mu,\tilde{y}} - \frac{1}{4}\left(-1\right)^{\tilde{y}}\left[S A_{\mu}\right]_{\tilde{y}}$ and in higher-order couplings to the emergent gauge field given by
\begin{align}
 \nonumber \mathcal{L}_{a-A} = & -\frac{K \left(2-\alpha\right) \left(-1\right)^y}{8\pi v}\left[\Delta^{T}a_{0}\right]_{y}\left[\Delta^{T}\Delta A_{0}\right]_{y} -\frac{v K \left(2+\a\right)\left(-1\right)^y}{8\pi}\left[\Delta^{T}a_{1}\right]_{y}\left[\Delta^{T}\Delta A_{1}\right]_{y} \\
 & -\frac{i K^2}{4\pi}\left[\Delta^{T}a_{0}\right]_{y}\left[\Delta^{T}\Delta A_{1}\right]_{y} + \frac{i\a}{16\pi}\left[\Delta^{T}a_{1}\right]_{y}\left[\Delta^{T}\Delta A_{0}\right]_{y}~.
\end{align}

\section{Gauge-theory calculations}
\label{app.gauge-calc}
\subsection*{Bosonic parton superfluid}
\label{app.gauge-calc_CondBosons}
To analyze the implications of a condensed bosonic parton, first recall that the gauge theory is of the form ${\cal L}=\sum \nolimits_{\sigma}{\cal L}_\sigma\left[b_\sigma,a_\sigma+\frac{\sigma}{2}\mathsf{A}_\sigma\right] + {\cal L}_\text{Maxwell}[a_\uparrow,a_\downarrow]$, where $\vec{\mathsf{A}}_{\tilde y} \equiv \bigl(\vec{A}_{\tilde y-1/2}+\vec{A}_{\tilde y+1/2}\bigr)/2 $ and we have endowed $\vec{a}$ and $\vec{\mathsf{A}}$ with the (redundant) spin label corresponding to the dual-wire parity. The two species couple only through the Maxwell term
\small
\begin{align}
 \mathcal{L}_{\text{Maxwell}} = \frac{\kappa}{4\pi} \left\{\frac{1}{\tilde{v}} \left[\left(a_{0,2\tilde{y},\down} - a_{0,2\tilde{y}-1,\up}\right)^2 + \left(a_{0,2\tilde{y}+1,\up} - a_{0,2\tilde{y},\down}\right)^2 \right] + \tilde{v} \left[\left(a_{1,2\tilde{y},\down} - a_{1,2\tilde{y}-1,\up}\right)^2 + \left(a_{1,2\tilde{y}+1,\up} - a_{1,2\tilde{y},\down}\right)^2 \right] \right\}~.
\end{align}
\normalsize
(Additional short-range interactions between opposite species can always be subsumed by adding higher-order  derivatives to the Maxwell term.) Condensation of $b_\downarrow$ corresponds to $\delta{\cal L}_{\down} = g_t \cos\left(\varphi_{2\tilde{y}+2,\down}-\varphi_{2\tilde{y},\down}\right)$ reaching strong coupling first. We introduce a dimensionless coupling constant with bare value $\tilde{g}_t=\pi d_0^2 g_t/v_B$ and replace \begin{align}
 \delta {\cal L}_{\down} = \frac{v_B \tilde{g}_t}{\pi d_0^2}\cos{(\left[\Delta_2 \varphi_\down\right]_{2\tilde{y}+1})} \rightarrow \frac{v_B}{2\pi d_*^2}\left[\Delta_2 \varphi_\down\right]_{2\tilde{y}+1}^2~.
\end{align}
Here, $d_*$ is the length scale at which the coupling constant renormalized to order unity. Upon integrating out $\theta_\downarrow$ the Lagrangian governing $\varphi_\downarrow$ is given by
\begin{align}
 \mathcal{L}_{\down} + \delta\mathcal{L}_{\down} =\frac{1}{2\pi u_B}\left|i\omega \varphi_{\down}-a_{0,\down} + \textstyle\frac{1}{2}\mathsf{A}_{0,\down}\right|^2 + \frac{v_B}{2\pi}\left|ik_x \varphi_{\down}-a_{1,\down} + \textstyle\frac{1}{2}\mathsf{A}_{1,\down}\right|^2 + \frac{v_B}{2\pi }\left|\Delta_{k_y} \varphi_{\down}\right|^2~,
\end{align}
where we have Fourier transformed using a two-wire unit cell. 
Further performing the Gaussian integral over $\varphi_\down$ we obtain
${\cal L}_\text{ind}\left[a_\down - \frac{1}{2}\mathsf{A}_\down\right] =\frac{1}{2\pi} (a_\downarrow^*-\frac{1}{2}\mathsf{A}_\downarrow^*)_\mu \Pi^{\text{ind}}_{\mu \nu} (a_\downarrow-\frac{1}{2}\mathsf{A}_\downarrow)_\nu$ with 
\begin{align}
 \Pi^{\text{ind}} = \frac{1 }{\omega^{2}+u_B v_Bk_x^{2}+ u_B v_B|\Delta_{k_y}|^2} \begin{pmatrix}
 v_B k_x^2 + v_B |\Delta_{k_y}|^2 & -v_B \omega k_x \\
 -v_B \omega k_x & v_B \omega^2 + u_B v_B^2 |\Delta_{k_y}|^2
 \end{pmatrix}~.
\end{align}
This induced action supplements the bare $\mathcal{L}_{\text{Maxwell}}$, which in Fourier space reads as
\begin{align}
 \mathcal{L}_{\text{Maxwell}}= \frac{\kappa}{4\pi} \left[ \frac{1}{\tilde{v}} \left( \left|a_{0,\down}-e^{-i d_0 k_y }a_{0,\up}\right|^{2}+\left|a_{0,\up}-a_{0,\down}\right|^{2}\right) + \tilde{v}\left(\left|a_{1,\down}-e^{-i d_0 k_y}a_{1,\up}\right|^{2}+\left|a_{1,\up}-a_{1,\down}\right|^{2}\right) \right]~.
\end{align} 
We now shift $a_{\mu,\sigma} \rightarrow a_{\mu,\sigma} + \frac{1}{2} \mathsf{A}_{\mu,\down}$ such that $\mathsf{A}_{\mu,\down}$ is eliminated from $\mathcal{L}_\text{ind}$. It instead enters $\mathcal{L}_\up$ and, at subleading order $|\Delta_{k_y}|^2\mathsf{A}_{\mu,\downarrow}$, the Maxwell term ${\cal L}_\text{Maxwell}$. Finally, performing the integral over $a_{\mu,\downarrow}$ and retaining only the leading orders in $\Delta_{k_y}$ we arrive at the effective action ${\cal L}_\text{eff} = {\cal L}_\uparrow[b_\uparrow,a_\uparrow+\frac{\mathsf{A}_\uparrow + \mathsf{A}_\downarrow}{2}]+\frac{1}{2\pi} \gamma_\mu a_{\mu,\up}^* \left[ \Pi^{(0)} + \Pi^{(2)} \right]_{\mu\nu} \gamma_\nu a_{\nu,\up}$ with 
\begin{align}
 \Pi^{(0)} = &\frac{1}{\omega^2+c_x^2 k_x^2 + c_y^2 \left|\Delta_{k_y}\right|^2 }
 \begin{pmatrix}
 c_x^2 k_x^2 + c_y^2 \left|\Delta_{k_y}\right|^2 & -c_x k_x \omega \\
 -c_x k_x \omega & c_y^2 \left|\Delta_{k_y}\right|^2 + \omega^2 
 \end{pmatrix}~, \\
 \Pi^{(2)} = & \frac{\left|\Delta_{k_y}\right|^{2}}{4\left[\omega^{2}+c_{1}^{2}k_x^{2}+c_{2}^{2}\left|\Delta_{k_y}\right|^{2}\right]}
 \begin{pmatrix}
 \frac{\kappa u_{B}}{\tilde{v}}\left[c_{1}^{2}k_x^{2}+c_{2}^{2}\left|\Delta_{k_y}\right|^{2}+\frac{\tilde{} u_{B}+\tilde{v}}{\kappa u_{B}}\omega^{2}\right] & c_{1}k_x\omega \\ 
 c_{1}k_x\omega & \frac{\kappa\tilde{v}}{v_{B}}\left[\frac{v_{B}+\kappa\tilde{v}}{\kappa\tilde{v}}c_{1}^{2}k_x^{2}+c_{2}^{2}\left|\Delta_{k_y}\right|^{2}+\omega^{2}\right]
 \end{pmatrix}~,
\end{align}
where $c_x^2 = \frac{v_B \tilde{v}\left(\kappa u_B + \tilde{v}\right)}{v_B+\kappa \tilde{v}}$, $\gamma_0 = \sqrt{\frac{\kappa}{\left(\kappa u_B + \tilde{v} \right)}}$, $c_y^2 = \frac{v_B}{\gamma_0^2}$, and $\gamma_1 = \gamma_0 c_x$. The analytically continued gauge-field propagator has poles at real frequencies $\omega = \pm \sqrt{\tilde{v}^{2}k_x^{2} + v_B u_{B}\left|\Delta_{k_y}\right|^{2}+\frac{4 v_B\tilde{v}}{\kappa d_*^2}}$. For any finite $d_*$, the gauge field $\vec{a}_\up$ is gapped; integrating it out results in ${\cal L}_{\up}[b_\up, \frac{1}{2} \mathsf{A}_\up + \frac{1}{2} \mathsf{A}_\down ]$, up to renormalization of parameters and short-range interactions. Crucially, $b_\up$ now couples to the external probing field with unit charge, i.e., carries spin 1. 

Finally, consider the case where the second parton also condenses, i.e., where $\delta\mathcal{L}_{\up} \sim \cos\left(\varphi_{2\tilde{y}+1,\up}-\varphi_{2\tilde{y}-1,\up}\right)$ flows to strong coupling as well. Treating ${\delta \cal L}_{\uparrow}$ in the same way as $\delta{\cal L}_{\downarrow}$ and integrating out $b_\uparrow$, we find
\begin{align}
 \mathcal{L}_{\text{Meissner}} =\frac{v_B}{2\pi } \bar{A}_{\mu} \left[\delta^{\mu\nu} - \frac{p^\mu p^\nu}{\vec p \cdot \vec p}\right]\bar{A}_{\nu} +\mathcal{O}(p^2)~,
\end{align}
where $\vec {\bar{A}} = \bigl(A_0 \frac{1}{\sqrt{u_B v_B}},\vect A\bigr)$ and $\vec{p} = \bigl(\frac{\omega}{\sqrt{v_B u_B}},~k_x,~k_y\bigr)$. This response to the probing field $\vec A$ implies, that the $U(1)$ spin-rotation symmetry is spontaneously broken.

\subsection*{Bosonic-parton quantum Hall state}
\label{app.gauge-calc_BIQH}
The Lagrangian for the quantum Hall state of bosonic partons, described in Sec.~\ref{subsubsec.BP_phases_QH-CSL}, is
\begin{align}
 \mathcal{L}_\text{QH} = \frac{i}{\pi}\partial_x \theta_{\tilde{y}} \left(\partial_\tau \varphi_{\tilde{y}} -a_{0,\tilde{y}}\right) + \frac{v_{B}}{2\pi} \left(\partial_{x}\varphi_{\tilde{y}} - a_{1,\tilde{y}}\right)^{2} + \frac{u_{B}}{2\pi}\left(\partial_{x}\theta_{\tilde{y}}\right)^{2} + g_{\text{QH}}\cos\left(S^T \Delta \varphi_{\tilde{y}}-2n\theta_{\tilde{y}}\right) + \mathcal{L}_{\text{Maxwell}}~,
\end{align}
with $\mathcal{L}_{\text{Maxwell}} = \frac{\kappa}{4\pi\tilde{v}}\left[\Delta a_{0}\right]_y^{2}+\frac{\kappa\tilde{v}}{4\pi}\left[\Delta a_{1}\right]_y^{2}$. 
To integrate out the bosonic partons, we introduce new conjugate variables that diagonalize the interwire interaction, i.e., $\bar{\theta}_{\tilde{y}}=\frac{1}{2}\left(S^T\Delta \varphi_{\tilde{y}}\right)+n\theta_{\tilde{y}}$ and $\bar{\varphi}_{\tilde{y}}=\frac{1}{n}\varphi_{\tilde{y}}$. In terms of $\bar \theta$ and $\bar \varphi$, the Lagrangian is 
\begin{align}
 \nonumber \mathcal{L}_\text{QH} = & \frac{i}{\pi}\partial_x \bar{\theta}_{\tilde{y}} \left( \partial_\tau \bar{\varphi}_{\tilde{y}}- a_{0,\tilde{y}}\right) + \frac{v_{B}}{2\pi}\left(n\partial_{x}\bar{\varphi}_{\tilde{y}}-a_{1,\tilde{y}}\right)^{2} + \frac{u_{B}}{2\pi n^{2}}\left(\partial_{x}\bar{\theta}_{\tilde{y}}\right)^{2} + g_{\text{QH}} \cos\left(2\bar{\theta}_{\tilde{y}}\right) \\
 & + \frac{i}{2\pi}S^{T}\Delta\partial_{x}\bar{\varphi}_{\tilde{y}}a_{0,\tilde{y}} + \frac{u_{B}}{8\pi}\left(S^{T}\Delta\partial_{x}\bar{\varphi}_{\tilde{y}}\right)^{2}-\frac{u_{B}}{2\pi n}\partial_{x}\bar{\theta}_{\tilde{y}}S^{T}\Delta\partial_{x}\bar{\varphi}_{\tilde{y}} + \mathcal{L}_{\text{Maxwell}}~.
\end{align}
We introduce the dimensionless coupling constant $\tilde{g}_{QH} = 8\pi d_0^2 g_{\text{QH}}/v_B$ and replace $\frac{v_B \tilde{g}_{QH}}{8\pi d_0^2}\cos\left(2\bar{\theta}_{\tilde{y}}\right)\rightarrow \frac{v_B}{2\pi d_*^2}\bar{\theta}_{\tilde{y}}^2$, where $d_*$ is the length scale at which it renormalizes to order unity. Performing the Gaussian integral over $\bar{\varphi}$ we find, to lowest order in derivatives, 
\begin{align}
 \mathcal{L}'_\text{QH} = & \frac{i}{\pi n}\bar{\theta}_{\tilde{y}}\left(\partial_{x}a_{0,\tilde{y}}-\partial_{\tau}a_{1,\tilde{y}}\right)+\frac{v_B}{2\pi d_*^2}\bar{\theta}_{\tilde{y}}^{2} + \frac{i}{2\pi n}\left[Sa_{1}\right]_y\left[\Delta a_{0}\right]_y + \frac{1}{2\pi v_B n^2}\left[\Delta a_{0}\right]_y^{2}+\frac{u_{B}}{2\pi n^{2}}\left[\Delta a_{1}\right]_y^{2}~.
\end{align}
Finally, we perform the integral over $\bar{\theta}$ and obtain the effective gauge theory 
\begin{align}
 \mathcal{L}_{\text{gauge}} = \frac{i}{2\pi n}\left[Sa_{1}\right]_y \left[\Delta a_{0}\right]_y + \left(\frac{\kappa}{4\pi\tilde{v}}+\frac{1}{2\pi v_B n^2}\right)\left[\Delta a_{0}\right]_y^{2}+\left(\frac{\kappa\tilde{v}}{4\pi}+\frac{u_{B}}{2\pi n^{2}}\right)\left[\Delta a_{1}\right]_y^{2} + \frac{d_*^2}{2\pi v_B n^2}\left(\partial_{x}a_{0,\tilde{y}}-\partial_{\tau}a_{1,\tilde{y}}\right)^2~.
\end{align}

\subsection*{Fermionic partons QSH}
\label{app.gauge-calc_QSH}
To assess the relevance of monopoles in the QSH phase of fermionic partons we compute their correlation function within the monopole-free theory. The corresponding Lagrangian was introduced in Sec.~\ref{subsubsec.FP_phases_QSH-AFM}; it is given by 
$\mathcal{L}_{\text{QSH}} =\mathcal{L}_{\text{matter}} +\mathcal{L}_{\text{gauge}}$ with
\begin{align}
 \mathcal{L}_{\text{matter}} = & \frac{i}{\pi}\partial_{x}\theta_{f,\tilde{y},\sigma}\left(\partial_\tau \varphi_{f,\tilde{y},\sigma} -a_{0,\tilde{y}}\right)+\frac{v_{F}}{2\pi}\left(\partial_{x}\varphi_{f,\tilde{y},\sigma}-a_{1,\tilde{y}}\right)^{2}+\frac{u_{F}}{2\pi}\left(\partial_{x}\theta_{f,\tilde{y},\sigma}\right)^{2} + \frac{\tilde{v}}{8\pi \kappa d_*^2} \left[S_2 \theta_{f,\sigma} - P \Delta_2 \varphi_{f,\sigma}\right]^2_{\tilde{y}} \\
 \mathcal{L}_{\text{gauge}} =& \frac{\kappa}{8\pi \tilde{v}} \left[\Delta a_{0}\right]_{y}^2 + \frac{\kappa\tilde{v}}{8\pi} \left[\Delta a_{1}\right]_{y}^2 + \frac{i}{4\pi} \alpha \left(-1\right)^{y} \left[\Delta a_1\right]_y \left[\Delta a_0\right]_y~,
\end{align}
where, as usual, $d_*$ is the length scale beyond which the quadratic approximation of the cosine is appropriate. (For the last term in $\mathcal{L}_{\text{gauge}}$, cf.~Appendix \ref{app.gauge_fermions}.) It is straightforward to integrate out the matter fields and obtain the induced gauge-field action. Upon expanding in frequencies and $x$-momenta small compared to the QSH gap, $E_{\text{QSH}} \equiv \sqrt{\frac{4 \tilde{v} v_F}{\kappa d_*^2}}$, we find
\begin{align}
\label{eq.gauge-calc_QSH_L-gauge-with-alt-0}
 \mathcal{L}_{\text{ind}} = &\frac{1}{8\pi v_{F}}\left[\Delta_{2}a_{0}\right]_{\tilde{y}}^{2}+\frac{u_{F}}{8\pi}\left[\Delta_{2}a_{1}\right]_{\tilde{y}}^{2}+\frac{\kappa d_*^2}{8\pi \tilde{v}}\left(\partial_{x}a_{0,\tilde{y}}-\partial_{\tau}a_{1,\tilde{y}}\right)^{2} + \frac{i}{4\pi}\left(-1\right)^{\tilde{y}}\left[S_{2}a_{1}\right]_{\tilde{y}} \left[\Delta_{2}a_{0}\right]_{\tilde{y}}+\mathcal{L}_{\text{ind}}'~.
\end{align}
Here $\mathcal{L}_{\text{ind}}'$ contains terms suppressed by at least two additional powers of $\Delta_2$. These contributions can easily be retained but do not affect any long-wavelength properties; we, therefore, discard them. In the absence of oscillatory terms $\sim (-1)^{\tilde y}$ in the total gauge-field action, ${\cal L}_\text{gauge}+{\cal L}_\text{ind}$, we approximate $\Delta_2 = 2 \Delta + \Delta^2\approx 2 \Delta$ for describing
long-wavelength properties. We thus obtain the effective gauge-field action
\begin{align}
\label{eq.gauge-calc_QSH_L-gauge}
 \mathcal{L}^{\text{eff}}_{\text{gauge}} = \frac{\kappa}{8\pi} \left\{\frac{1}{\tilde{v}}\gamma^{2}\left[\Delta a_{0}\right]^{2}_y+\tilde{v}\eta^2\left[\Delta a_{1} \right]^{2}_y+d_{*}^{2}\frac{1}{\tilde{v}}\left(\partial_x a_{0,\tilde{y}}-\partial_\tau a_{1,\tilde{y}}\right)^{2} \right\}~,
\end{align}
where $\gamma^2 =1+\frac{4\tilde{v}}{\kappa v_{F}}$ and $\eta^2 =1+\frac{4u_{F}}{\kappa\tilde{v}}$. 
Oscillatory terms $\sim (-1)^{\tilde y}$ couple gauge fields at $y$ momenta $k_y$ and $k_y+\pi/d_*$. We obtain the long-wavelength action in that case, by integrating out modes at momenta $|k_y| > \pi/2d_*$. Returning to real space, we again find $ \mathcal{L}^{\text{eff}}_{\text{gauge}}$, but with modified parameters $\gamma^{2}=\frac{\kappa^{2}+\left(2-\alpha\right)^{2}}{\kappa^{2}}+\frac{4\tilde{v}}{\kappa v_{F}}$, $\eta^{2}=\frac{\kappa^{2}+\left(2+\alpha\right)^{2}}{\kappa^{2}}+\frac{4u_{F}}{\kappa\tilde{v}}$.

To compute the correlation function of a probing monopole in this theory, we supplement $\mathcal{L}^\text{eff}_{\text{gauge}}$ by the monopole action of Eq.~\eqref{eq.FP_symm+theory_monopoles2} and integrate out $a_\mu$. At leading order in $\Delta_{k_y}$, we find
\begin{align}
\label{eq.gauge-calc_QSH_L-M-eff}
 \mathcal{L}_{\mathcal{M},\text{eff}} = \left[\frac{\tilde{v}}{2\pi\kappa}+\frac{\tilde{v}}{2\pi\kappa{\gamma}^{2}}\frac{{\eta}^{2}\tilde{v}^{2}\left|\Delta_{k_y}\right|^{2}+\omega^{2}}{{\eta}^{2}\tilde{v}^{2}\left|\Delta_{k_y}\right|^{2}+\omega^{2}+\frac{{\eta}^2}{{\gamma}^{2}}\tilde{v}^{2}k_x^{2}}\right]k_x^{2}\left|\phi_{\mathcal{M}}\right|^{2}+ \frac{g_{\mathcal{M}}}{2} \left[\mathcal{M}_{-\vec{k}}\mathcal{M}_{\vec{k}} + \mathcal{M}^\dagger_{-\vec{k}}\mathcal{M}^\dagger_{\vec{k}}\right]~.
\end{align}
The monopole-monopole correlation function, evaluated within a monopole-free background $g_{\cal M}=0$, is given by
\begin{align}
\label{eq.gauge-calc_QSH_corr}
 C^\mathcal{M}_{\vect{R}} = \exp\left\{-\frac{\kappa d_*}{8\pi^2\tilde{v}}\frac{{\gamma}^{2}}{{\gamma}^{2}+1} \int_{\vect{k},\omega} \left[2-2\cos\left(\vect{k}\cdot\vect{R}\right)\right]\left[\frac{1}{k_x^2} + \mathcal{O}\left(1\right)\right] \right\} \sim
 \begin{cases}
 e^{-\frac{4}{\pi}\frac{{\gamma}^{2}}{{\gamma}^{2}+1} \left(\frac{x}{d_*}\right)^2} & y=0~, \\
 e^{-\frac{2}{\pi}\sqrt{\frac{\kappa\tilde{v}}{v_{F}}}\frac{{\gamma}^{2}}{{\gamma}^{2}+1}f\left(y\right)\frac{L}{d_*}} & y \neq 0~,
\end{cases}
\end{align}
where $L$ is the wire length, and the $\omega$ and $k_x$ integrals are cut off by $E_{\text{QSH}}$ and $v_F^{-1}E_{\text{QSH}}$, respectively. The function $f\left(y\right)\equiv 1- \frac{\sin\left(\pi y/2\right)}{\pi y/2}$ is bounded from below by $1-2/\pi \approx 0.36$ for non-zero $y$. At long distances, this correlation function decays faster than exponential. Monopoles described by ${\cal M}$ are thus strongly irrelevant.

We now turn to dressed monopoles, which were introduced in Sec.~\ref{subsubsec.FP_phases_QSH-AFM} via the short-hand $\mathcal{M}_\text{dressed} \sim f_\downarrow^\dagger f_\uparrow \mathcal{M}$. Their explicit expression is
\begin{align}
 \mathcal{M}_{\text{dressed},\vect r} = \begin{cases}
 \mathcal{M}_{\vect{r}}f^\dagger_{\vect{r} - \frac{1}{2}\hat{y},\up,L} f_{\vect{r} + \frac{1}{2}\hat{y},\down,L} & y \text{ even}~, \\
 \mathcal{M}_{\vect{r}}f^\dagger_{\vect{r} + \frac{1}{2}\hat{y},\up,R} f_{\vect{r} - \frac{1}{2}\hat{y},\down,R} & y \text{ odd}~.
 \end{cases}
\end{align}
It is instructive to relate its correlation function to configurations of the emergent gauge field $\vec{a}$. We therefore write
\begin{align}
 C^{\mathcal{M}}_{\text{dressed},\vect{r}-\vect{r'}} = \left< \mathcal{M}^\dagger_{\text{dressed},\vect{r}} \mathcal{M}_{\text{dressed},\vect{r}'}\right> \equiv \frac{\int\mathcal{D}\left[f_{\sigma},a,\phi_{\mathcal{M}}\right]\mathcal{M}^\dagger_{\text{dressed},\vect{r}} \mathcal{M}_{\text{dressed},\vect{r}'} e^{-\mathcal{S_{\mathcal{M}}-\mathcal{S}_{\text{QSH}}}}}{\int\mathcal{D}\left[f_{\sigma},a,\phi_{\mathcal{M}}\right]e^{-\mathcal{S_{\mathcal{M}}-\mathcal{S_{\text{QSH}}}}}}~,
\end{align}
and perform the integrals over both the fermions and the monopole field $\phi_\mathcal{M}$. To lowest order in $k_x$, we find 
\begin{align}
\label{eq.cmdressed}
 C^{\mathcal{M}}_{\text{dressed},\vect{R}} = & \exp\left\{-\frac{\kappa d_*}{16\pi^2\tilde{v}}{\gamma}^{2}\int_{\vect{k},\omega}\frac{2-2\cos\left(\vect{k}\cdot\vect{R}\right)}{k_x^{2}}\left[1-\frac{\kappa}{4\pi\tilde{v}}{\gamma}^{2}d_*^2\bigl\langle \left|\varepsilon_2\right|^{2}\bigr\rangle_0 \right]\right\}~,
\end{align}
where $\left<\ldots\right>_0$ is evaluated with respect to Eq.~\eqref{eq.gauge-calc_QSH_L-gauge} and $\varepsilon_2 = \Delta a_0/d_*$ is the $y$-component of the emergent electric field in the $a_2=0$ gauge. Consequently, the long-distance behavior of the dressed monopole-monopole correlation function matches the form obtained by evaluating Eq.~\eqref{eq.PC_gauge_monopole-corr} with the appropriate singular configuration, $\vec{a}^{\mathcal{M}}$ (cf.~Fig.~\ref{fig:monopoles}). Explicitly, we find
\begin{align}
 C^{\mathcal{M}}_{\text{dressed},\vect{R}} \sim \exp\left[\frac{\kappa {\eta} d_*}{4 \sqrt{x^2 + \left(\frac{d_* y}{{\gamma}}\right)^2}}\right]~.
\end{align}

We conclude by pointing out that $C^{\mathcal{M}}$ can also be expressed in the form of Eq.~\eqref{eq.cmdressed} but with $\gamma^2$ replaced by $\gamma^2 -\frac{4 \tilde v}{\kappa v_F}$. This apparently innocuous change spoils the cancellation between the bare $k_x^{-2}$ singularity and the $\langle|\varepsilon_2|^2 \rangle_0$ term, which requires its prefactor to match the one in ${\cal L}^\text{eff}_\text{gauge}$---changing one but not the other results in qualitatively different long-distance behavior [cf.~Eq.~\eqref{eq.FP_phases_QSH-AFM_naive-monopole-corr}, where we discarded oscillatory terms and hence found Eq.~\eqref{eq.cmdressed} with $\gamma=1$]. The presence or absence of the oscillatory terms in the total gauge-field action, ${\cal L}_\text{gauge}+{\cal L}_\text{ind}$, changes both of these factors equally and, therefore, does not qualitatively affect the monopole-monopole correlation function.

\end{widetext}

\end{document}